\documentclass[10pt,journal,compsoc]{IEEEtran}

\usepackage{amsmath,amsfonts}
\usepackage{array}
\usepackage[caption=false,font=normalsize,labelfont=sf,textfont=sf]{subfig}
\usepackage{textcomp}
\usepackage{stfloats}
\usepackage{url}
\usepackage{verbatim}
\usepackage{graphicx}
\usepackage{cite}
\usepackage[linesnumbered, ruled, vlined, commentsnumbered, boxed]{algorithm2e}
\SetKwRepeat{Do}{do}{while}
\usepackage{graphicx}
\usepackage{amsthm}
\usepackage{amssymb}
\usepackage{float}
\usepackage{verbatim}
\usepackage{tipa}
\usepackage{color}
\usepackage{latexsym}
\usepackage{subfig}
\usepackage{multirow}
\usepackage{capt-of}
\usepackage{threeparttable}
\usepackage{xcolor}
\usepackage{tipa}
\SetKw{Continue}{continue}
\SetKw{Break}{break}

\newtheorem{theorem}{Theorem}
\newtheorem{definition}{Definition}
\newtheorem{example}{Example}

\begin{document}

\title{
A Unified and Scalable Algorithm Framework of User-Defined Temporal $(k,\mathcal{X})$-Core Query
}

\author{Ming~Zhong,
        Junyong~Yang,
        Yuanyuan~Zhu,
        Tieyun~Qian,
        Mengchi~Liu,
        and~Jeffrey~Xu~Yu%
\thanks{M. Zhong, J. Yang, Y. Zhu and T. Qian are with the School
of Computer Science, Wuhan University.
E-mail: \{clock, thomasyang, yyzhu, qty\}@whu.edu.cn}%
\thanks{M. Liu is with the South China Normal University.
E-mail: liumengchi@scnu.edu.cn}%
\thanks{J. X. Yu is with the Chinese University of Hong Kong.
E-mail: yu@se.cuhk.edu.hk}%
\thanks{Manuscript received April 19, 2005; revised August 26, 2015.}}

\IEEEtitleabstractindextext{%
\begin{abstract}
Querying cohesive subgraphs on temporal graphs (e.g., social network, finance network, etc.) with various conditions has attracted intensive research interests recently. In this paper, we study a novel Temporal $(k,\mathcal{X})$-Core Query (TXCQ) that extends a fundamental Temporal $k$-Core Query (TCQ) proposed in our conference paper by optimizing or constraining an arbitrary metric $\mathcal{X}$ of $k$-core, such as size, engagement, interaction frequency, time span, burstiness, periodicity, etc. Our objective is to address specific TXCQ instances with conditions on different $\mathcal{X}$ in a unified algorithm framework that guarantees scalability. For that, this journal paper proposes a taxonomy of measurement $\mathcal{X}(\cdot)$ and achieve our objective using a two-phase framework while $\mathcal{X}(\cdot)$ is time-insensitive or time-monotonic. Specifically, Phase 1 still leverages the query processing algorithm of TCQ to induce all distinct $k$-cores during a given time range, and meanwhile locates the ``time zones'' in which the cores emerge. Then, Phase 2 conducts fast local search and $\mathcal{X}$ evaluation in each time zone with respect to the time insensitivity or monotonicity of $\mathcal{X}(\cdot)$. By revealing two insightful concepts named tightest time interval and loosest time interval that bound time zones, the redundant core induction and unnecessary $\mathcal{X}$ evaluation in a zone can be reduced dramatically. Our experimental results demonstrate that TXCQ can be addressed as efficiently as TCQ, which achieves the latest state-of-the-art performance, by using a general algorithm framework that leaves $\mathcal{X}(\cdot)$ as a user-defined function.
\end{abstract}

}

\maketitle

\IEEEdisplaynontitleabstractindextext

%
\IEEEpeerreviewmaketitle

\IEEEraisesectionheading{\section{Introduction}\label{sec:introduction}}

%
%
%
%

\begin{figure}[t!]
\centering
\includegraphics[width=\linewidth]{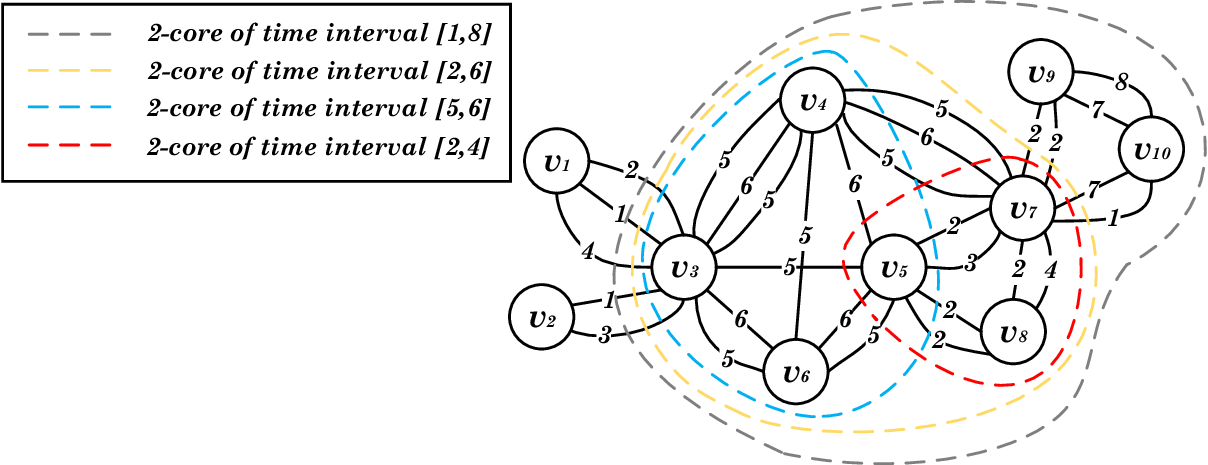}
\caption{A running example of temporal graph.}\label{fig:temporalgraph}
\end{figure}

	
\IEEEPARstart{D}iscovering cohesive subgraphs or communities from temporal graphs has great values in many application scenarios, thereby drawing intensive research interests~\cite{wu2015core, li2018persistent, galimberti2018mining, chu2019online, ma2019efficient, qin2020periodic, bai2020efficient, li2021efficient, yu2021querying, tang2022reliable, yang2023tcq} in recent years. Here, a temporal graph refers to an undirected multigraph in which each edge has a timestamp to indicate when it occurred, as illustrated in Fig~\ref{fig:temporalgraph}. For example, consider a finance graph consisting of bank accounts as vertices and fund transfer transactions between accounts as edges with natural timestamps. For applications such as anti-money-laundering, we would like to search communities like $k$-cores that contain a known suspicious account and emerge within a specific time interval of events like the FIFA World Cup, and investigate the associated accounts.

There are typically two categories of $k$-core studies on temporal graphs. The first one is to find primitive $k$-cores from a kind of projection of the temporal graph over a given time interval, such as span-core~\cite{galimberti2018mining}, historical $k$-core~\cite{yu2021querying}, and temporal $k$-core~\cite{wu2015core, yang2023tcq}. Their query semantics is relatively simple and general, and the efficiency or scalability of solution is the research highlight. The second one creates an elaborate $k$-core definition with a time-relevant metric, such as interaction frequency~\cite{wu2015core,bai2020efficient}, persistence~\cite{li2018persistent}, burstiness~\cite{chu2019online}, periodicity~\cite{qin2020periodic}, continuity~\cite{li2021efficient} and reliability~\cite{tang2022reliable}, and addresses the new problems with dedicated solutions.
	
In this paper, motivated by the curiosity of whether querying various elaborate $k$-cores on temporal graphs can be addressed uniformly in a general algorithm framework that guarantees scalability, we propose a fundamental Temporal $k$-Core Query (TCQ) and an extended Temporal $(k,\mathcal{X})$-Core Query (TXCQ), where $k$ is an ordinary cohesiveness threshold and $\mathcal{X}(\cdot)$ represents any other reasonable measurement of (temporal) $k$-core. In a nutshell, given a temporal graph $\mathcal{G}$ and a time interval $[Ts,Te]$, TCQ aims to find $k$-cores from the projection of $\mathcal{G}$ over each subinterval of $[Ts,Te]$, and TXCQ additionally requires the $\mathcal{X}$ values of $k$-cores are optimal or satisfy a given constraint. The study on TCQ and TXCQ is significant for two reasons.

Firstly, TCQ generalizes the previous Historical $k$-Core Query (HCQ)~\cite{yu2021querying} by inducing $k$-cores from the projected graphs over each subinterval of $[Ts,Te]$ but not only $[Ts,Te]$ itself, so that the query semantics is more flexible. Because, users usually do not know the exact time interval of targeted historical $k$-core in real-world applications. Thus, it is more reasonable to assume that users can only offer a flexible time interval and need to induce cores from all its subintervals. For example, for detecting money laundering by soccer gambling during the FIFA World Cup, the $k$-cores emerging during a few of hours around one of the matches are more valuable than a large $k$-core emerging over the whole month. Actually, TCQ represents a group of HCQ, and HCQ can be seen as a special case of TCQ.

Secondly, TXCQ further extends TCQ to unify many existing or even potential elaborate $k$-core queries. A variety of metrics of $k$-cores have been investigated, as shown in Table~\ref{tab:attr}. Most of their semantics are still meaningful when querying within a given time interval, and can be exactly or similarly expressed by TXCQ. For example, find $k$-cores with the optimal engagement or interaction frequency during a specific period is a reasonable and nontrivial problem, since for a set of vertices in a core their engagement or interaction frequency changes dynamically within different time intervals. Thus, it saves great research efforts if we can address most elaborate $k$-core queries uniformly in an algorithm framework like ``Swiss army knife'' whose scalability can be guaranteed without regard to the complexity of $\mathcal{X}(\cdot)$.

\begin{figure}[t!]
\centering
\includegraphics[width=\linewidth]{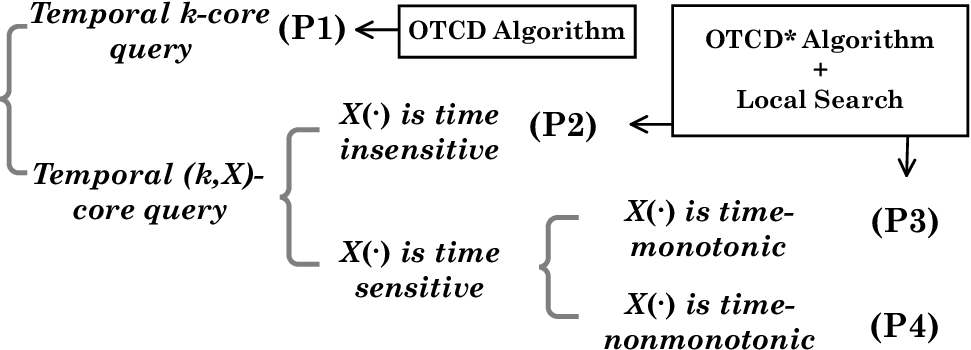}
\caption{A taxonomy of temporal $k$-core query models.}\label{fig:tcqstr}
\end{figure}

Let us explain the above points of perspective by the following example.

\begin{example}\label{exm:1}
As illustrated in Fig~\ref{fig:temporalgraph}, given a time interval [1,8], HCQ only returns the largest core marked by the grey dashed line. In contrast, our TCQ returns four cores marked by dashed lines with different colors. These cores may reveal various insights unseen by the largest one. For example, some cores like red and blue that emerge during short periods may be caused by special events. Also, some persistent or periodic cores may be found. These elaborate cores can be represented by TXCQ alternatively, in which a query condition on an arbitrary metric of core can be defined by users. For example, we can implement $\mathcal{X}(\cdot)$ to measure the time span of core, and set a TXCQ instance to find the core with the shortest time span, so that the blue core that emerges during [5,6] will be returned. 
\end{example}



In our previous conference paper~\cite{yang2023tcq}, we present an Optimized Temporal Core Decomposition (OTCD) algorithm to deal with TCQ. It runs a decremental TCD procedure over all subintervals of a given time interval, which always induces a temporal $k$-core from the previously induced temporal $k$-core except the initial one. Moreover, it adopts a subinterval pruning technique based on an intuitive concept named Tightest Time Interval (TTI), which leverages the properties of TTI to predict which subintervals will induce duplicated cores. Lastly, an in-memory data structure named Temporal Edge List (TEL) is proposed for implementing TCD and TTI-based pruning efficiently.

In this journal version, we further study the processing of TXCQ. According to our taxonomy shown in Fig~\ref{fig:tcqstr}, there are three types (i.e., P2, P3 and P4) of TXCQ, and two of them (i.e., P2 and P3) can be addressed by extending OTCD algorithm proposed for TCQ (i.e., P1). Then, we present a two-phase framework to deal with specific TXCQ instances of P2 and P3 uniformly. Phase 1 algorithm still follows OTCD algorithm to induce $k$-cores, and meanwhile locates the ``time zones'' in which the cores emerge by both TTI and Loosest Time Interval (LTI). For each time zone, Phase 2 algorithm revisit it and conducts local search to find the subintervals whose cores satisfy a given query condition on metric $\mathcal{X}$. Most importantly, the local search will not violate the scalability of OTCD by leveraging the time insensitivity or monotonicity of $\mathcal{X}(\cdot)$.

Our contributions are summarized as follows.

\begin{itemize}
\item We formalize general time-range cohesive subgraph query problems on ubiquitous temporal graphs, namely, TCQ and TXCQ. Many previous typical $k$-core query models on temporal graphs can be equivalently represented by them.
\item To deal with TCQ, we propose a novel TCD algorithm, and optimize it with TTI-based pruning. The optimized algorithm OTCD is scalable in terms of the span of query time interval for reducing both ``intra-core'' and ``inter-core'' redundant computation significantly. Moreover, we propose TEL to implement OTCD algorithm efficiently in physical level.
\item To deal with TXCQ, we extend the TCQ solution to a general two-phase framework, which leaves $\mathcal{X}(\cdot)$ as a user-defined function. The framework uses TTI and LTI to locate the ``time zone'' of each distinct $k$-core, and only evaluates the $\mathcal{X}$ values for necessary subintervals in each zone with respect to the distribution of $\mathcal{X}$ value, thereby still guaranteeing the scalability.
\item Lastly, we evaluate the efficiency and effectiveness of our algorithms on real-world datasets. The experimental results demonstrate that TXCQ can be addressed as efficiently as TCQ, which outperforms the existing approaches by at least three orders of magnitude.
\end{itemize}
	
The rest of this paper is organized as follows. Section~\ref{sec:problem} formalizes the data model and two query models, namely, TCQ and TXCQ. Section~\ref{sec:dwtcq} and~\ref{sec:dwtxcq} present our approaches to deal with TCQ and TXCQ respectively. Section~\ref{sec:expr} introduces the experimental evaluation. Section~\ref{sec:relwork} summarizes the related work. Lastly, Section~\ref{sec:conc} concludes our work.

\begin{table}
    \centering
    \caption{Mathematical notation and acronym.}
    \begin{tabular}{cc}
    \hline
      $\mathcal{G}$, $\mathcal{G}_{[ts,te]}$ & a temporal graph and its projected graph over $[ts,te]$\\
      $\mathcal{V}$, $\mathcal{V}_{[ts,te]}$ & the vertex sets of $\mathcal{G}$ and $\mathcal{G}_{[ts,te]}$ respectively\\
      $\mathcal{E}$, $\mathcal{E}_{[ts,te]}$ & the edge sets of $\mathcal{G}$ and $\mathcal{G}_{[ts,te]}$ respectively\\
      $\mathcal{C}^k(\mathcal{G})$ & the $k$-core in $\mathcal{G}$\\
      $\mathcal{T}_{[ts,te]}^k$ & the temporal $k$-core $\mathcal{C}^k(\mathcal{G}_{[ts,te]})$ of $[ts,te]$ (in $\mathcal{G}$)\\
      $\mathcal{X}(\mathcal{T}_{[ts,te]}^{k})$ & a user-defined measurement of $\mathcal{T}_{[ts,te]}^k$ \\
      $\mathcal{B}_{[ts,te]}^k$ & the time zone of temporal $k$-core with TTI $[ts,te]$\\
      $\mathcal{R}_{[ts,te]}^{[ts',te']}$ & the rectangle bounded by TTI $[ts,te]$ and LTI $[ts',te']$\\
      \hline
      TCQ & \underline{T}emporal k-\underline{C}ore \underline{Q}uery\\
      TXCQ & \underline{T}emporal $(k, \underline{\mathcal{X}})$-\underline{C}ore \underline{Q}uery\\
      TTI & \underline{T}ightest \underline{T}ime \underline{I}nterval\\
      LTI & \underline{L}oosest \underline{T}ime \underline{I}nterval\\
      TEL & \underline{T}emporal \underline{E}dge \underline{L}ist\\
      TL,SL,DL & \underline{T}ime \underline{L}ist, \underline{S}ource \underline{L}ist, \underline{D}estination \underline{L}ist in TEL\\
      TCD & \underline{T}emporal \underline{C}ore \underline{D}ecomposition operation/algorithm\\
      OTCD & \underline{O}ptimized \underline{TCD} algorithm\\
      OTCD* & \underline{OTCD} with rectangle pruning and time zone location\\
      LS & \underline{L}ocal \underline{S}earch including TI-LS, TMO-LS and TMC-LS\\
    \hline
    \end{tabular}
    \label{tab:notation}
\end{table}

\section{Preliminary}\label{sec:problem}

\subsection{Data Model}\label{sec:datamodel}
	
A \textit{temporal graph} is normally an undirected graph $\mathcal{G} = (\mathcal{V}, \mathcal{E})$ with parallel temporal edges. Each temporal edge $(u, v, t) \in \mathcal{E}$ is associated with a timestamp $t$ that indicates when the interaction happened between the vertices $u, v \in \mathcal{V}$. For example, the temporal edges could be fund transfer transactions between bank accounts in a finance graph. Without loss of generality, we use continuous integers that start from 1 to denote timestamps. Fig~\ref{fig:temporalgraph} illustrates a temporal graph as our running example.
	
In particular, given a time interval $[ts, te]$, we define the \textit{projected graph} of $\mathcal{G}$ over $[ts, te]$ as $\mathcal{G}_{[ts, te]} = (\mathcal{V}_{[ts, te]}, \mathcal{E}_{[ts, te]})$, where $\mathcal{V}_{[ts, te]} = \mathcal{V}$ and $\mathcal{E}_{[ts, te]} = \{(u, v, t) | (u, v, t)\in \mathcal{E}, t\in[ts, te]\}$.

\begin{table*}[t]
    \centering
    \caption{Typical measurements $\mathcal{X}(\cdot)$ of (temporal) $k$-core, which marked by * do not strictly adhere to the original definitions due to inconsistent models. We denote by $\mathcal{V}_{[ts,te]}^{k}$ the vertices of $\mathcal{T}_{[ts,te]}^{k}$, $\mathcal{E}_{[ts,te]}^{k}$ the edges of $\mathcal{T}_{[ts,te]}^{k}$, $\mathcal{N}_{[ts,te]}^{k}(v)$ the neighbor vertices of $v$ in $\mathcal{T}_{[ts,te]}^{k}$, $\mathcal{N}_{[ts,te]}(v)$ the neighbor vertices of $v$ in $\mathcal{G}_{[ts,te]}$, and $p$ a parameter.}
    \label{tab:attr}
    \begin{tabular}{ccc}
    \hline
    Metric & Measurement $\mathcal{X}(\mathcal{T}_{[ts,te]}^{k})$ & Category\\
    \hline
    size~\cite{yao2021efficient} & $|\mathcal{V}_{[ts,te]}^{k}|$ & time-insensitive\\
    frequency~\cite{bai2020efficient, wu2015core} & $\min\{|\{(u,v,t)\in \mathcal{E}_{[ts,te]}^{k}\}|\ |u,v\in \mathcal{V}_{[ts,te]}^{k}\}$ & time-insensitive\\
    time span & $\max\{|t-t'|\ |(u,v,t),(u',v',t')\in \mathcal{E}_{[ts,te]}^{k}\}$ & time-insensitive\\
    persistence*~\cite{li2018persistent} & $\max\{(te'-ts')-(te''-ts'')\ | [ts',te']\in \mathcal{T}_{[ts,te]}^{k}$.LTI, $[ts'',te'']=\mathcal{T}_{[ts,te]}^{k}$.TTI\} & time-insensitive\\
    periodicity*~\cite{qin2020periodic} & $\max\{n |$ there exist TTIs $[ts_1,te_1], \cdots, [ts_n,te_n]$ with $ts_{i+1} - te_i \geqslant p$ and $\mathcal{V}^k_{[ts_i,te_i]} = \mathcal{V}_{[ts,te]}^{k}$\} & time-insensitive\\
    growth rate & $|\mathcal{V}_{[ts,te]}^{k}|/(te-ts)$ & time-monotonic\\
    burstiness*~\cite{chu2019online} & $\sum_{v\in \mathcal{V}^k_{[ts,te]}}|\mathcal{N}_{[ts,te]}^{k}(v)|/(te-ts)$ & time-monotonic\\
    engagement~\cite{zhang2020exploring} & $\min\{\mathcal{N}_{[ts,te]}^{k}(v) / \mathcal{N}_{[ts,te]}(v) | v\in \mathcal{V}_{[ts,te]}^{k}\}$ & time-monotonic\\
    \hline
    \end{tabular}
\end{table*}	

\subsection{Query Model}\label{sec:querymodel}

\subsubsection{Temporal $k$-Core Query (TCQ)}
 
For revealing communities in graphs, the $k$-core query is widely adopted. Given an undirected graph $G$ and an integer $k$, $k$-core is the maximal induced subgraph of $G$ in which all vertices have degrees at least $k$, which is denoted by $\mathcal{C}^{k}(G)$. The \textit{coreness} of a vertex $v$ in a graph $G$ is the largest value of $k$ such that $v$ belongs to $\mathcal{C}^{k}(G)$.

In this paper, we propose a novel query model called Temporal $k$-Core Query (TCQ) that generalizes the previous Historical $k$-Core Query (HCQ)~\cite{yu2021querying}. Both TCQ and HCQ are to find $k$-cores from $\mathcal{G}$ by a given time interval. The main difference is that the query time interval $[Ts, Te]$ of TCQ is a range but not fixed query condition like HCQ. In TCQ, $Ts$ and $Te$ are the minimum start time and maximum end time of inducing $k$-cores respectively, and thereby $k$-cores induced by each subinterval $[ts, te] \subseteq [Ts, Te]$ are all potential results of TCQ. Specifically, TCQ will return $\mathcal{C}^{k}(\mathcal{G}_{[ts,te]})$ (note that, the degree is the number of neighbor vertices but not neighbor edges) as \textit{temporal $k$-cores}. We denote by $\mathcal{T}^k_{[ts,te]}(\mathcal{G})$ a temporal $k$-core that appears over $[ts,te]$ on $\mathcal{G}$. 
	
The formal definition of TCQ is as follows.

\begin{definition}[Temporal $k$-Core Query]\label{def:tcq}
    For a temporal graph $\mathcal{G}$, given an integer $k$ and a time interval $[Ts, Te]$, return all distinct $\mathcal{T}_{[ts, te]}^{k}(\mathcal{G}) = \mathcal{C}^{k}(\mathcal{G}_{[ts,te]})$ with $[ts, te] \subseteq [Ts, Te]$.
\end{definition}
	
Note that, TCQ only returns the distinct temporal $k$-cores that are not identical to each other, since multiple subintervals of $[Ts,Te]$ may induce the same subgraph of $\mathcal{G}$. When the context is self-evident, $\mathcal{T}_{[ts,te]}^{k}(\mathcal{G})$ is abbreviated as $\mathcal{T}_{[ts,te]}^{k}$.

\subsubsection{Temporal $(k,\mathcal{X})$-Core Query (TXCQ)}

Many $k$-core query models also consider other metrics than the cohesiveness $k$, as shown in Table~\ref{tab:attr}. For example, the size of $k$-core may be preferred to be small, as the target communities have limited members. Therefore, we further extend the TCQ model and propose a Temporal $(k,\mathcal{X})$-Core Query (TXCQ) model as follows. 

\begin{definition}[Temporal $(k,\mathcal{X})$-Core Query]\label{def:TXCQ}
For a temporal graph $\mathcal{G}$, given an integer $k$, a time interval $[Ts,Te]$ and an arbitrary measurement $\mathcal{X}(\cdot)$ of $k$-core, return all temporal $k$-cores $\mathcal{T}_{[ts,te]}^{k}$ with $[ts,te] \subseteq [Ts,Te]$, while $\mathcal{X}(\mathcal{T}_{[ts,te]}^{k})$ is optimal or satisfy a specific constraint.
\end{definition}

In TXCQ, $\mathcal{X}(\cdot)$ can represent any reasonable measurement of $\mathcal{T}_{[ts,te]}^{k}$, such as those listed in Table~\ref{tab:attr}. Moreover, TXCQ generally has a certain query condition to optimize or constrain the $\mathcal{X}$ values of result temporal $k$-cores. Formally, for each result temporal $k$-core $\mathcal{T}_{[ts,te]}^{k}$, there does not exist $[ts',te'] \subseteq [Ts,Te]$ such that $\mathcal{X}(\mathcal{T}_{[ts',te']}^{k}) \succ \mathcal{X}(\mathcal{T}_{[ts,te]}^{k})$ and $[ts,te] \neq [ts',te']$, or $\mathcal{X}(\mathcal{T}_{[ts,te]}^{k}) \succcurlyeq \sigma$, where $\succ$/$\succcurlyeq$ denotes the superiority of $\mathcal{X}$ value and $\sigma$ denotes a threshold. The notations and acronyms are listed in Table~\ref{tab:notation}.

\section{Dealing with TCQ}\label{sec:dwtcq}

We start the journey to a unified query processing framework of TXCQ from dealing with TCQ. In this section, we briefly introduce a scalable TCQ algorithm proposed in our conference paper~\cite{yang2023tcq}, which reduces both ``intra-core'' and ``inter-core'' redundant computation significantly. 

\subsection{TCD Algorithm}\label{sec:tcdalgo}
\subsubsection{Temporal Core Decomposition (TCD)}\label{sec:tcd}
Firstly, we introduce Temporal Core Decomposition (TCD) as a basic operation on temporal graphs, which is derived from the traditional \textit{core decomposition}~\cite{batagelj2003m} on ordinary graphs. TCD refers to a two-step operation of inducing a temporal $k$-core $\mathcal{T}^k_{[ts,te]}$ of a fixed time interval $[ts,te]$ from a temporal graph $\mathcal{G}$. The first step is \textit{truncation}: remove temporal edges with timestamps not in $[ts, te]$ from $\mathcal{G}$, namely, induce the projected graph $\mathcal{G}_{[ts,te]}$. The second step is \textit{decomposition}: iteratively peel vertices with degree less than $k$ and the edges linked to them together. The correctness of TCD is as intuitive as that of core decomposition.

An excellent property of TCD operation is that, it can induce a temporal $k$-core $\mathcal{T}^k_{[ts,te]}$ from another temporal $k$-core $\mathcal{T}^k_{[ts',te']}$ with $[ts,te] \subset [ts',te']$, because $\mathcal{T}^k_{[ts,te]}$ is surely a subgraph of $\mathcal{T}^k_{[ts',te']}$ (see proof in~\cite{yang2023tcqcompl}).

\begin{figure}[t!]
\centering
\includegraphics[width=\linewidth]{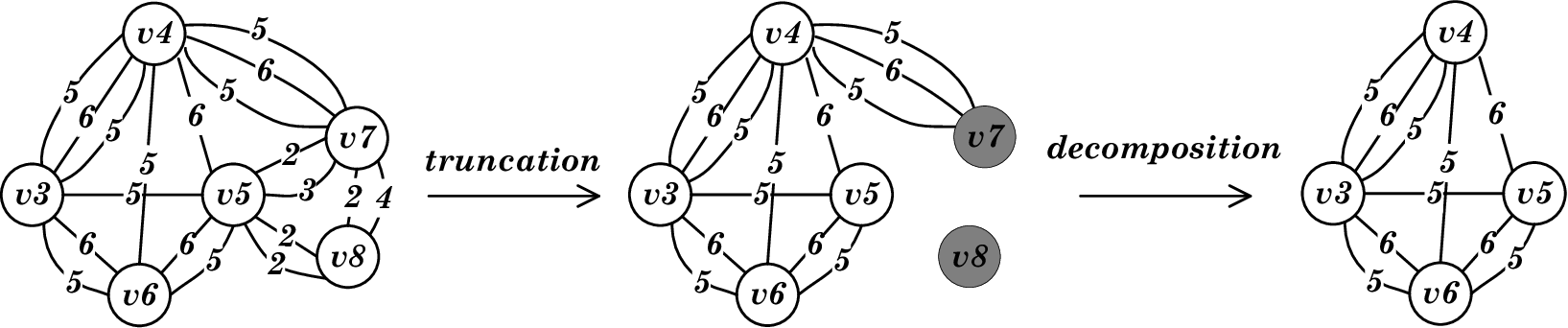}
\caption{Temporal core decomposition from $\mathcal{T}_{[2,6]}^{2}$ to $\mathcal{T}_{[5,6]}^{2}$.}\label{fig:tcdstep}
\end{figure}

For example, Figure~\ref{fig:tcdstep} illustrates the procedure of TCD from $\mathcal{T}^{2}_{[2,6]}$ to $\mathcal{T}^{2}_{[5,6]}$ on our running example graph in Fig~\ref{fig:temporalgraph}. The edges with timestamps not in $[5,6]$ (marked by dashed lines) are firstly removed from $\mathcal{T}_{[2,6]}^{2}$ by truncation, which results in the decrease of degrees of vertices $v_5$, $v_7$ and $v_8$. Then, the vertices with degree less than 2 (marked by dark circles), namely, $v_7$ and $v_8$ are further peeled by decomposition, together with their edges. The remaining temporal graph is $\mathcal{T}_{[5,6]}^{2}$.
	
\subsubsection{TCD Algorithm}\label{sec:tcd-basic}
We propose a TCD algorithm to address TCQ by using the above TCD operation. In general, given a TCQ instance, the TCD algorithm enumerates each subinterval of $[Ts,Te]$ in a particular order, so that the temporal $k$-cores of each subinterval are induced decrementally from previously induced temporal $k$-cores except the initial one.
	
\begin{figure}[t!]
\centering
\includegraphics[width=\linewidth]{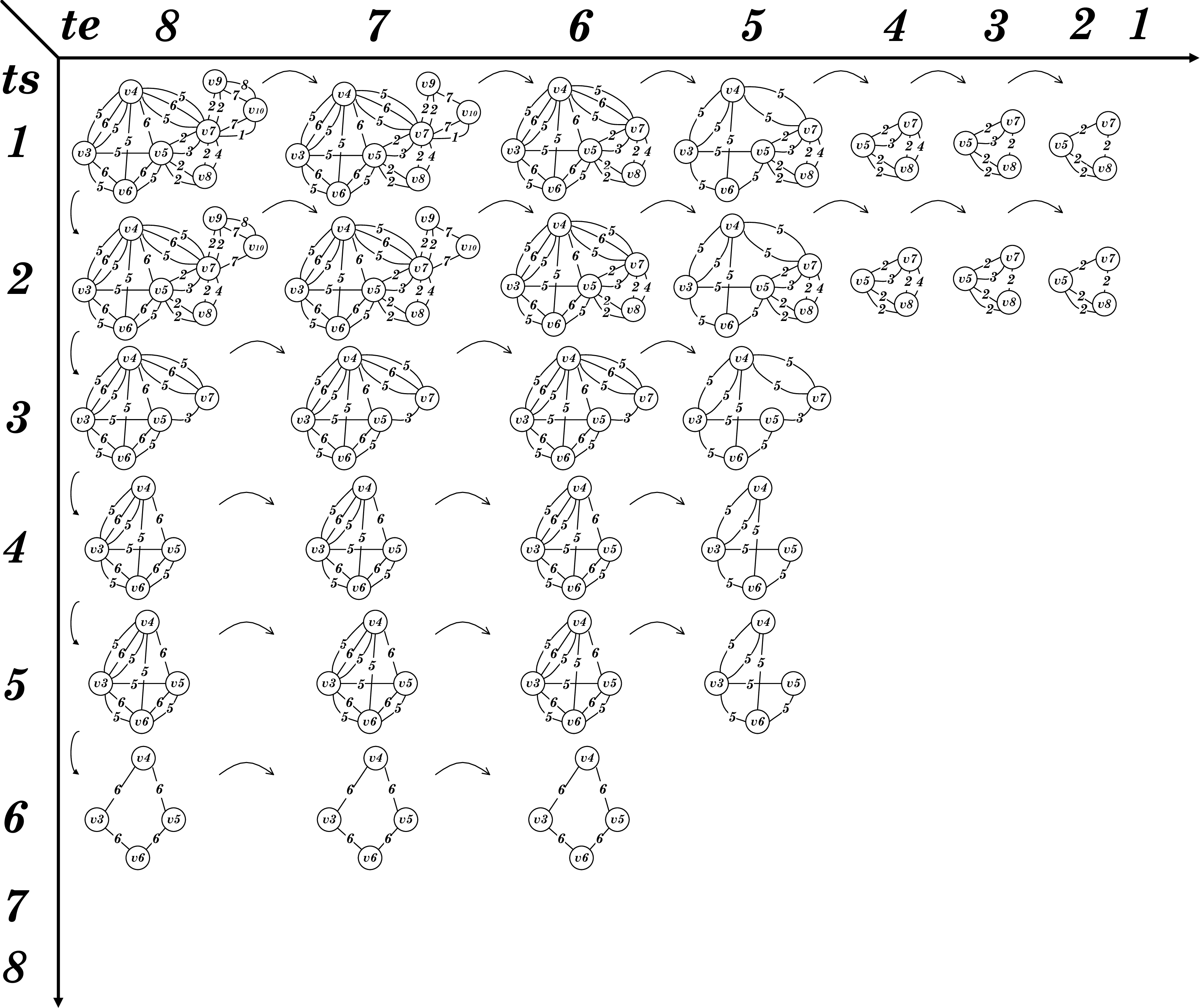}
\caption{A demonstration of TCD algorithm for finding temporal 2-cores of time interval [1,8].}\label{fig:tcdprocedure}
\end{figure}

Specifically, we enumerate a subinterval $[ts,te]$ of $[Ts,Te]$ as follows. Initially, let $ts = Ts$ and $te = Te$. It means we induce the largest temporal $k$-core $\mathcal{T}^k_{[Ts,Te]}$ at the beginning. Then, we will anchor the start time $ts = Ts$ and decrease the end time $te$ from $Te$ until $ts$ gradually. As a result, we can always leverage TCD to induce the temporal $k$-core of current subinterval $[ts,te]$ from the previously induced temporal $k$-core of $[ts,te+1]$ but not from $\mathcal{G}_{[ts,te]}$ or even $\mathcal{G}$. Whenever the value of $te$ is decreased to $ts$, the value of $ts$ will be increased to $ts+1$ until $ts = Te$, and the value of $te$ will be reset to $Te$. Then, we induce $\mathcal{T}_{[ts+1,te]}^k$ from $\mathcal{T}_{[ts,te]}^k$, and start over the decremental TCD procedure.
Fig~\ref{fig:tcdprocedure} gives a demonstration of TCD algorithm for finding temporal 2-cores of time interval [1,8] on our running example graph. 

\subsection{OTCD Algorithm}\label{sec:opt}
	
\begin{figure*}
\centering
\subfloat[Without pruning.]{\label{subfig:withoutpruning}
\includegraphics[width=0.305\textwidth]{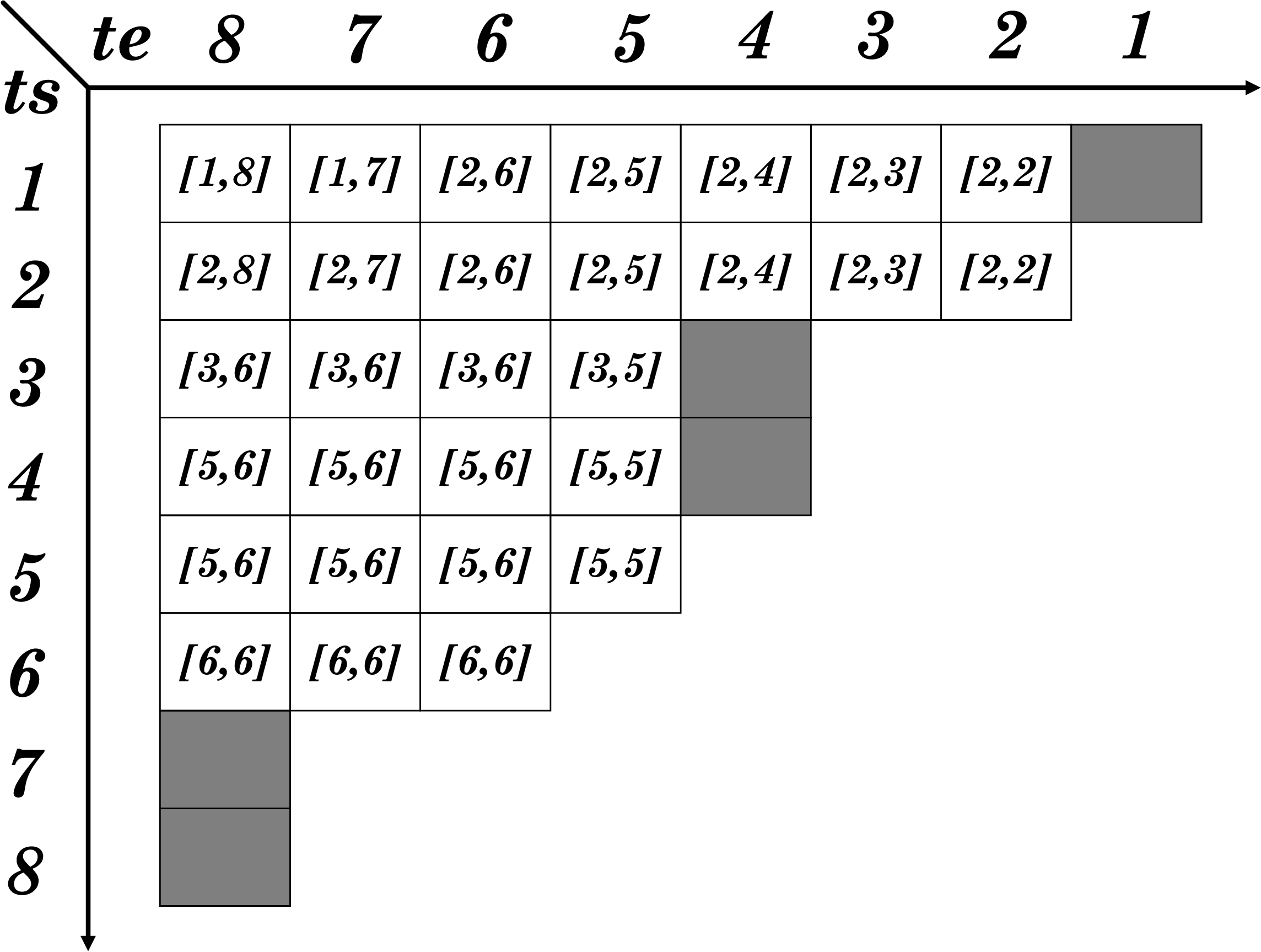}}
\subfloat[With pruning.]{\label{subfig:withpruning}
\includegraphics[width=0.6\textwidth]{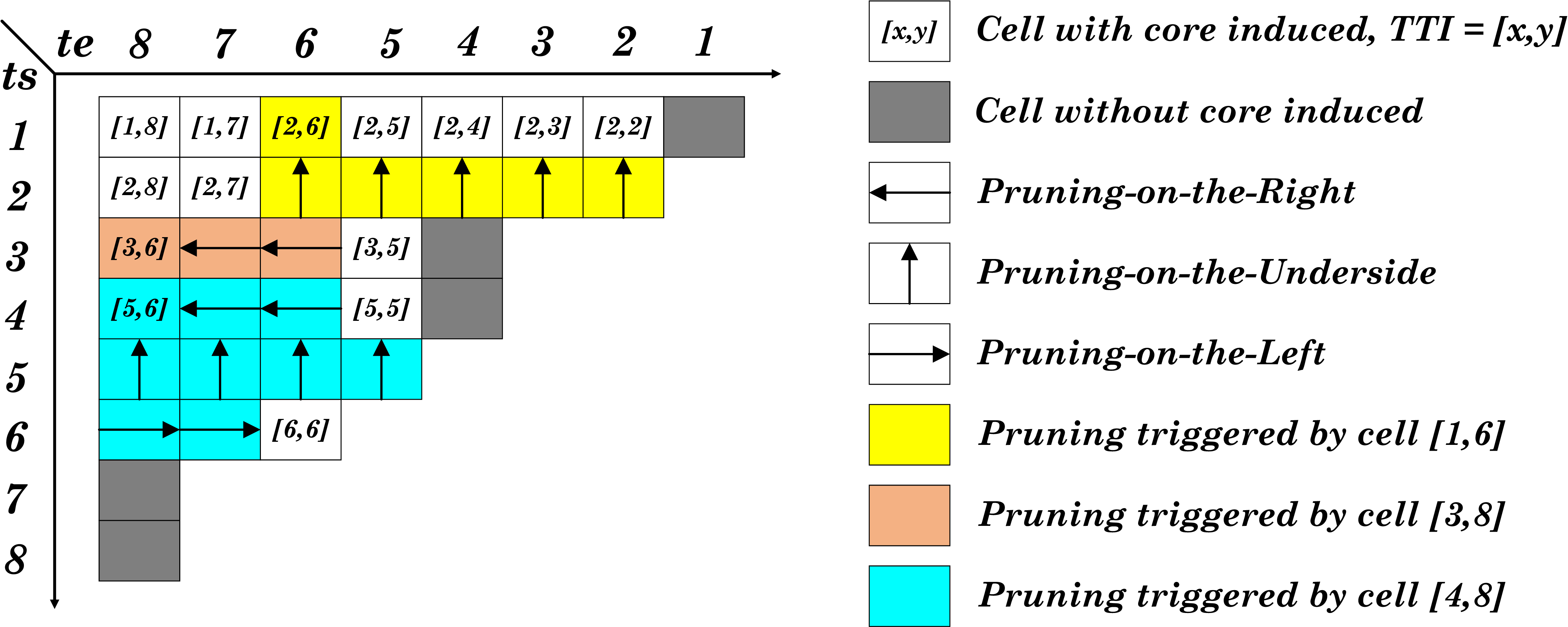}}
\caption{Examples of subinterval pruning based on tightest time interval.}\label{fig:tcdproceduretil}
\end{figure*}
	
\subsubsection{Tightest Time Interval (TTI)}\label{sec:tti}
We have such an observation, a temporal $k$-core of $[ts,te]$ may only contain edges with timestamps in a subinterval $[ts',te'] \subset [ts,te]$, since the edges in $[ts,ts')$ and $(te',te]$ have been removed by core decomposition. For example, consider a temporal $k$-core $\mathcal{T}_{[4,8]}^{2}$ illustrated in Fig~\ref{fig:tcdprocedure}. We can see that it does not contain edges with timestamps 4, 7 and 8. As a result, if we continue to induce $\mathcal{T}_{[4,7]}^{2}$ from $\mathcal{T}_{[4,8]}^{2}$ and to induce $\mathcal{T}_{[4,6]}^{2}$ from $\mathcal{T}_{[4,7]}^{2}$, the returned temporal $k$-cores remain unchanged. The sameness of temporal $k$-cores induced by different subintervals inspires us to optimize TCD algorithm by directly pruning subintervals in advance.
	
For that, we propose the concept of Tightest Time Interval (TTI) for temporal $k$-cores. Given a temporal $k$-core of $[ts,te]$, its TTI refers to the minimal time interval $[ts',te']$ that can induce an identical temporal $k$-core to $\mathcal{T}^k_{[ts,te]}$, namely, there is no subinterval of $[ts',te']$ that can induce an identical temporal $k$-core to $\mathcal{T}^k_{[ts,te]}$. We formalize the definition of TTI as follows.
	
	\begin{definition}[Tightest Time Interval]\label{def:tti}
		Given a temporal $k$-core $\mathcal{T}_{[ts,te]}^{k}$, its tightest time interval $\mathcal{T}_{[ts,te]}^{k}$.TTI is $[ts',te']$, if and only if\\
		1) $\mathcal{T}_{[ts',te']}^{k}$ is an identical temporal $k$-core to $\mathcal{T}_{[ts,te]}^{k}$;\\
		2) there does not exist $[ts'',te''] \subset [ts',te']$, such that $\mathcal{T}_{[ts'',te'']}^{k}$ is an identical temporal $k$-core to $\mathcal{T}_{[ts,te]}^{k}$.
	\end{definition}

The computation and properties of TTI that are essential to TCD algorithm optimization are presented in~\cite{yang2023tcq}. Fig~\ref{subfig:withoutpruning} abstracts Fig~\ref{fig:tcdprocedure} as a schedule table of subinterval enumeration. For example, the cell in row 1 and column 6 represents a subinterval $[1,6]$, in which $[2,6]$ is the TTI of $\mathcal{T}^2_{[1,6]}$. In particular, the grey cells indicate that the temporal $k$-cores of the corresponding subintervals do not exist. Fig~\ref{subfig:withoutpruning} clearly reveals that TCD algorithm suffers from inducing a number of identical temporal $k$-cores (with the same TTIs).

\subsubsection{TTI-Based Pruning Rules}
	
The main idea of optimizing TCD algorithm is to predict the induction of identical temporal $k$-cores by leveraging TTI, thereby skipping the corresponding subintervals during the enumeration. Specifically, whenever a temporal $k$-core of $[ts,te]$ is induced, we evaluate its TTI as $[ts',te']$. If $ts'>ts$ or/and $te'<te$, it is triggered that a number of subintervals on the schedule can be pruned in advance (see proof in~\cite{yang2023tcqcompl}). According to different relations between $[ts,te]$ and $[ts',te']$, our pruning technique can be categorized into three rules which are not mutually exclusive.
	
\textbf{Rule 1: Pruning-on-the-Right (PoR)}.
If TTI $[ts',te']$ in the current cell $[ts,te]$ meets $te'<te$, the following cells in this row from $[ts,te-1]$ until $[ts,te']$ will be skipped.

For example, Fig~\ref{subfig:withpruning} illustrates two instances of PoR (the cells in orange and blue colors with left arrow). When $\mathcal{T}_{[3,8]}^{2}$ has been induced, we evaluate its TTI as $[3,6]$, and thus PoR is triggered. PoR immediately excludes the following two cells $[3,7]$ and $[3,6]$ from the schedule. As a proof, we can see the TTIs in these two cells are both $[3,6]$ in Fig~\ref{subfig:withoutpruning}.
 
\textbf{Rule 2: Pruning-on-the-Underside (PoU)}.
Moreover, if $ts'> ts$, for each row $r\in [ts+1,ts']$, the cells $[r,te]$, $[r,te-1]$, $\cdots$, $[r,r]$ will be skipped.

For example, Fig~\ref{subfig:withpruning} illustrates two PoU instances (the cells in yellow and blue colors with up arrow). On enumerating the cell $[1,6]$, since the contained TTI is $[2,6]$, the cells $[2,6]$, $\cdots$, $[2,2]$ are pruned by PoU, because the TTIs in these cells are the same as the cells $[1,6]$, $\cdots$, $[1,2]$ respectively, though the TTIs of cells $[1,5]$, $\cdots$, $[1,2]$ have not been evaluated.
	
\textbf{Rule 3: Pruning-on-the-Left (PoL)}.
Lastly, if both $ts'> ts$ and $te' < te$, for each row $r\in [ts'+1,te']$, the cells $[r,te]$, $[r,te-1]$, $\cdots$, $[r,te'+1]$ will also be skipped, besides the cells pruned by PoR and PoU.

For example, Fig~\ref{subfig:withpruning} illustrates a PoL instance (the cells in blue color with right arrow). On enumerating the cell $[4,8]$, PoL is triggered since the contained TTI is $[5,6]$. Then, the cells $[6,8]$ and $[6,7]$ are pruned by PoL because the TTIs contained in them are the same as the cell $[6,6]$ on the right of them. PoL is more tricky than PoU because the cells are pruned for containing the same TTIs as other cells that are scheduled to traverse after them by TCD algorithm. Note that, the cell $[4,8]$ triggers all three kinds of pruning. In fact, a cell may trigger PoL only, PoU only, or all three rules.

\subsubsection{Optimized TCD Algorithm}\label{sec:otcd}

Compared with TCD algorithm, the improvement of Optimized TCD (OTCD) algorithm is simply to conduct a pruning operation whenever a temporal $k$-core has been induced. Specifically, we evaluate the TTI of this temporal $k$-core, check each pruning rule to determine if it is triggered, and prune the specific subintervals on the schedule in advance. In this way, OTCD algorithm only performs TCD operations that are necessary for returning all distinct answers to a given TCQ instance. Conceptually, the new pruning operation of optimized algorithm eliminates the ``inter-core'' redundant computation, and the original TCD operation eliminates the ``intra-core'' redundant computation.

As illustrated in Fig~\ref{subfig:withpruning}, OTCD algorithm completely eliminates repeated inducing of identical temporal $k$-cores, namely, each distinct temporal $k$-core is induced exactly once during the whole procedure. It means, the real computational complexity of OTCD algorithm is the summation of complexity for inducing each distinct temporal $k$-core but not the temporal $k$-core of each subinterval of $[Ts, Te]$. Therefore, we say OTCD algorithm is scalable with respect to the query time interval $[Ts, Te]$. For many real-world datasets, the span of $[Ts, Te]$ could be very large, while there exist only a limited number of distinct temporal $k$-cores over this period, so that OTCD algorithm can still process the query efficiently.

\subsection{Implementation}

\begin{figure*}
\centering
\includegraphics[width=0.8\textwidth]{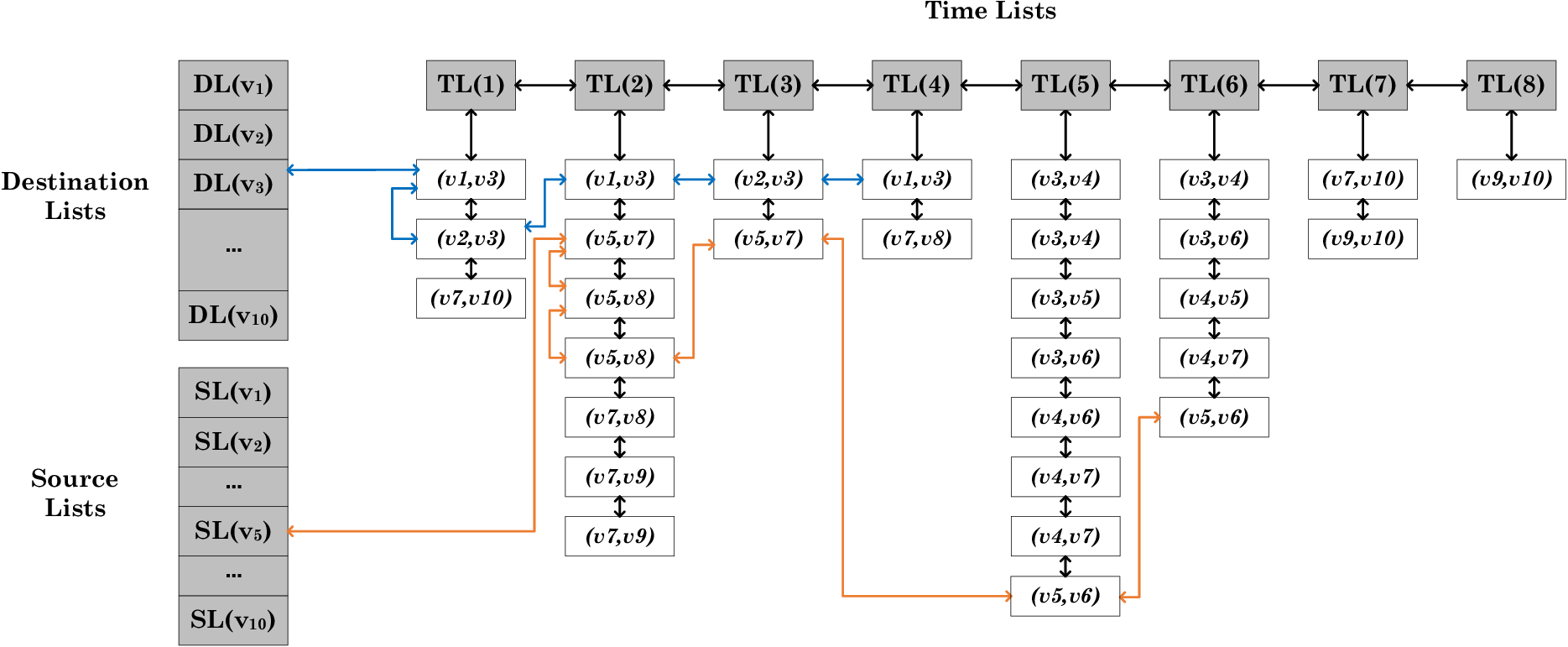}
\caption{The conceptual illustration of the partial temporal edge list of our running example graph.}\label{fig:telultra}
\end{figure*}

We propose a novel data structure called Temporal Edge List (TEL) for representing an arbitrary temporal graph (including temporal $k$-cores that are also temporal graphs), which is both the input and output of TCD operation. Conceptually, TEL($\mathcal{G}$) preserves a temporal graph $\mathcal{G}=(\mathcal{V},\mathcal{E})$ by organizing its edges in a 3-dimension space, each dimension of which is a set of bidirectional linked lists, as illustrated in Fig~\ref{fig:telultra}. The first dimension is time, namely, all edges in $\mathcal{E}$ are grouped by their timestamps. Each group is stored as a bidirectional linked list called Time List (TL), and TL($t$) denotes the list of edges with a timestamp $t$. Then, TEL($\mathcal{G}$) uses a bidirectional linked list, in which each node represents a timestamp in $\mathcal{G}$, as a timeline in ascending order to link all TLs, so that some temporal operations can be facilitated. Moreover, the other two dimensions are source vertex and destination vertex respectively. We use a container to store the Source Lists (SL) or Destination Lists (DL) for each vertex $v \in \mathcal{V}$, where SL($v$) or DL($v$) is a bidirectional linked list that links all edges whose source or destination vertex is $v$. Actually, an SL or DL is an adjacency list of the graph, by which we can retrieve the neighbor vertices and edges of a given vertex efficiently. Given a temporal graph $\mathcal{G}$, TEL($\mathcal{G}$) is built in memory by adding its edges iteratively. For each edge $(u, v, t) \in \mathcal{E}$, it is only stored once, and TL($t$), SL($u$) and DL($v$) will append its pointer at the tail respectively.
	
Fig~\ref{fig:telultra} illustrates a partial TEL of our example graph. The SLs and DLs other than SL($v_5$) and DL($v_3$) are omitted for conciseness. 
Note that, the order of two vertices in an edge is to avoid duplication and does not affect the functionality of TEL.
The superiority of TEL is summarized in~\cite{yang2023tcq}.


\section{Dealing with TXCQ}\label{sec:dwtxcq}
In this section, we propose a scalable algorithm framework that addresses TXCQ with diverse conditions uniformly, so that many existing and even potential ``elaborate'' TCQ instances can be solved easily and efficiently by extending the methods in \cite{yang2023tcq}. 

\subsection{Taxonomy of TXCQ}\label{sec:tax}

Firstly, let us consider two interesting properties of measurement $\mathcal{X}(\cdot)$ of temporal $k$-cores.

\begin{definition}[Time Sensitivity]\label{def:ts}
Given any two identical temporal $k$-core $\mathcal{T}_{[ts,te]}^{k}$ and $\mathcal{T}_{[ts',te']}^{k}$ with $[ts,te] \neq [ts',te']$, the time sensitivity of $\mathcal{X}(\cdot)$ depends on the equality of $\mathcal{X}(\mathcal{T}_{[ts,te]}^{k})$ and $\mathcal{X}(\mathcal{T}_{[ts',te']}^{k})$. Specifically, $\mathcal{X}(\cdot)$ is time-insensitive if $\mathcal{X}(\mathcal{T}_{[ts,te]}^{k})$ = $\mathcal{X}(\mathcal{T}_{[ts',te']}^{k})$ is guaranteed, and is time-sensitive otherwise.
\end{definition}

Intuitively, $\mathcal{X}(\cdot)$ is time-insensitive revealing that the $\mathcal{X}$ value of a temporal $k$-core $\mathcal{T}_{[ts,te]}^{k}$ is determined totally by the structure of $\mathcal{T}_{[ts,te]}^{k}$, without regard to the specific time interval $[ts,te]$ or its projected graph $\mathcal{G}_{[ts,te]}$. For example, the measurements of size, frequency and time span listed in Table~\ref{tab:attr} are all time-insensitive.

For time-sensitive $\mathcal{X}(\cdot)$, it can be further categorized by monotonicity.

\begin{definition}[Time Monotonicity]\label{def:tm}
Given any two identical temporal $k$-core $\mathcal{T}_{[ts,te]}^{k}$ and $\mathcal{T}_{[ts',te']}^{k}$ with $[ts,te] \subset [ts',te']$, the time monotonicity of $\mathcal{X}(\cdot)$ depends on whether $\mathcal{X}(\mathcal{T}_{[ts,te]}^{k})$ is always better or worse than $\mathcal{X}(\mathcal{T}_{[ts',te']}^{k})$. Specifically, $\mathcal{X}(\cdot)$ is time-monotonic if $\mathcal{X}(\mathcal{T}_{[ts,te]}^{k}) \succcurlyeq$ (or $\preccurlyeq$) $\mathcal{X}(\mathcal{T}_{[ts',te']}^{k})$ is guaranteed, and is time-nonmonotonic otherwise.
\end{definition}

Intuitively, time monotonicity indicates that the $\mathcal{X}$ value of a specific temporal $k$-core becomes better or worse monotonically with the expanding (or shrinking) of time interval, as long as its structure remains unchanged. For example, the measurements of growth rate, engagement and burstiness listed in Table~\ref{tab:attr} are all time-monotonic. 

With respect to these two properties of $\mathcal{X}(\cdot)$, TXCQ can be generally divided into three disjoint categories, namely, time-insensitive, time-monotonic and time-nonmonotonic, as shown in Fig~\ref{fig:tcqstr}. We develop a unified framework to address both time-insensitive and time-monotonic TXCQ, and leave the time-nonmonotonic TXCQ as an open problem.

\subsection{Two-Phase Framework}\label{sec:overview}

\begin{figure*}
\centering
\subfloat[LTI and TTI in $\mathcal{B}_{[t4,t5]}^{k}$.]{\label{subfig:ltizone}
\includegraphics[width=0.32\textwidth]{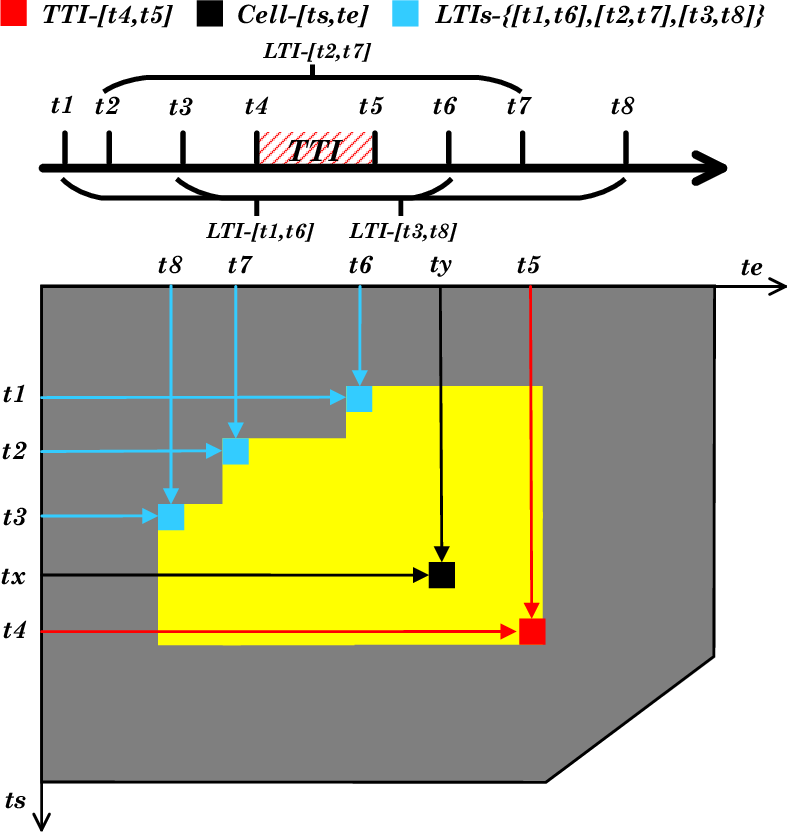}}
\subfloat[Rectangle areas in $\mathcal{B}_{[t4,t5]}^{k}$.]{\label{subfig:rectarea}
\includegraphics[width=0.32\textwidth]{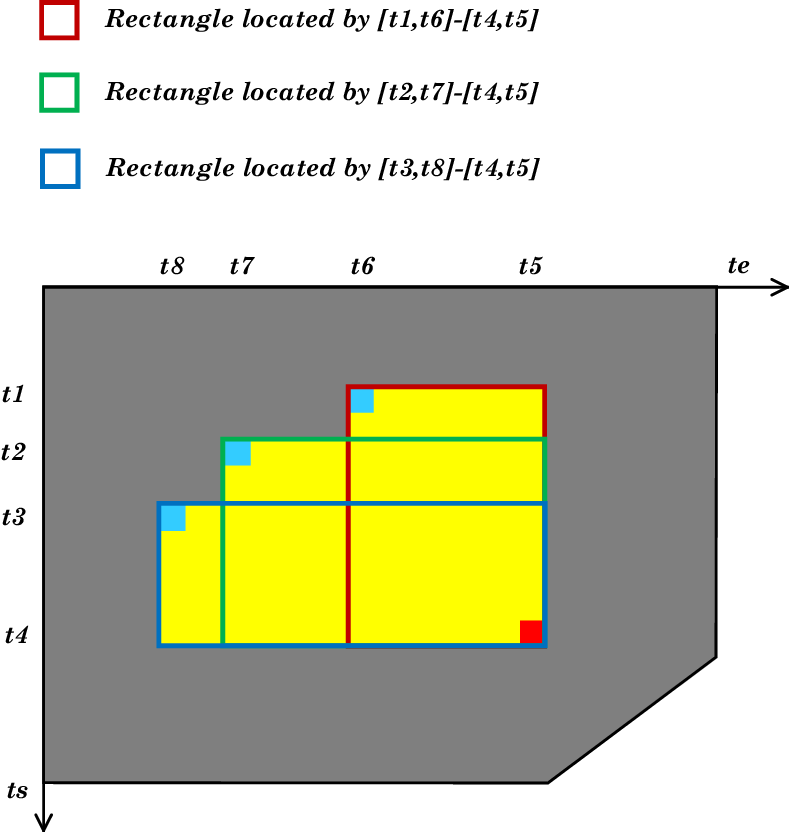}}
\subfloat[Demonstration of TMC-LS.]{\label{subfig:gpv}
\includegraphics[width=0.32\textwidth]{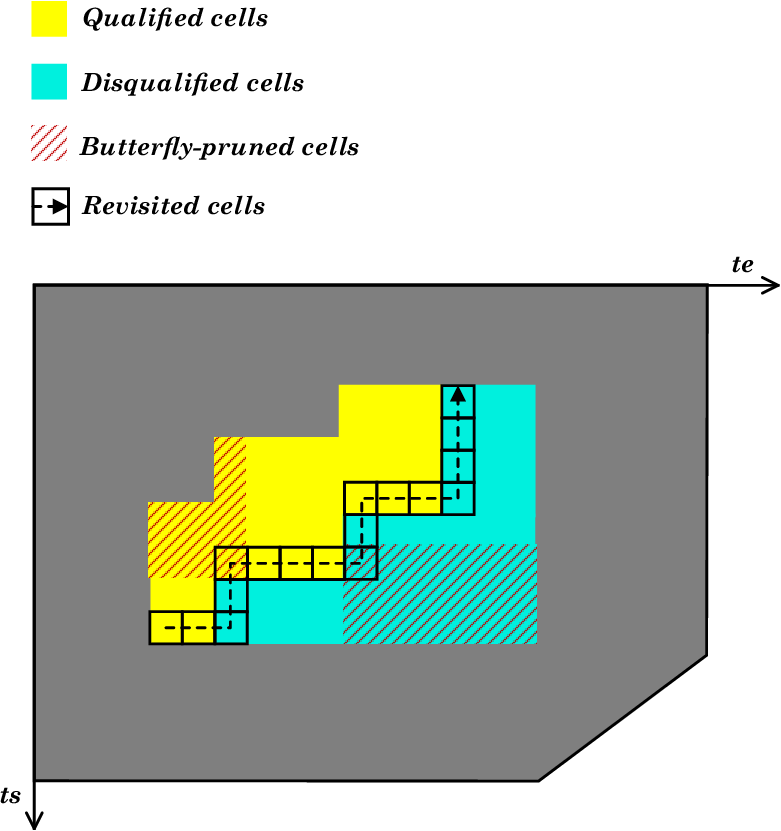}}
\caption{Abstract illustrative examples of the TTI, LTIs, rectangles, decision boundary and butterfly pruning in a time zone.}\label{fig:timezone}
\end{figure*}

The processing of TXCQ is inherently different from TCQ. When dealing with TCQ, for the subintervals that will induce an identical temporal $k$-core, we prune them as earlier as possible except at least one of them. However, for time-sensitive TXCQ, we still need to evaluate the $\mathcal{X}$ values of temporal $k$-cores of all subintervals with respect to the specific subintervals and their projected graphs. Thus, we cannot prune subintervals directly for TXCQ.

A straightforward way is to execute TCD algorithm but not conduct full TCD operation for each subinterval. Instead, we can skip the decomposition if the current subinterval can be pruned by PoR or PoU in advance because the temporal $k$-core that will be induced has been induced before. After the TCD operation for each subinterval, we check whether the $\mathcal{X}$ value of induced temporal $k$-core satisfies the given query condition. However, such an algorithm called TCD* does not preserve the scalability of OTCD algorithm for still enumerating all subintervals.

Therefore, we propose a two-phase framework to achieve scalable processing of both time-insensitive and time-monotonic TXCQ. In the first phase, we still use a revised OTCD algorithm to find all distinct temporal $k$-cores, and meanwhile all subintervals of the query time interval $[Ts,Te]$ are partitioned into a set of disjoint ``time zones''. In the second phase, each time zone will be revisited and we use dedicated local search algorithms to find the qualified subintervals in a time zone for each category of TXCQ.

\subsubsection{Phase 1: Core Induction and Zone Location}\label{sec:ph1}

Consider a set of subintervals that induce an identical temporal $k$-core whose TTI is $[ts,te]$, which is denoted by $\mathcal{B}_{[ts,te]}^{k}$. We have an intuitive and interesting observation: the subintervals in any possible $\mathcal{B}_{[ts,te]}^{k}$ are distributed in a contiguous \emph{time zone}, as illustrated by the yellow ``stair'' area of schedule table in Fig~\ref{subfig:ltizone}. Then, the monotonicity of $\mathcal{X}(\cdot)$ can be leveraged to prune subintervals during the search in each zone (see Section~\ref{sec:ph2}). To prove the correctness of our observation, we propose a new concept in contrast to TTI, namely, Loosest Time Interval (LTI).
 
\begin{definition}[Loosest Time Interval]\label{def:lti}
   Given a temporal $k$-core $\mathcal{T}_{[ts,te]}^{k}$, a time interval $[ts',te']$ is its loosest time interval if and only if\\
		1) $\mathcal{T}_{[ts',te']}^{k}$ is an identical temporal $k$-core to $\mathcal{T}_{[ts,te]}^{k}$;\\
		2) there does not exist $[ts'',te''] \supset [ts',te']$, such that $\mathcal{T}_{[ts'',te'']}^{k}$ is an identical temporal $k$-core to $\mathcal{T}_{[ts,te]}^{k}$.
\end{definition}

Intuitively, an LTI of $\mathcal{T}_{[ts,te]}^{k}$ represents a maximal subinterval that can induce a temporal $k$-core identical to $\mathcal{T}_{[ts,te]}^{k}$. In other words, expanding either boundary of LTI will lead to a structurally different temporal subgraph, like shrinking either boundary of TTI. Unlike the TTI that is unique for each distinct temporal $k$-core (see proof in~\cite{yang2023tcqcompl}), a temporal $k$-core can have multiple LTIs. We denote by $\mathcal{T}_{[ts,te]}^{k}$.LTI the set of LTIs of $\mathcal{T}_{[ts,te]}^{k}$. It is easy to prove that the intervals in $\mathcal{T}_{[ts,te]}^{k}$.LTI are partially overlapped and contain $\mathcal{T}_{[ts,te]}^{k}$.TTI. For example, Fig~\ref{subfig:ltizone} illustrates a time zone of $\mathcal{B}_{[t4,t5]}^{k}$. For any subinterval $[tx,ty]$ located in this zone, we have $\mathcal{T}_{[tx,ty]}^{k}$.TTI = $[t4,t5]$, and $\mathcal{T}_{[tx,ty]}^{k}$.LTI = $\{[t1,t6], [t2,t7], [t3,t8]\}$. For brevity, we call $[ts,te]$ of $\mathcal{B}_{[ts,te]}^{k}$ and $\mathcal{T}_{[ts,te]}^{k}$.LTI as the TTI and LTIs of the time zone $\mathcal{B}_{[ts,te]}^{k}$ respectively. 

It is worthy to note that, the TTI and LTIs actually determine the boundary of time zone $\mathcal{B}^k_{[ts,te]}$, namely, we can infer all subintervals in a time zone by its TTI and LTIs. For ease of understanding, we can divide a time zone into a set of rectangles, each of which is located by an LTI as its top-left corner and the TTI as its bottom-right corner. For example, the time zone in Fig~\ref{subfig:ltizone} is divided into the three rectangles in Fig~\ref{subfig:rectarea}.

The following theorem presents our findings of time zone formally.

\begin{theorem}\label{theo:rect}
For $\mathcal{B}_{[ts,te]}^{k}$ with LTIs $\{[ts',te']\}$, we denote by $\mathcal{R}^{[ts',te']}_{[ts,te]}=\{[ts'',te''] | ts''\in [ts',ts], te''\in [te,te']\}$ a set of subintervals that forms a rectangle, and have $\mathcal{B}_{[ts,te]}^{k}$ = $\bigcup \mathcal{R}^{[ts',te']}_{[ts,te]}$.
\end{theorem}

For each $\mathcal{R}^{[ts',te']}_{[ts,te]}$ that represents a rectangle, it is certainly that the TCD algorithm enumerates the subinterval $[ts',te']$ at the top-left corner first, so that the rest subintervals can be pruned (but will be revisited in the second phase). Thus, we design the following LTI-based pruning rule to both optimize TCD algorithm like TTI-based pruning rules and locate time zones.

\textbf{Rule 4: Rectangle-Pruning}. When a cell $[ts,te]$ with TTI $[ts',te']$ is visited in the TCD procedure, a rectangle-pruning is triggered when $[ts',te'] \subset [ts,te]$. Specifically, the other cells $\{[r,c]\}$ with $r \in [ts,ts']$ and $c \in [te',te]$ will be skipped, and $[ts,te]$ will be recorded as an LTI of $\mathcal{B}_{[ts',te']}^{k}$. 

Consider the time zone $\mathcal{B}_{[t4,t5]}^{k}$ illustrated in Fig~\ref{subfig:ltizone} again. The LTI $[t1,t6]$ will be visited first, and the TTI of temporal $k$-core induced by it is $[t4,t5] \subset [t1,t6]$. The pair of LTI and TTI locate a new rectangle $\mathcal{R}^{[t1,t6]}_{[t4,t5]}$ marked by the red box in Fig~\ref{subfig:rectarea}. Then, all other subintervals in the rectangle are pruned and $[t1,t6]$ is recorded as the LTI of $\mathcal{B}_{[t4,t5]}^{k}$. Similarly, the LTIs $[t2,t7]$ and $[t3,t8]$ will trigger rectangle-pruning, which safely and completely prunes other subintervals in $\mathcal{B}_{[t4,t5]}^{k}$, and locates the time zone with TTI together. As a result, only LTIs will be enumerated and their temporal $k$-cores will be induced during the procedure. If a time zone has only one LTI, there is even no redundant induction at all.

Algorithm~\ref{alg:phase1} gives the pseudo code of Phase 1 algorithm called OTCD*. 
Compared with OTCD that only returns all distinct temporal $k$-cores, OTCD* also returns the TTI and LTIs for each of them as the boundary of time zone.
Note that, the physical implementation of TCD operation is very efficient by using TEL. There are only two TELs (i.e., $\mathcal{T}$ and $\mathcal{T}'$) needed in the memory to run OTCD*. Moreover, the rectangle-pruning (line~\ref{lin:rp}) and the jump of pruned cells (line~\ref{lin:jump}) are implemented efficiently with some simple tricks. 
Obviously, the time complexity of OTCD* is correlated to the number of LTIs in $\mathcal{G}_{[Ts,Te]}$ and the average cost of TCD operations that induce the temporal $k$-core of LTIs.

 	\begin{algorithm}[t!]
		\DontPrintSemicolon
		\KwIn{$\mathcal{G}$, $k$, $[Ts,Te]$}
            \KwOut{\{($\mathcal{T}$, $\mathcal{T}$.TTI, $\mathcal{T}$.LTI)\} for each distinct temporal $k$-core $\mathcal{T}$ = $\mathcal{T}^k_{[ts,te]}(\mathcal{G})$ with $[ts,te]\subseteq [Ts,Te]$}
            $\mathcal{T}'\leftarrow$ TCD($\mathcal{G}$, $k$, $[Ts,Te]$)\;
            \For{$ts\leftarrow Ts$ \KwTo $Te$}{
                \While{there is cell left in this row}{
                    jump to next not pruned $[ts,te]$ from $te=Te$ until $ts$\;\label{lin:jump}
                    \If{$[ts,te]$ is the first cell visited}{
                        $\mathcal{T}\leftarrow$TCD($\mathcal{T}'$, $k$, $[ts,te]$)\;
                        \If{$te$ = $Te$}{
                            $\mathcal{T}'\leftarrow \mathcal{T}$\;
                        }
                    }
                    \Else{$\mathcal{T}\leftarrow$ TCD($\mathcal{T}$, $k$, $[ts,te]$)}
                    $[ts',te']\leftarrow \mathcal{T}$.TTI\;
                    \If{$[ts',te']\subset [ts,te]$}{
                        rectangle-pruning\;\label{lin:rp}
                    }
                    collect ($\mathcal{T}$, $[ts',te']$, $\{[ts,te]\}$)\;
                }
            }
		\caption{Phase 1 (OTCD*).}\label{alg:phase1}
	\end{algorithm}

\subsubsection{Phase 2: Zone Revisit and Local Search}\label{sec:ph2}

With the output of Phase 1 algorithm, our Phase 2 algorithm will search each returned time zone of $\mathcal{B}^k_{[ts,te]}$ respectively to find the result temporal $k$-cores whose $\mathcal{X}$ values are optimal or satisfy the constraint. Obviously, evaluate $\mathcal{X}(\mathcal{T}^k_{[ts',te']})$ for each $[ts',te'] \in \mathcal{B}^k_{[ts,te]}$ is a straighforward but not scalable method. Instead, we propose three local search algorithms to handle time-insensitive, time-monotonic optimizing and time-monotonic constraining TXCQ respectively. They have one thing in common, namely, only the necessary subintervals in each $\mathcal{B}^k_{[ts,te]}$ will be revisited.

\textbf{Time-Insensitive Local Search (TI-LS)}. It is the simplest case. Since $\mathcal{X}(\cdot)$ is time-insensitive and all subintervals in $\mathcal{B}^k_{[ts,te]}$ induce an identical temporal $k$-core, we only need to evaluate $\mathcal{X}(\mathcal{T}^k_{[ts,te]})$ for each $\mathcal{B}^k_{[ts,te]}$. If $\mathcal{X}(\mathcal{T}^k_{[ts,te]})$ is not globally optimal or does not satisfy the constraint, all subintervals in $\mathcal{B}^k_{[ts,te]}$ are discarded. Thus the time complexity of TI-LS is the same as that of $\mathcal{X}$ evaluation.

\textbf{Time-Monotonic Optimizing Local Search (TMO-LS)}. According to Definition~\ref{def:tm}, it is surely that for each $\mathcal{B}^k_{[ts,te]}$ the optimal $\mathcal{X}$ value can only be achieved by LTI or TTI, which depends on $\mathcal{X}(\cdot)$ is monotonically increasing or decreasing. Specifically, for each $\mathcal{B}_{[ts,te]}^{k}$ with LTIs $\{[ts',te']\}$, we only evaluate $\mathcal{X}(\mathcal{T}^k_{[ts,te]})$ if $\mathcal{X}(\cdot)$ is monotonically decreasing, or only evaluate $\mathcal{X}(\mathcal{T}^k_{[ts',te']})$ for all LTIs. Lastly, the LTIs or TTIs with the globally optimal $\mathcal{X}$ value will be returned.
Thus the time overhead of TMO-LS is at most $|\{[ts',te']\}|$ times of $\mathcal{X}$ evaluation.

\textbf{Time-Monotonic Constraining Local Search (TMC-LS)}. It is the most challenging case, which normally returns the temporal $k$-cores with $\mathcal{X}$ values better than a given threshold $\sigma$. An insightful observation is that, due to the time-monotonicity of $\mathcal{X}(\cdot)$, the distribution of $\mathcal{X}$ values in a time zone can be leveraged to avoid unnecessary computation. As illustrated in Fig~\ref{subfig:gpv}, there is a ``decision boundary'' between the subintervals satisfying the constraint and the other subintervals. Thus, the key to address this kind of TXCQ is searching along the boundary.

	\begin{algorithm}[t!]
		\DontPrintSemicolon
		\KwIn{($\mathcal{T}$, $[ts',te']$, $\{[ts,te]\}$), $\mathcal{X}(\cdot)$, $\sigma$}
		  \KwOut{all subintervals $[ts'',te'']\in$ $\mathcal{B}_{[ts',te']}^{k}$ with $\mathcal{X}(\mathcal{T}_{[ts'',te'']}^{k})\succcurlyeq$$\sigma$}
            $ts''\leftarrow ts'$, $te''\leftarrow \max\{te\}$\;\label{lin:start}
            \While{$ts''\geqslant \min\{ts\}$ and $te''\geqslant te'$}{\label{lin:end}
                \If{$[ts'',te'']\notin \mathcal{B}_{[ts',te']}^{k}$}{\label{lin:notbelong}
                    $te''\leftarrow te$ of next LTI, or break if there is none\;\label{lin:next}
                }
                \If{$\mathcal{X}(\mathcal{T}_{[ts'',te'']}^{k})\succcurlyeq \sigma$}{\label{lin:better}
                    collect $[r,te'']\in \mathcal{B}_{[ts',te']}^{k}$ with $r\leqslant ts''$\;\label{lin:collect}
                    $te''\leftarrow te''-1$\;\label{lin:right}
                }
                \Else{
                    $ts''\leftarrow ts''-1$\;\label{lin:up}
                }
            }
		\caption{Phase 2 (TMC-LS)}\label{alg:phase2}
	\end{algorithm}

Without loss of generality, we assume $\mathcal{X}(\cdot)$ is monotonically increasing, since the monotonically decreasing case can be solved by a mirrored process. Then, we have the following theorem.

\begin{theorem}\label{thm:boundary}
Given $\mathcal{B}^k_{[ts,te]}$, for a subinterval $[ts',te'] \in \mathcal{B}^k_{[ts,te]}$, if it is qualified, namely, $\mathcal{X}(\mathcal{T}^k_{[ts',te']}) \succcurlyeq \delta$, we have $\mathcal{X}(\mathcal{T}^k_{[r,c]}) \succcurlyeq \delta$ for any subinterval $[r,c] \in \mathcal{B}^k_{[ts,te]}$ with $r \leqslant ts'$ and $c \geqslant te'$, and if it is unqualified, namely, $\mathcal{X}(\mathcal{T}^k_{[ts',te']}) \prec \delta$, we have $\mathcal{X}(\mathcal{T}^k_{[r,c]}) \prec \delta$ for any subinterval $[r,c] \in \mathcal{B}^k_{[ts,te]}$ with $r \geqslant ts'$ and $c \leqslant te'$.
\end{theorem}

Theorem~\ref{thm:boundary} implies that, due to the time-monotonicity of $\mathcal{X}(\cdot)$, expanding a qualified subinterval will result in another qualified subinterval, and shrinking an unqualified subinterval will result in another unqualified subinterval, as long as the time interval is still in the same time zone.

Based on Theorem~\ref{thm:boundary}, we can optimize TMC-LS in a time zone of $\mathcal{B}^k_{[ts,te]}$ by using a logical ``butterfly'' pruning. Whenever a qualified subinterval is enumerated, we prune all the subintervals on its top-left side in the zone. In contrast, whenever an unqualified subinterval is enumerated, we prune all the subintervals on its bottom-right side in the zone. For example, the two areas marked by red diagonal lines are pruned, as illustrated in Fig~\ref{subfig:gpv}.

The pseudo code of TMC-LS algorithm is presented in Algorithm~\ref{alg:phase2}. The algorithm takes a triple returned by Algorithm~\ref{alg:phase1} as input. In the time zone of $\mathcal{B}^k_{[ts',te']}$, it searches from the bottom-left cell $[ts',\max\{te\}]$ (line~\ref{lin:start}) until arriving the top-right cell $[\min\{ts\},te']$ (line~\ref{lin:end}). For each revisited cell $[ts'',te'']$, we evaluate $\mathcal{X}(\mathcal{T}^k_{[ts'',te'']})$, and if the value is better than $\sigma$, all cells on the top of $[ts'',te'']$ are qualified and collected (lines~\ref{lin:better}-\ref{lin:collect}). The next cell to be revisited is the right cell of $[ts'',te'']$ if it is qualified (line~\ref{lin:right}) or the upper cell (line~\ref{lin:up}). Besides, if the revisited cell gets out of the time zone, we jump back to the next possibly qualified cell directly (lines~\ref{lin:notbelong}-\ref{lin:next}). In this procedure, the butterfly-pruning is performed implicitly, and only the cells along the ``decision boundary'' have to be revisited, as illustrated in Fig~\ref{subfig:gpv}. 
Thus the time overhead of TMC-LS is bounded by $p+q$ times of $\mathcal{X}$ evaluation, where $p$ and $q$ are the maximum width and height of rectangles in the given time zone.
As a result, TMC-LS still preserves the scalability.


\section{Experiment}\label{sec:expr}
In this section, we conduct experiments to verify both efficiency and effectiveness of the proposed algorithm on a Windows machine with Intel Core i7 2.20GHz CPU and 64GB RAM. The algorithms are implemented through C++ Standard Template Library. 

	

\noindent\textbf{Dataset}. We choose temporal graphs with different sizes, time spans and domains for our experiments from SNAP~\cite{snapnets} and KONECT~\cite{kunegis2013konect}. The basic statistics of these graphs are given in Table~\ref{table:dataset}. All timestamps are normalized to integers in seconds.
	
	\begin{table}[t]
		\centering
		\caption{Datasets.}\label{table:dataset}
		\begin{tabular}{lccc}
			\hline
			Name (\underline{abbreviation}) & $|\mathcal{V}|$ & $|\mathcal{E}|$ & Span(day)\\
			\hline
			\underline{CollegeMsg} & 1.8K & 20K & 193\\
			\underline{email}-Eu-core-temporal & 0.9K & 332K & 803\\
			sx-\underline{mathoverflow} & 24.8K & 506K & 2350\\
			sx-\underline{stackoverflow} & 2.6M & 63.5M & 2774\\\hline
			\underline{Youtube} & 3.2M & 9.4M & 226\\
			\underline{Flickr} & 2.3M & 33M & 198\\
			\underline{DBLP} & 1.8M & 29.5M & 17532\\
			\hline
		\end{tabular}
	\end{table}

\noindent\textbf{Algorithm}. For TCQ tests, we compare iPHC-Query~\cite{yang2023tcq}, TCD and OTCD. For TXCQ tests, we choose PHC* and TCD* as baselines, and compare the two-phase algorithm with corresponding local search strategies (named OTCD*+LS) with them.

\noindent\textbf{Query}. We choose the twenty TCQ instances in~\cite{yang2023tcqcompl} with specific $\mathcal{G}$, $k$ and $[Ts, Te]$ as benchmark. For comprising time-insensitive and time-monotonic TXCQ instances, we choose size~\cite{yao2021efficient} and engagement~\cite{zhang2020exploring} as the additional metrics $\mathcal{X}$ respectively to extend the TCQ instances. As a result, we have twenty minimum, most engaged and engagement-constrained ($\sigma$ = 0.6) temporal $k$-core queries respectively.

\subsection{Efficiency}

\subsubsection{TCQ Efficiency}

\begin{figure}[t]
		\centering
            \captionsetup[subfloat]{labelfont=scriptsize,textfont=scriptsize}
		\subfloat[CollegeMsg]{\label{subfig:collegemsg-col}
			\includegraphics[width=0.5\linewidth]{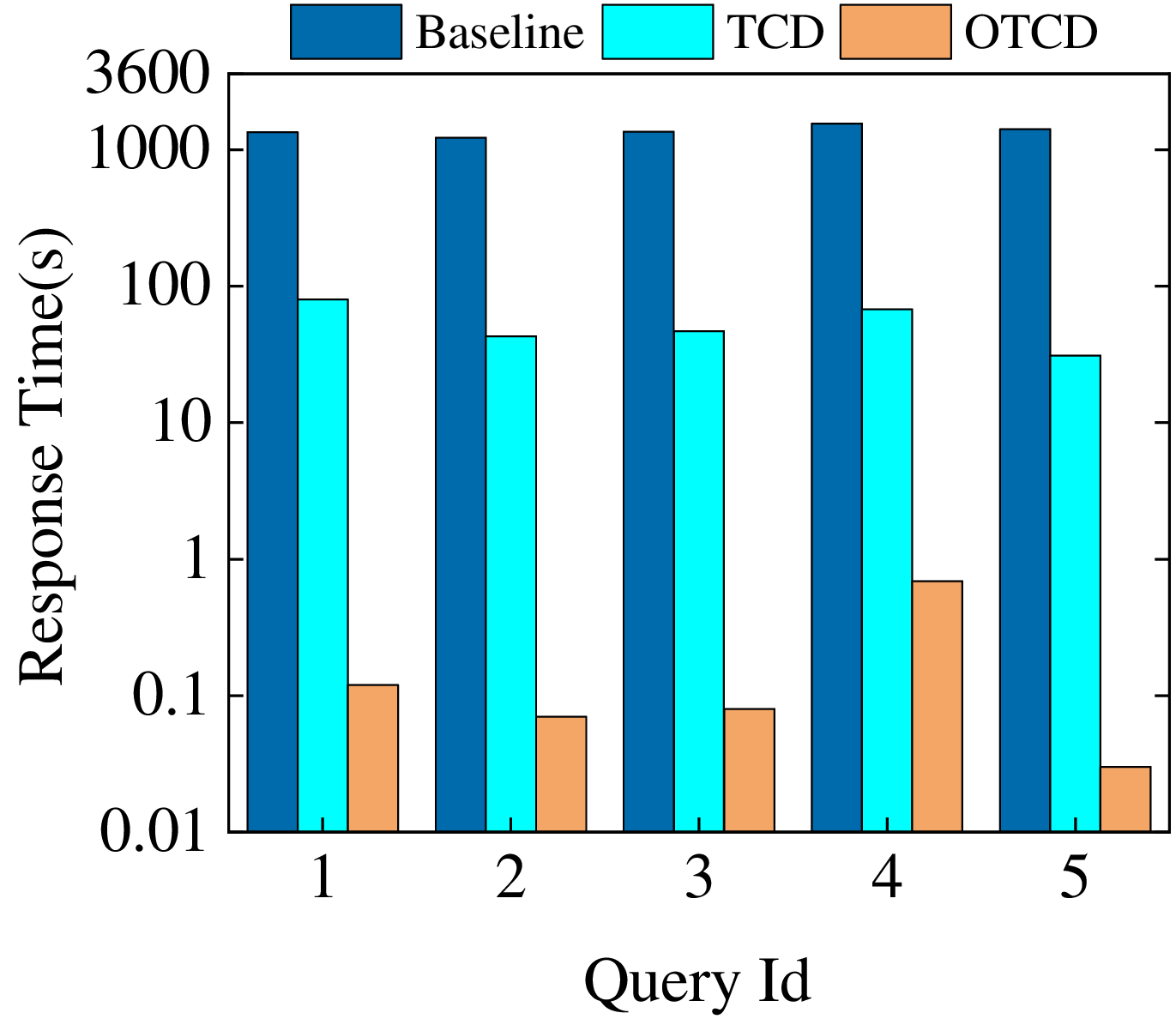}}
		\subfloat[email-Eu-core-temporal]{\label{subfig:email-Eu-core-temporal-colume}
			\includegraphics[width=0.5\linewidth]{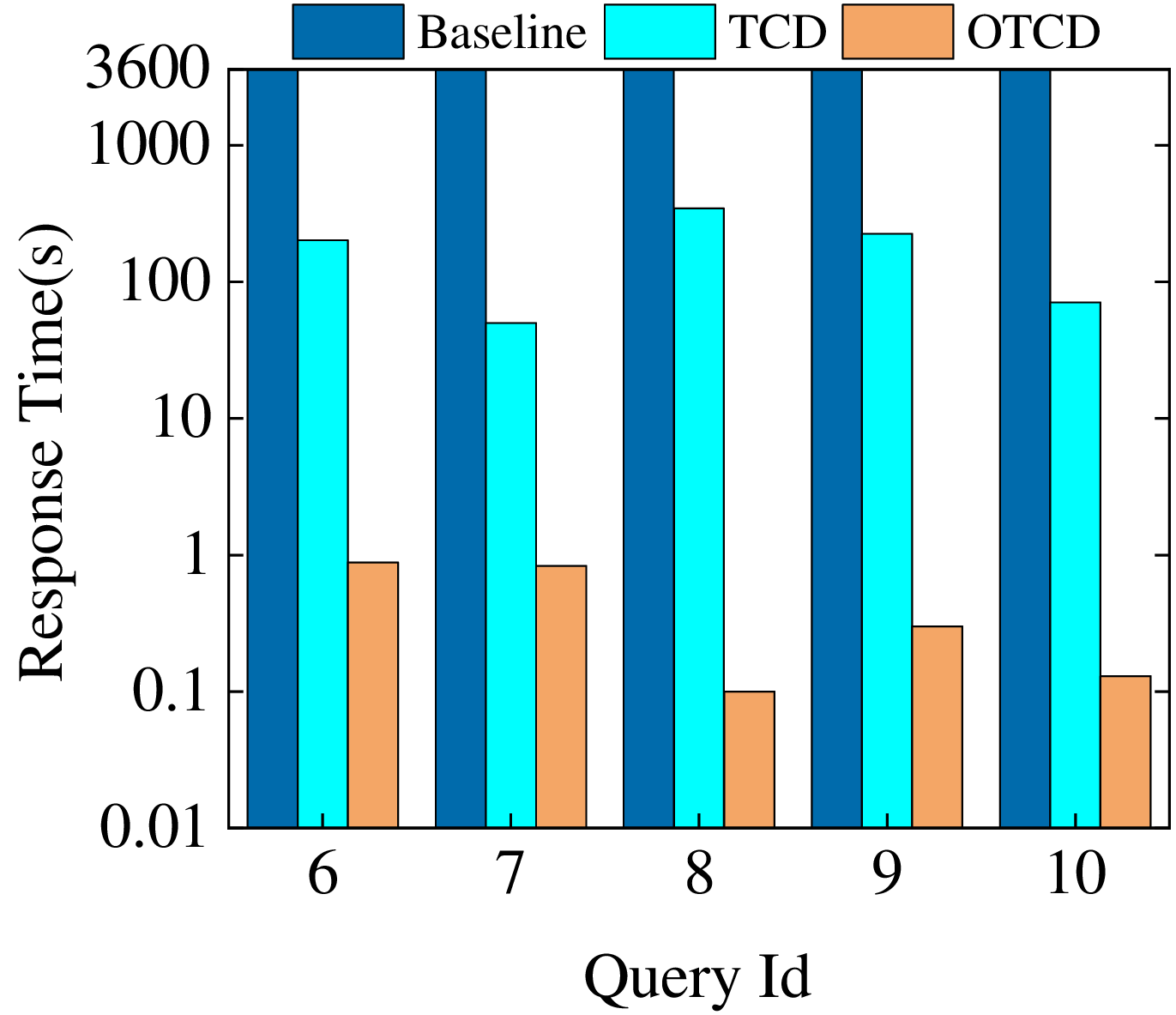}}
		\newline
		\subfloat[sx-mathoverflow]{\label{subfig:sx-mathoverflow-colume}
			\includegraphics[width=0.5\linewidth]{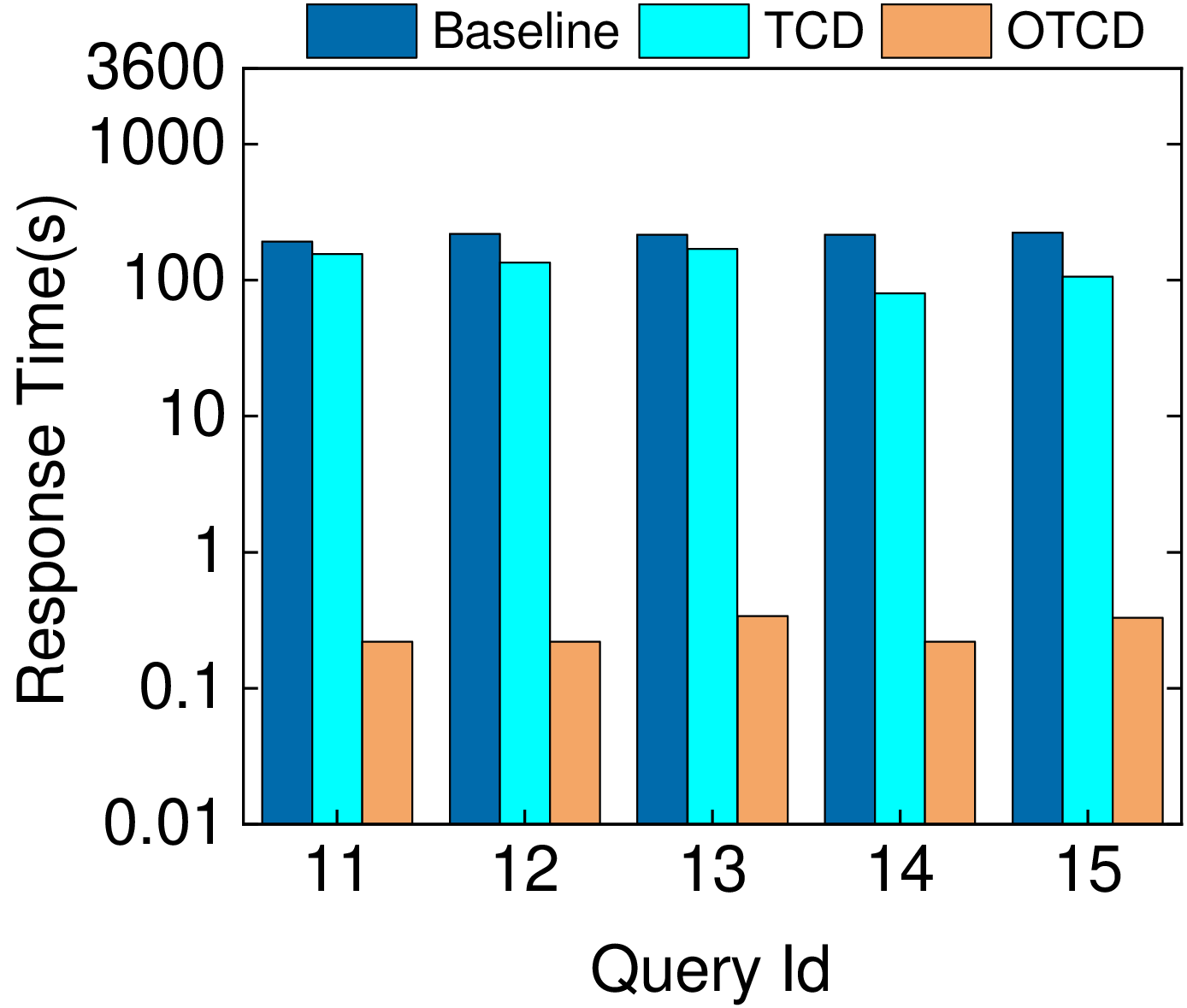}}
		\subfloat[sx-stackoverflow]{\label{subfig:sx-stackoverflow-colume}
			\includegraphics[width=0.5\linewidth]{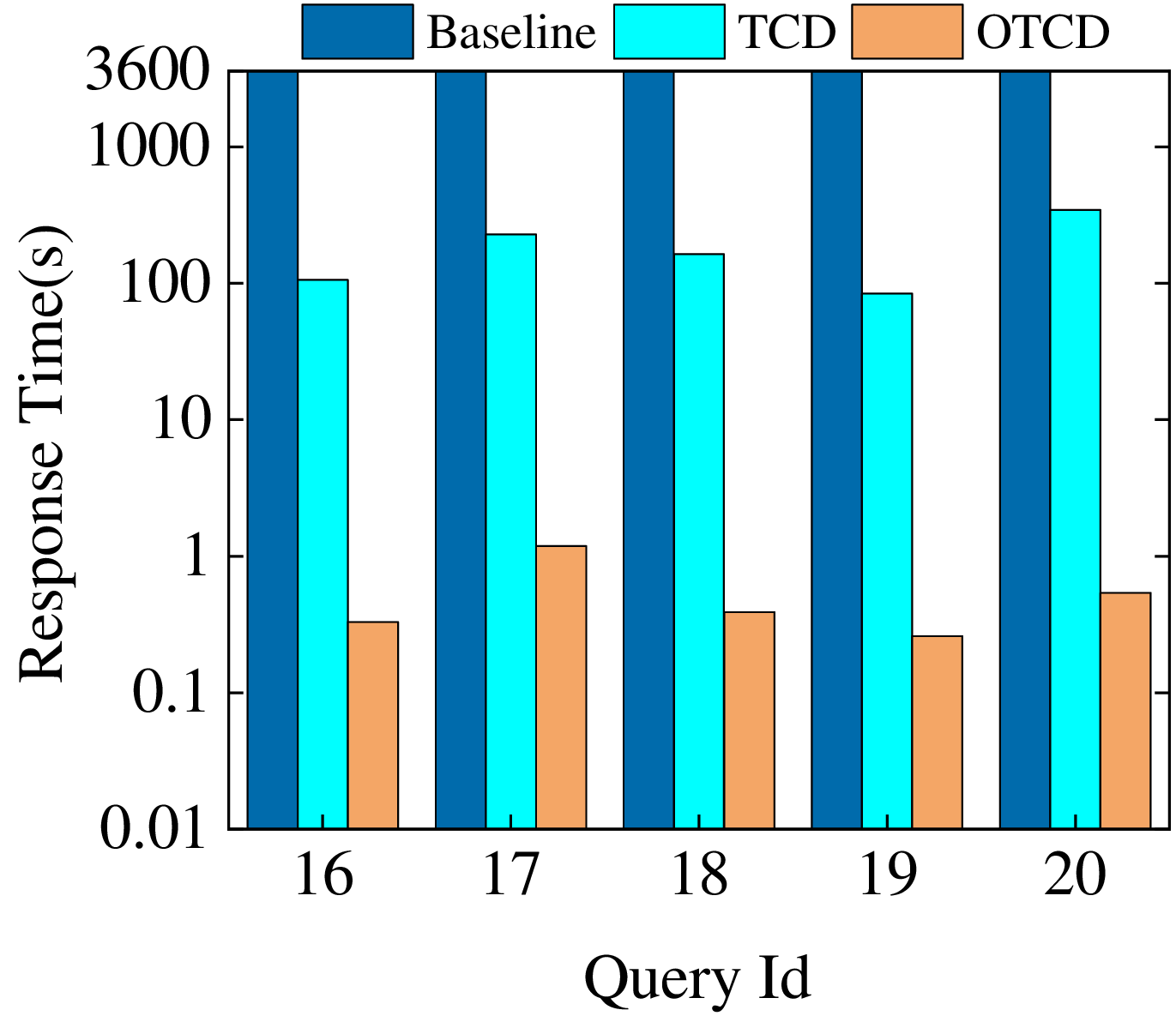}}
		
		\caption{The response time of selected queries on SNAP graphs.}\label{fig:colchart}
		
	\end{figure}

	
Figure~\ref{fig:colchart} compares the response time of Baseline (iPHC-Query), TCD and OTCD algorithms for each selected query respectively, which clearly demonstrates the efficiency of our algorithm. Firstly, TCD performs better than baseline for all twenty queries due to the physical efficiency of TEL, though they both decrementally or incrementally induce temporal $k$-cores. Specifically, TCD spends around 100 sec for each query. In contrast, baseline spends more than 1000 sec on CollegeMsg and even cannot finish within an hour on two other graphs, though it uses a precomputed index. Furthermore, OTCD runs two or three orders of magnitude faster than TCD, and only spends about 0.1-1 sec for each query, which verifies the effectiveness of our pruning method based on TTI. 
	
	\begin{figure}[t]
		\centering
		\includegraphics[width=\linewidth]{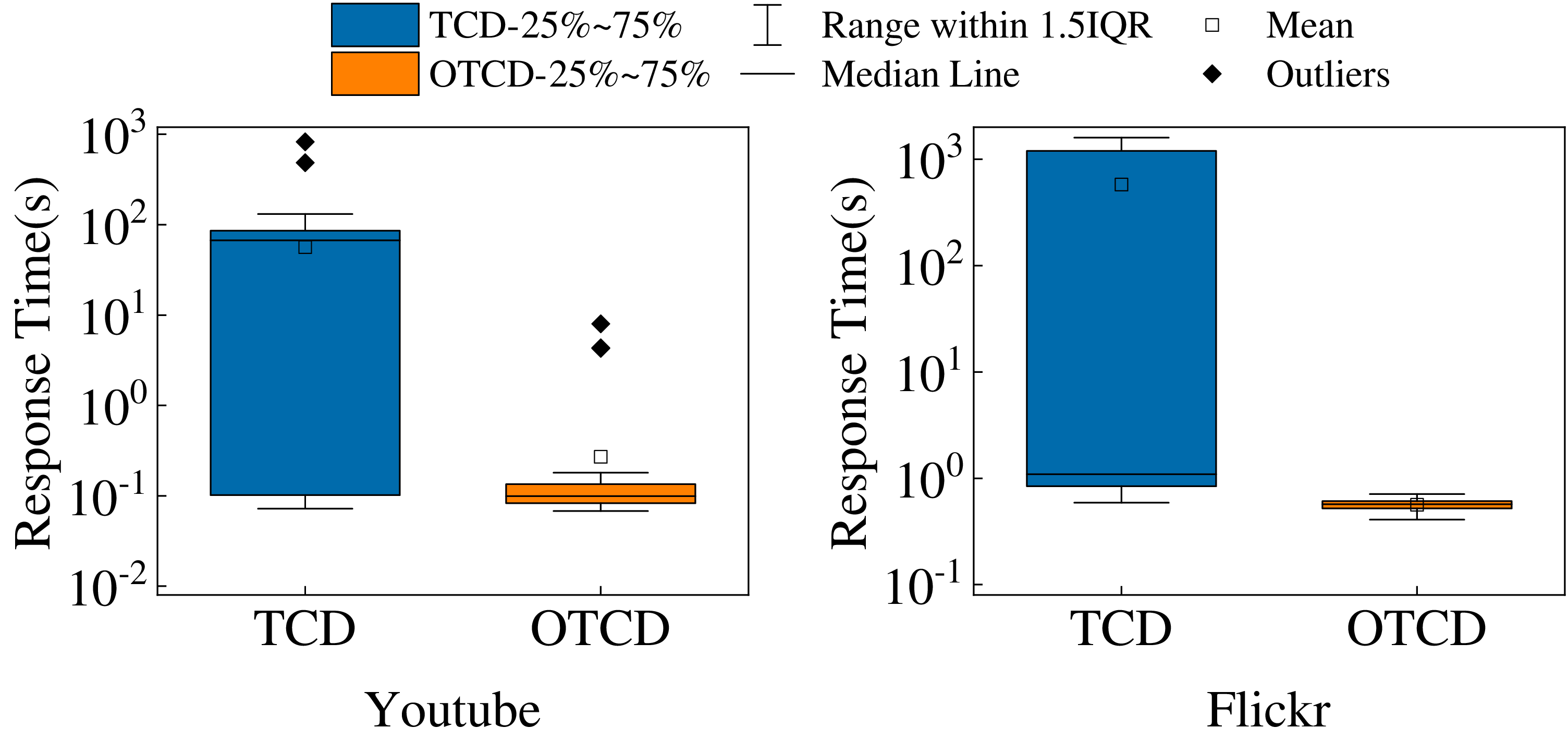}
		\caption{The statistical distribution of response time of random queries on
KONECT graphs.}\label{fig:box}
	\end{figure}
 
		\begin{table}[t]
		\centering
		\caption{Effect of pruning rules.}\label{table:prune}
		\begin{tabular}{c|ccc|ccc|c}
			\hline 
            \multirow{2}{*}{id} & 
            \multicolumn{3}{|c|}{Triggered Times} &
            \multicolumn{4}{|c}{Pruned Cell Percentage (\%)} \\
            \cline{2-8}
            &  PoR & PoU & PoL & PoR & PoU & PoL & Total\\
			\hline
			1 & 54 & 72 & 2 & 0.02 & 72 & 23.6 & 95.62\\
			6 & 2 & 4 & 1 & 0.01 & 51.8 & 32.1 & 83.91\\
			11 & 8 & 10 & 1 & 0.04 & 57.1 & 24.5 & 81.64\\
            16 & 5 & 9 & 1 & 0.04 & 56.9 & 33.5 & 90.44\\
			\hline
		\end{tabular}
	\end{table}

To compare the effect of three pruning rules in OTCD algorithm, Table~\ref{table:prune} lists their triggered times and the percentage of subintervals pruned by them for several queries respectively. PoR and PoU are triggered frequently because their conditions are more easily to be satisfied. However, PoR actually contributes pruned subintervals much less than the other two. Because it only prunes subintervals in the same row, and in contrast, PoU and PoL can prune an ``area'' of subintervals. Overall, the three pruning rules can achieve significant optimization effect together by enabling OTCD algorithm to skip more than 80 percents of subintervals.

To evaluate the stability of our approach, we conduct statistical analysis of one hundred valid random queries on two new graphs, namely, Youtube and Flickr. We visualize the distribution of response time of TCD and OTCD algorithms for these random queries as boxplots, which are shown by Figure~\ref{fig:box}. The boxplots demonstrate that the response time of OTCD varies in a very limited range, which indicates that the OTCD indeed performs stable in practice. The outliers represent some queries that have exceptionally more results, which can be seen as a normal phenomenon in reality. They may reveal that many communities of the social networks are more active during the period.

\begin{table}
    \centering
    \caption{Memory consumption of (O)TCD algorithm.}
    \begin{tabular}{lc}
    \hline
    Dataset & Process Memory (GB)\\
    \hline
    CollegeMsg & 0.02\\
    sx-mathoverflow & 0.06\\
    Youtube & 1.7\\
    DBLP & 3.1\\
    Flickr & 3.5\\
    sx-stackoverflow & 6.5\\
    \hline
    \end{tabular}
	\label{tab:procmem}
\end{table}

Moreover, Table~\ref{tab:procmem} reports the process memory consumption for different datasets, which depends on the size of TEL mostly. We can observe that, 1) for the widely-used graphs like Youtube, DBLP, Flickr and stackoverflow, several gigabytes of memory are needed for storing TEL, which is acceptable for the ordinary hardware; and 2) for the very large graphs with billions of edges, the size of TEL is hundreds of gigabytes approximately, which would require the distributed memory cluster like Spark.
	
To verify the scalability of our method with respect to the query parameters, we test the three algorithms with varing minimum degree $k$ and time span (namely, $Te-Ts$) respectively.

\begin{figure}[t!]
		\centering
            \captionsetup[subfloat]{labelfont=scriptsize,textfont=scriptsize}
		\subfloat[CollegeMsg]{\label{subfig:collegemsg-k}
			\includegraphics[width=0.33\linewidth]{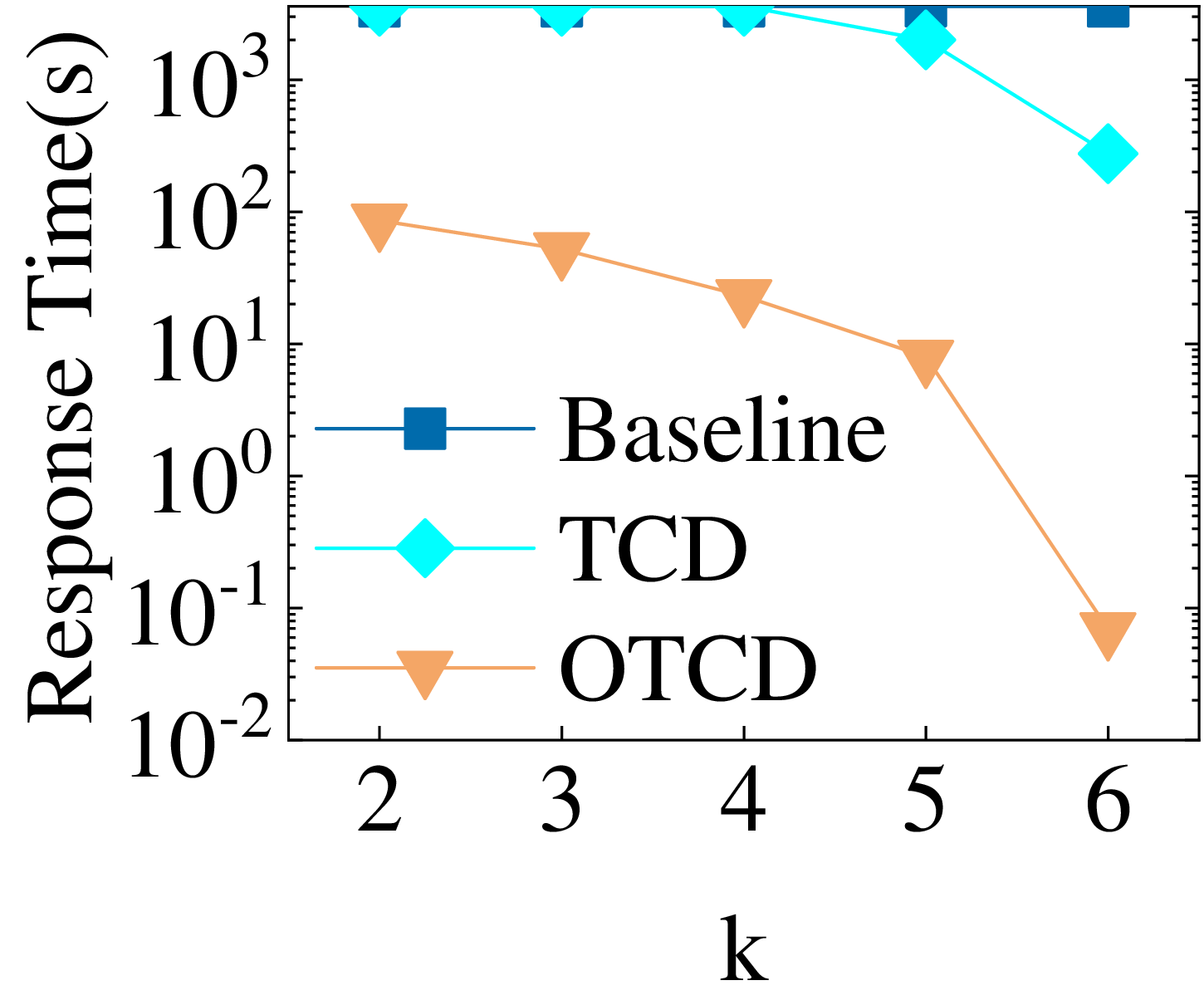}}
		\subfloat[sx-mathoverflow]{\label{subfig:sx-mathoverflow-k}
			\includegraphics[width=0.33\linewidth]{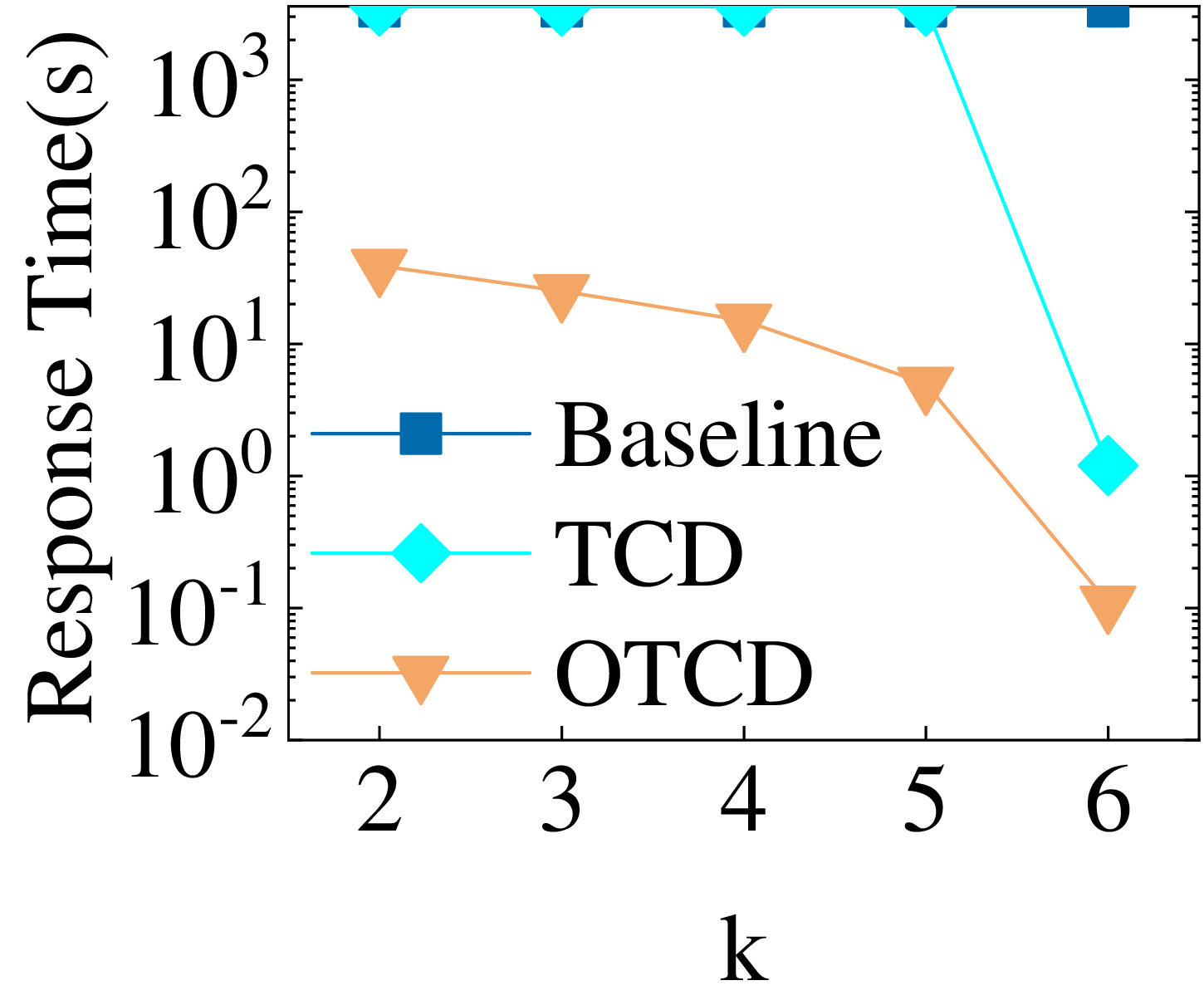}}
		\subfloat[sx-stackoverflow]{\label{subfig:sx-stackoverflow-k}
			\includegraphics[width=0.33\linewidth]{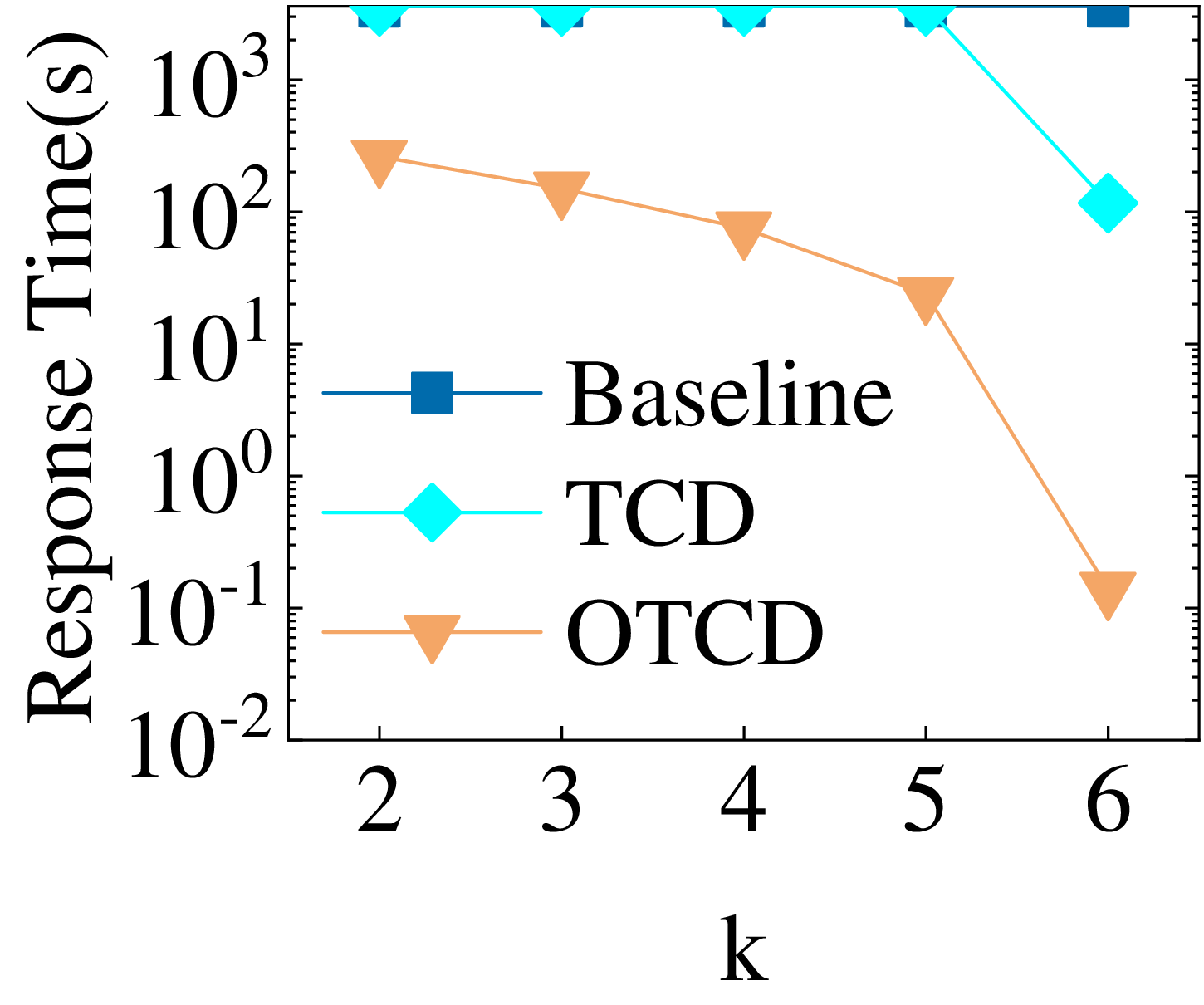}}
		\caption{The TCQ response time with respect to varying $k$.}\label{fig:kchart}
	\end{figure}

\begin{figure}[t!]
		\centering
            \captionsetup[subfloat]{labelfont=scriptsize,textfont=scriptsize}
		\subfloat[CollegeMsg]{\label{subfig:collegemsg-kn}
			\includegraphics[width=0.33\linewidth]{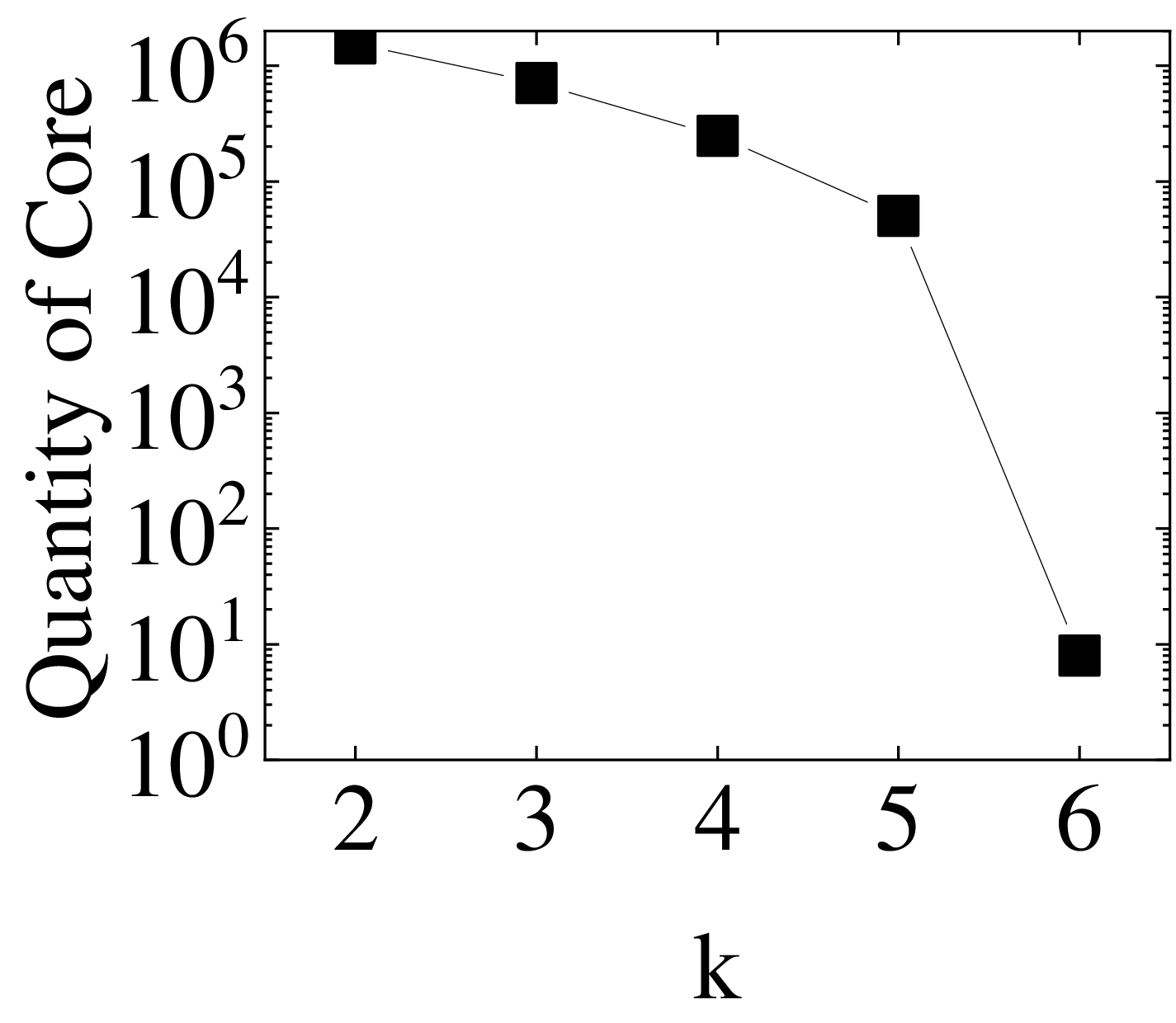}}
		\subfloat[sx-mathoverflow]{\label{subfig:sx-mathoverflow-kn}
			\includegraphics[width=0.33\linewidth]{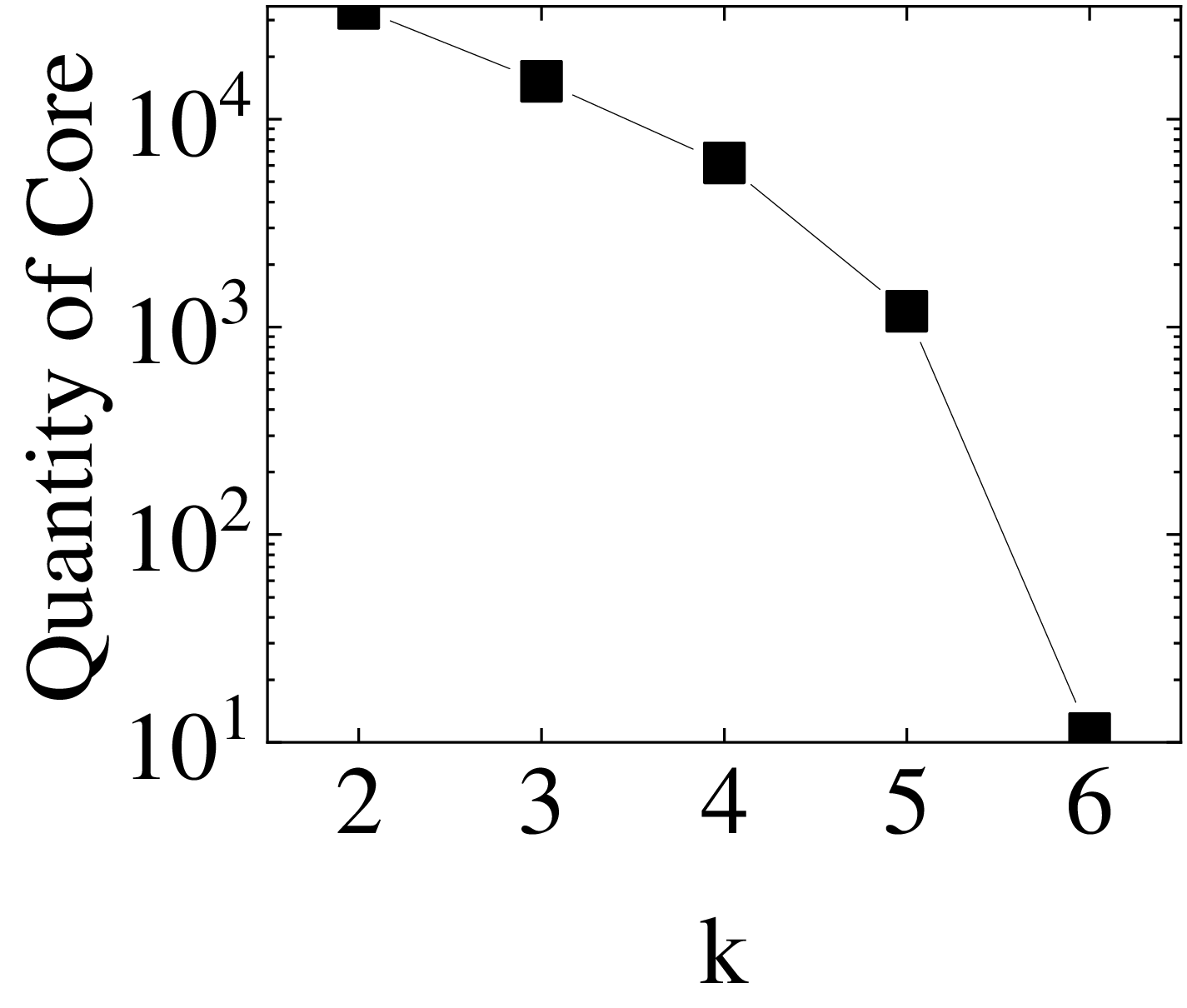}}
		\subfloat[sx-stackoverflow]{\label{subfig:sx-stackoverflow-kn}
			\includegraphics[width=0.33\linewidth]{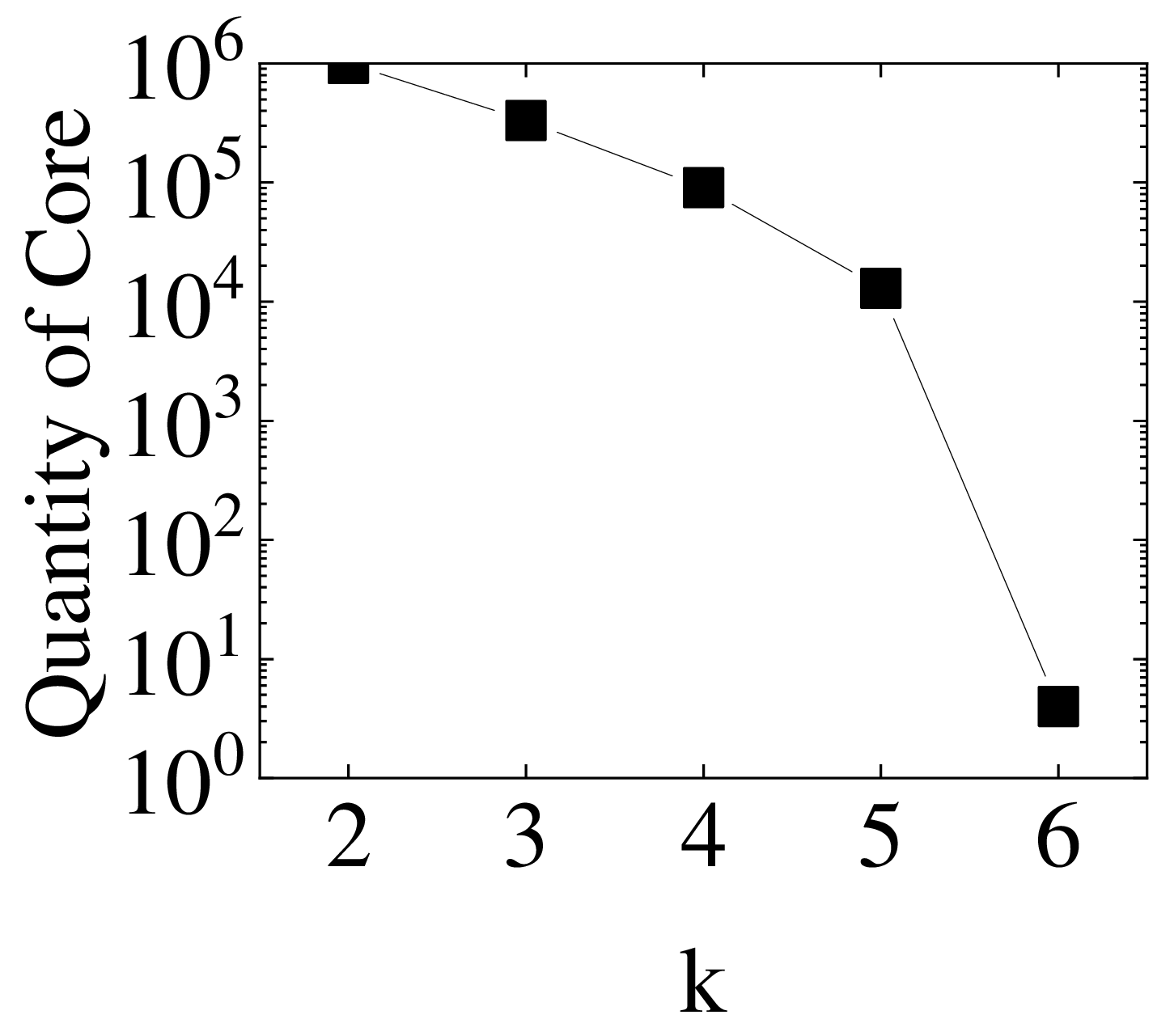}}
		\caption{The distinct temporal $k$-core number with respect to varying $k$.}\label{fig:kchart2}
	\end{figure}

We select a typical query with span fixed and $k$ ranging from 2 to 6 for different graphs. The response time of tested algorithms are presented in Figure~\ref{fig:kchart}, from which we have an important observation against common sense. That is, different from core decomposition on non-temporal graphs, when the value of $k$ increases, the response time of TCD and OTCD algorithms decreases gradually. For OTCD, the behind rationale is clear, namely, its time cost is only bounded by the scale of results, which decreases sharply with the increase of $k$. To support the claim, Figure~\ref{fig:kchart2} shows the trend of the amount of result cores changing with $k$. Intuitively, a greater value of k means a stricter constraint and thereby filters out some less cohesive cores. We can see the trend of runtime decrease for OTCD in Figure~\ref{fig:kchart} is almost the same as the trend of core amount decrease in Figure~\ref{fig:kchart2}, which also confirms the scalability of OTCD algorithm. For TCD, the behind rationale is complicated, since it enumerates all subintervals and each single decomposition is costlier with a greater value of $k$. It is just like peeling an onion layer by layer, which has less layers with a greater value of $k$, so that the total computational cost for core induction become less.

\begin{figure}[t!]
            \centering
            \captionsetup[subfloat]{labelfont=scriptsize,textfont=scriptsize}
		\subfloat[CollegeMsg]{\label{subfig:CollegeMsg-Span}
			\includegraphics[width=0.33\linewidth]{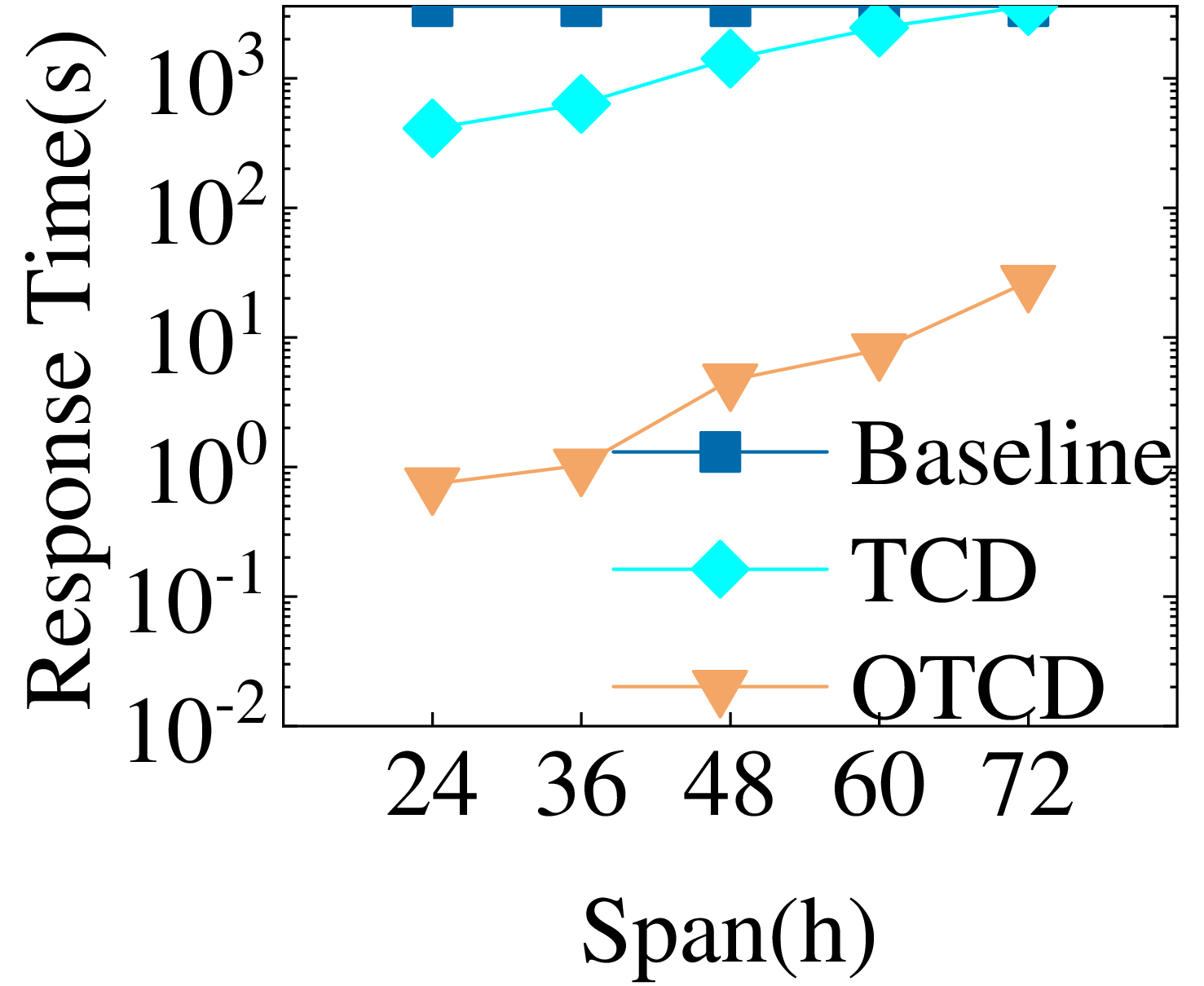}}
		\subfloat[sx-mathoverflow]{\label{subfig:sx-mathoverflow-Span}
			\includegraphics[width=0.33\linewidth]{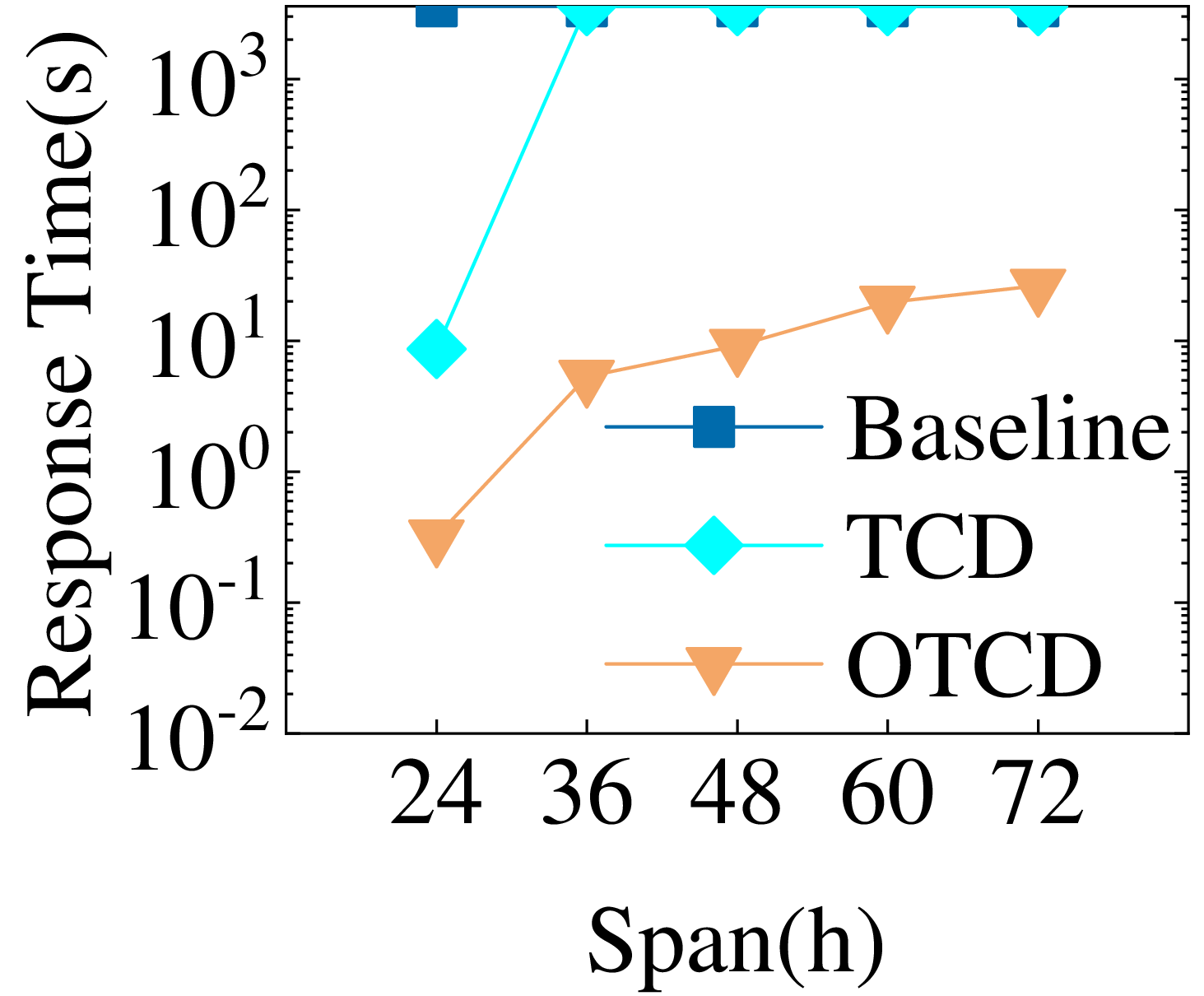}}
		\subfloat[sx-stackoverflow]{\label{subfig:sx-stackoverflow-Span}
			\includegraphics[width=0.33\linewidth]{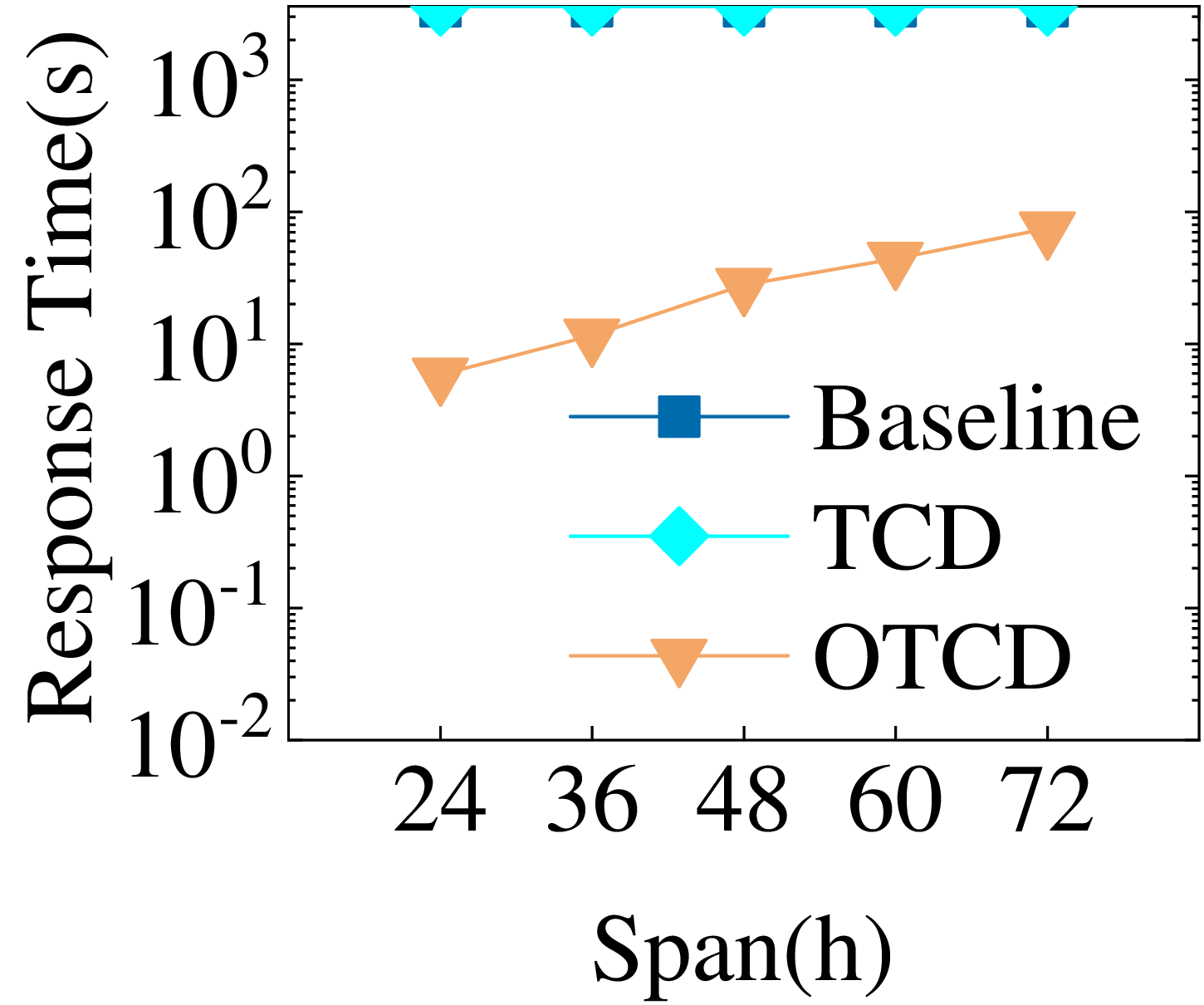}}
		\caption{The TCQ response time with respect to varying time span.}\label{fig:spanchart}	
	\end{figure}

Similar to the test of $k$, we also evaluate the scalability of different algorithms when the query time span increases. The results are presented in Figure~\ref{fig:spanchart}. Although the number of subintervals increases quadratically, the response time of OTCD still increases moderately following the scale of query results. In contrast, TCD runs dramatically slower when the query time span becomes longer.

The above results demonstrate that the efficiency of OTCD is not sensitive to the change of query parameters. Moreover, for a large graph with a long time span like Youtube, we test OTCD algorithm by querying temporal 10-cores over the whole time span. The result is, to find all 19,146 temporal 10-cores within 226 days, the OTCD algorithm spent about 55 minutes, which is acceptable for such a ``full graph scan'' task.
	
\subsubsection{TXCQ Efficiency}

\begin{figure*}[t!]
\centering
\captionsetup[subfloat]{labelfont=scriptsize,textfont=scriptsize}
    \subfloat[TI, college]{\label{subfig:college-i}
        \includegraphics[width=0.16\textwidth]{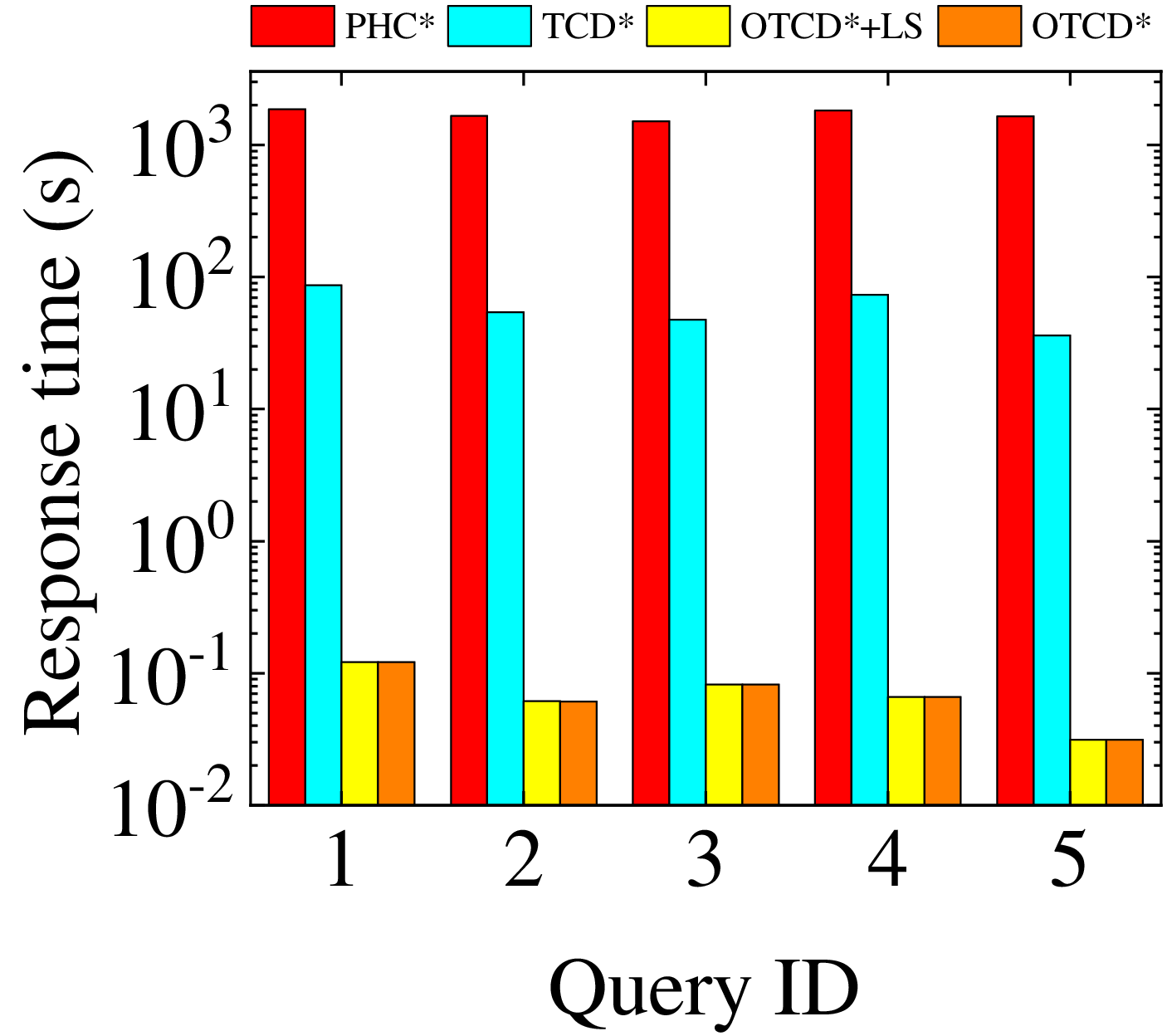}}
    \subfloat[TI, email]{\label{subfig:email-i}
        \includegraphics[width=0.16\textwidth]{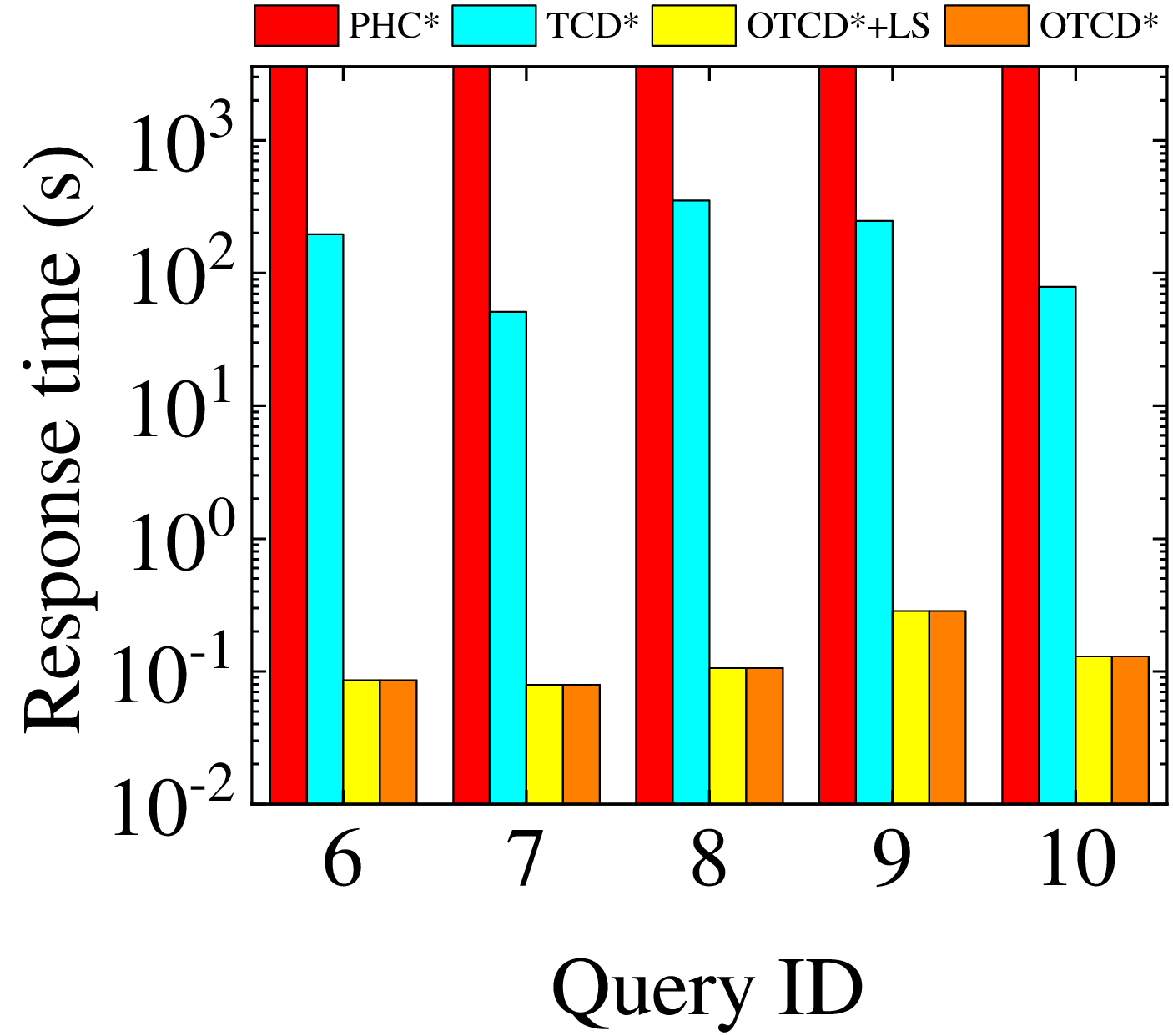}}
    \subfloat[TI, math]{\label{subfig:math-i}
        \includegraphics[width=0.16\textwidth]{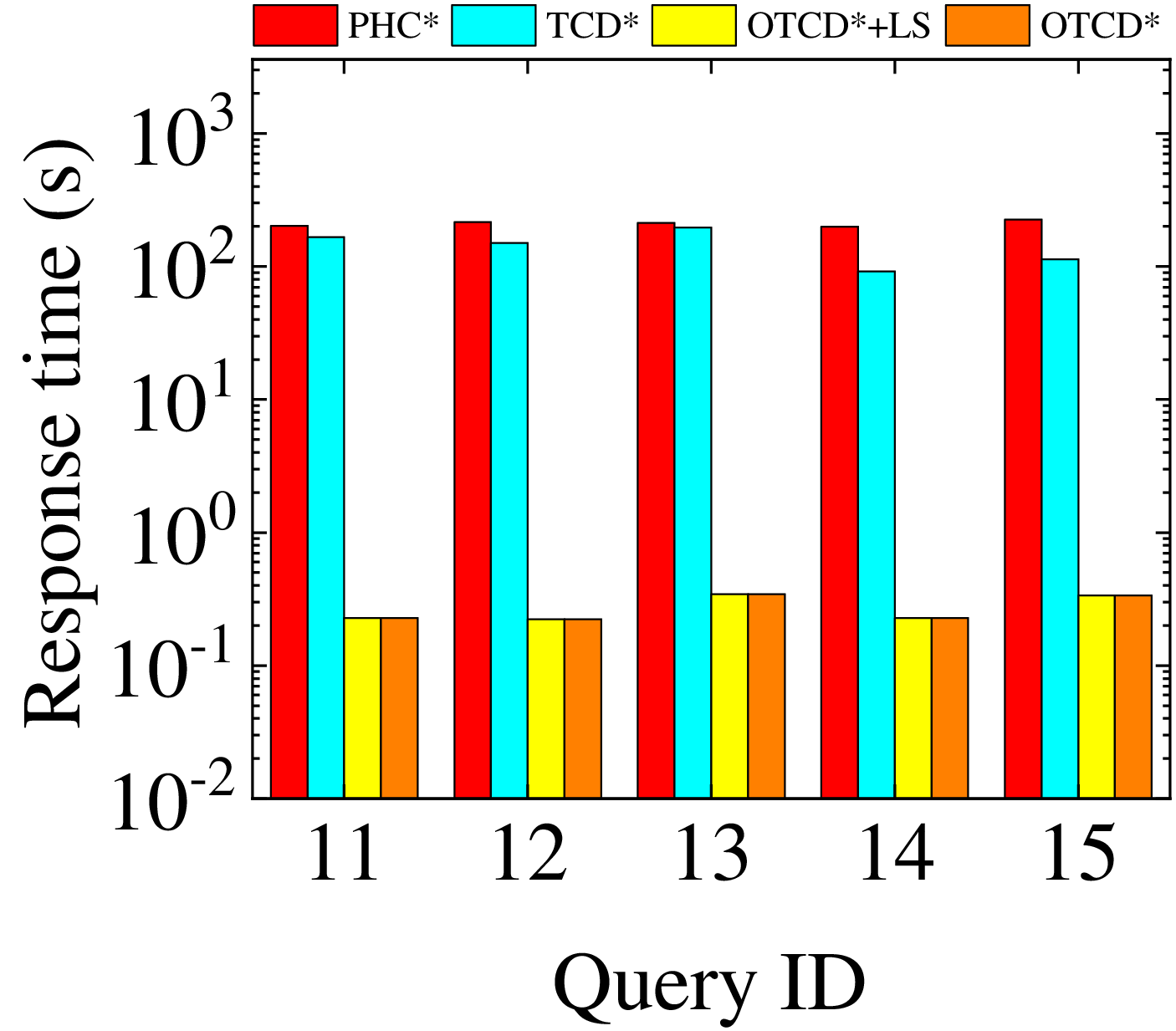}}
    \subfloat[TI, stack]{\label{subfig:stack-i}
        \includegraphics[width=0.16\textwidth]{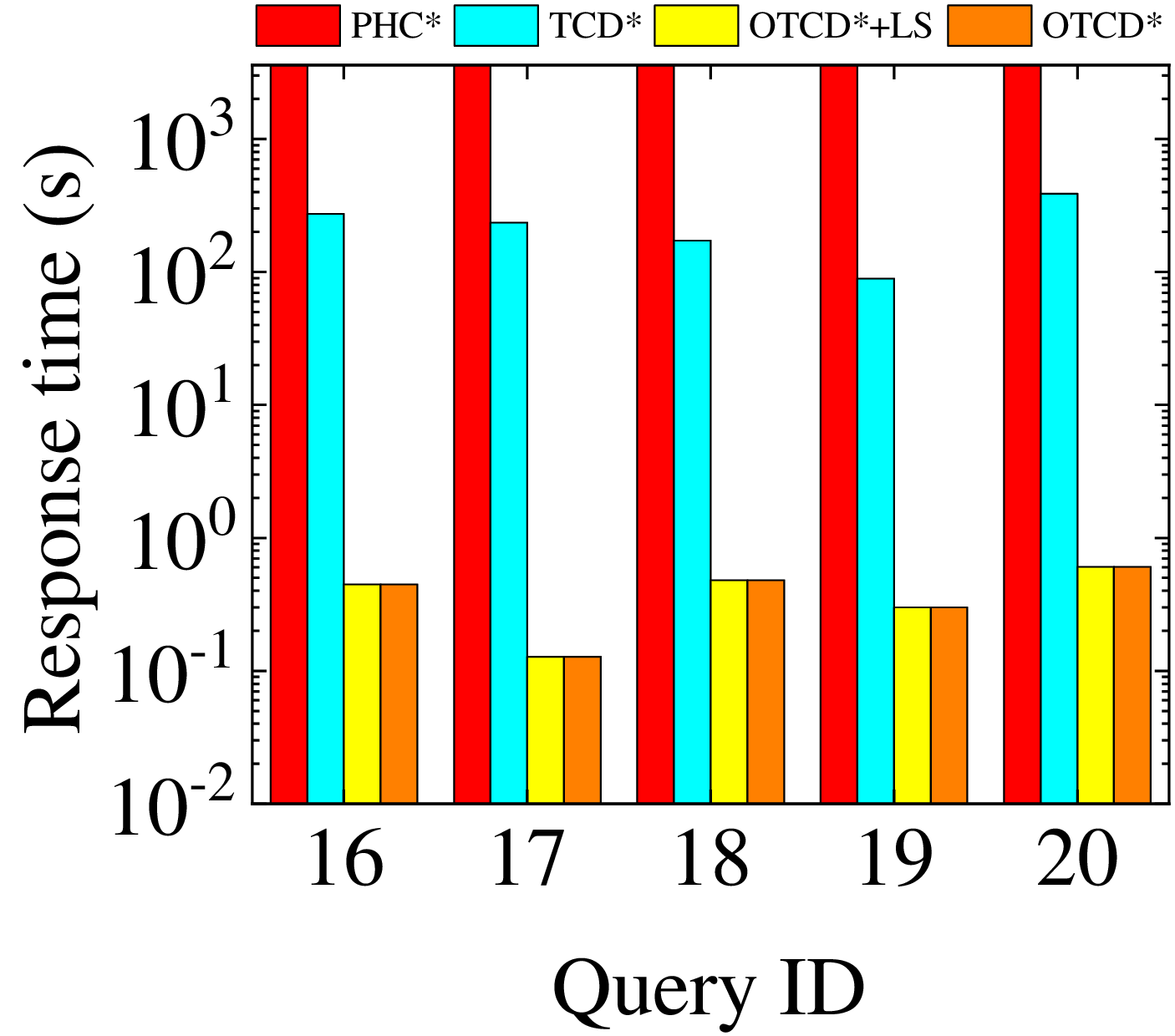}}
    \subfloat[TMO, college]{\label{subfig:college-o}
        \includegraphics[width=0.16\textwidth]{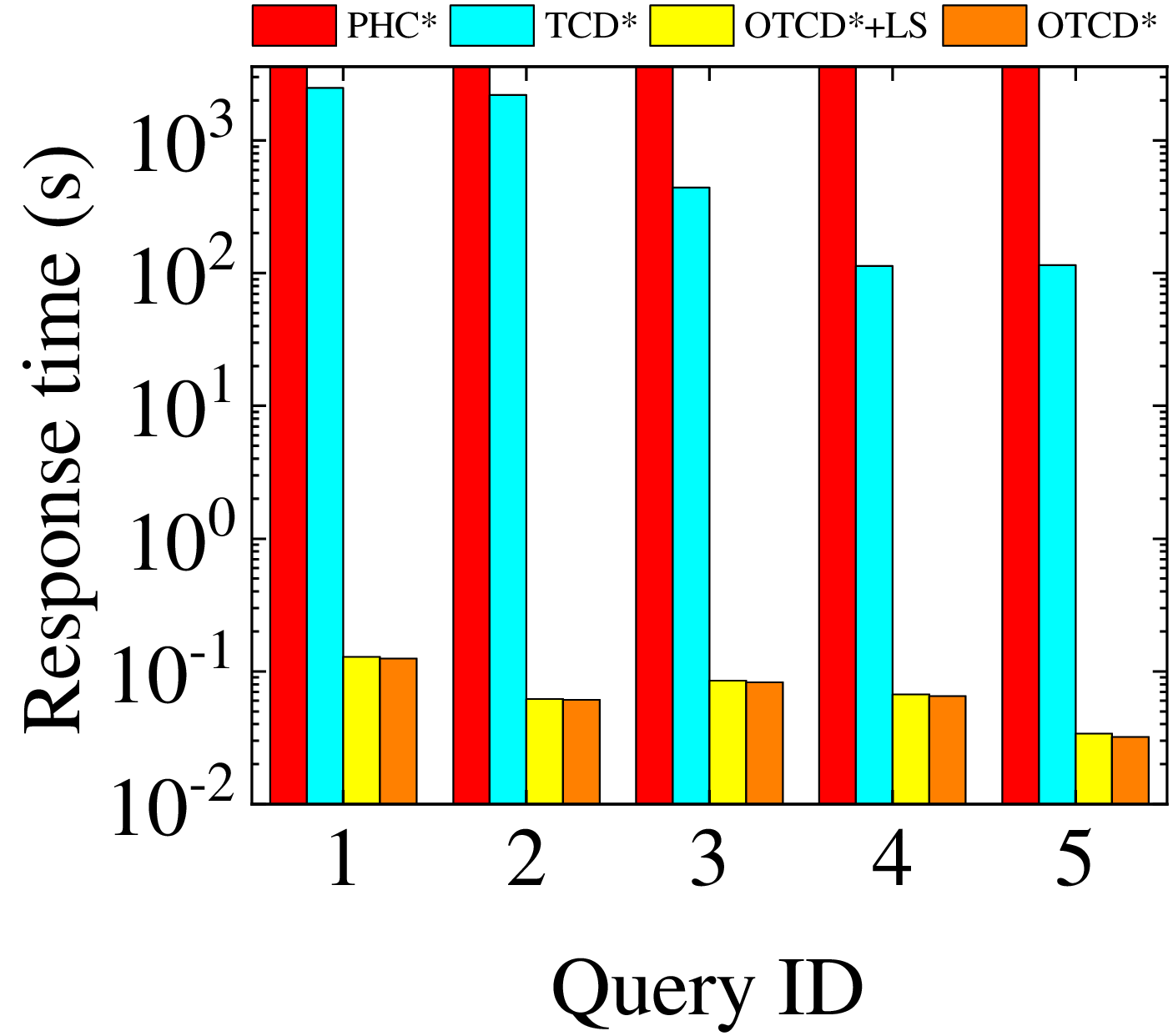}}
    \subfloat[TMO, email]{\label{subfig:email-o}
        \includegraphics[width=0.16\textwidth]{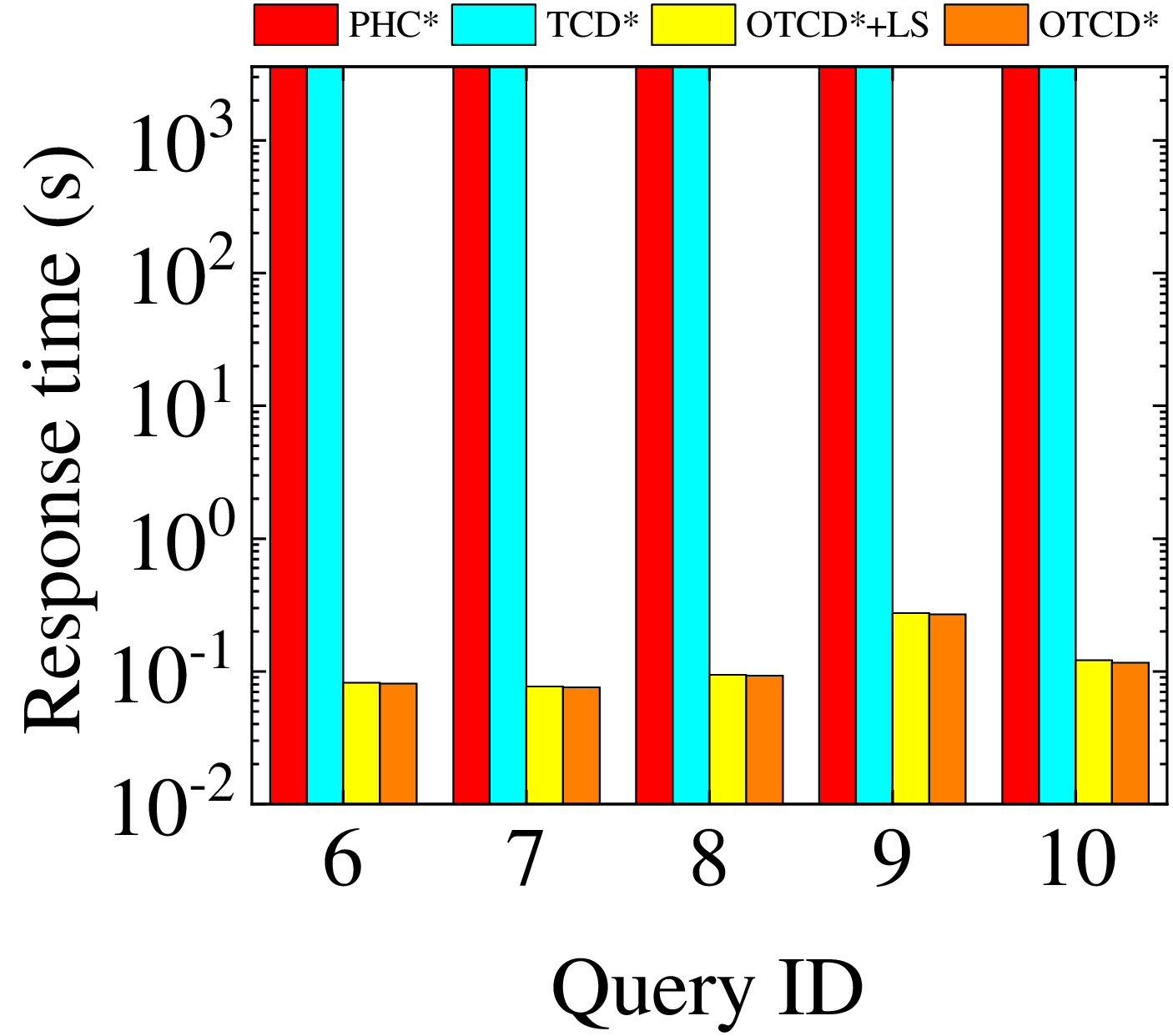}}
        \newline
    \subfloat[TMO, math]{\label{subfig:math-o}
        \includegraphics[width=0.16\textwidth]{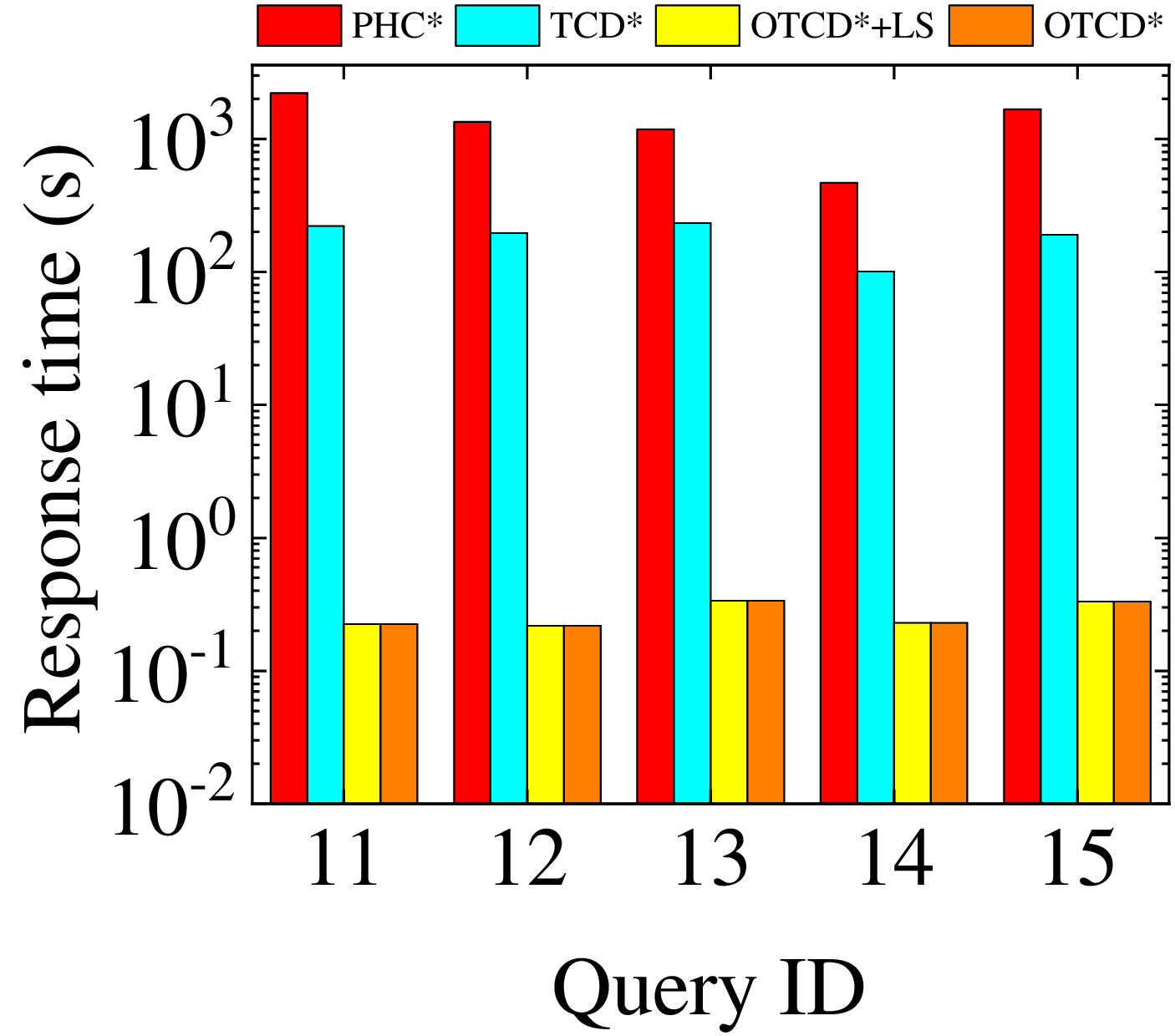}}
    \subfloat[TMO, stack]{\label{subfig:stack-o}
        \includegraphics[width=0.16\textwidth]{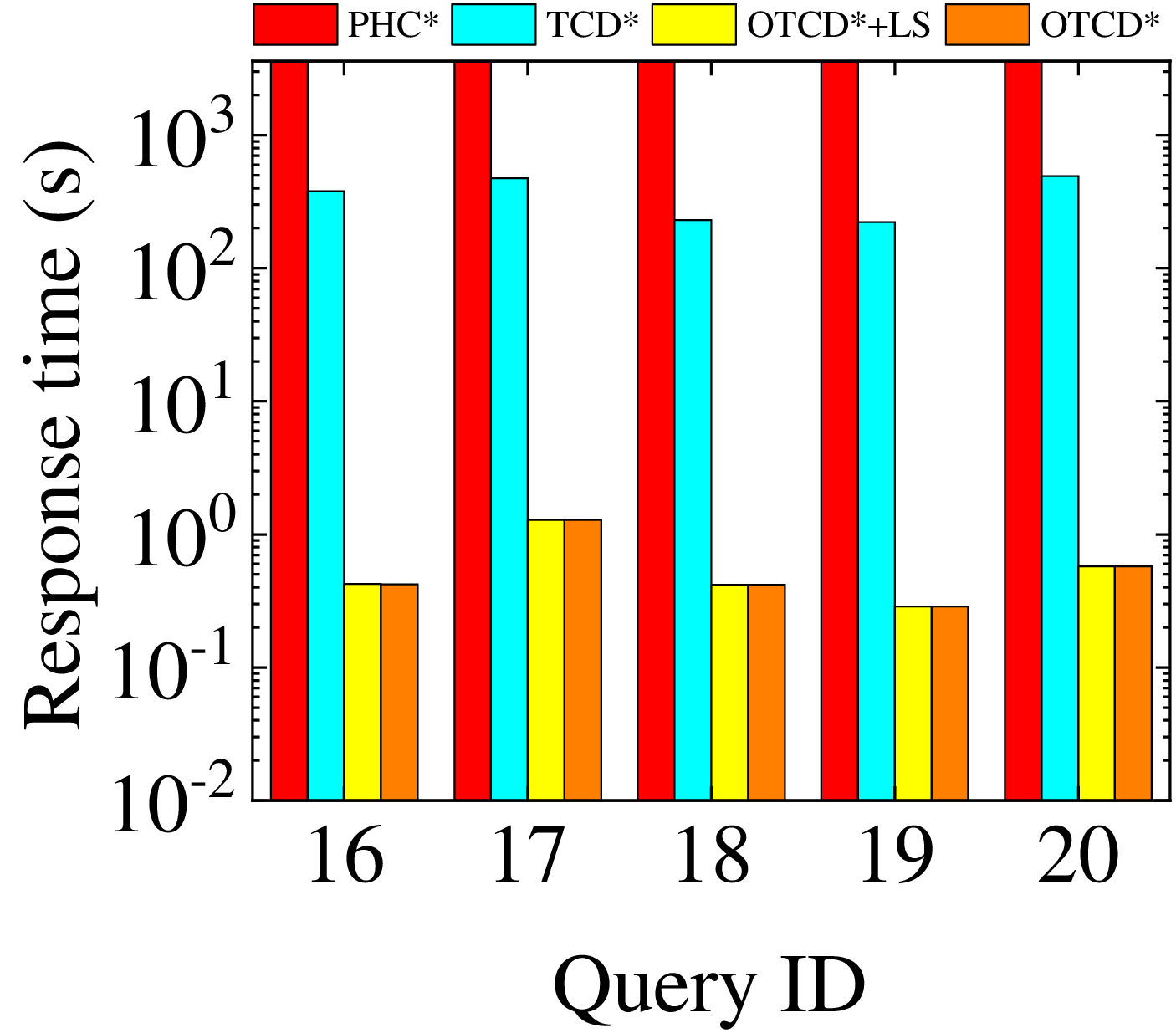}}
    \subfloat[TMC, college]{\label{subfig:college-c}
        \includegraphics[width=0.16\textwidth]{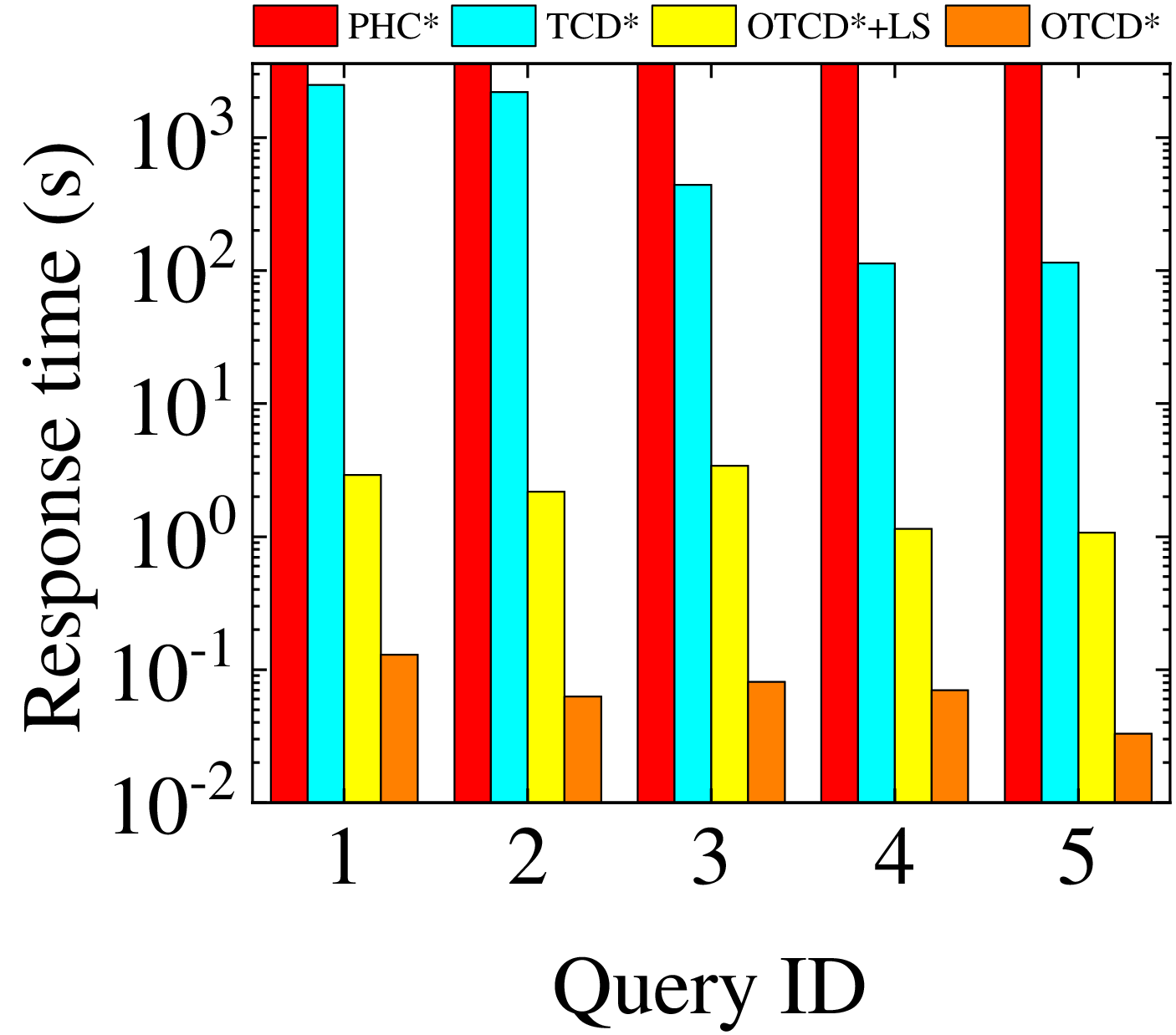}}
    \subfloat[TMC, email]{\label{subfig:email-c}
        \includegraphics[width=0.16\textwidth]{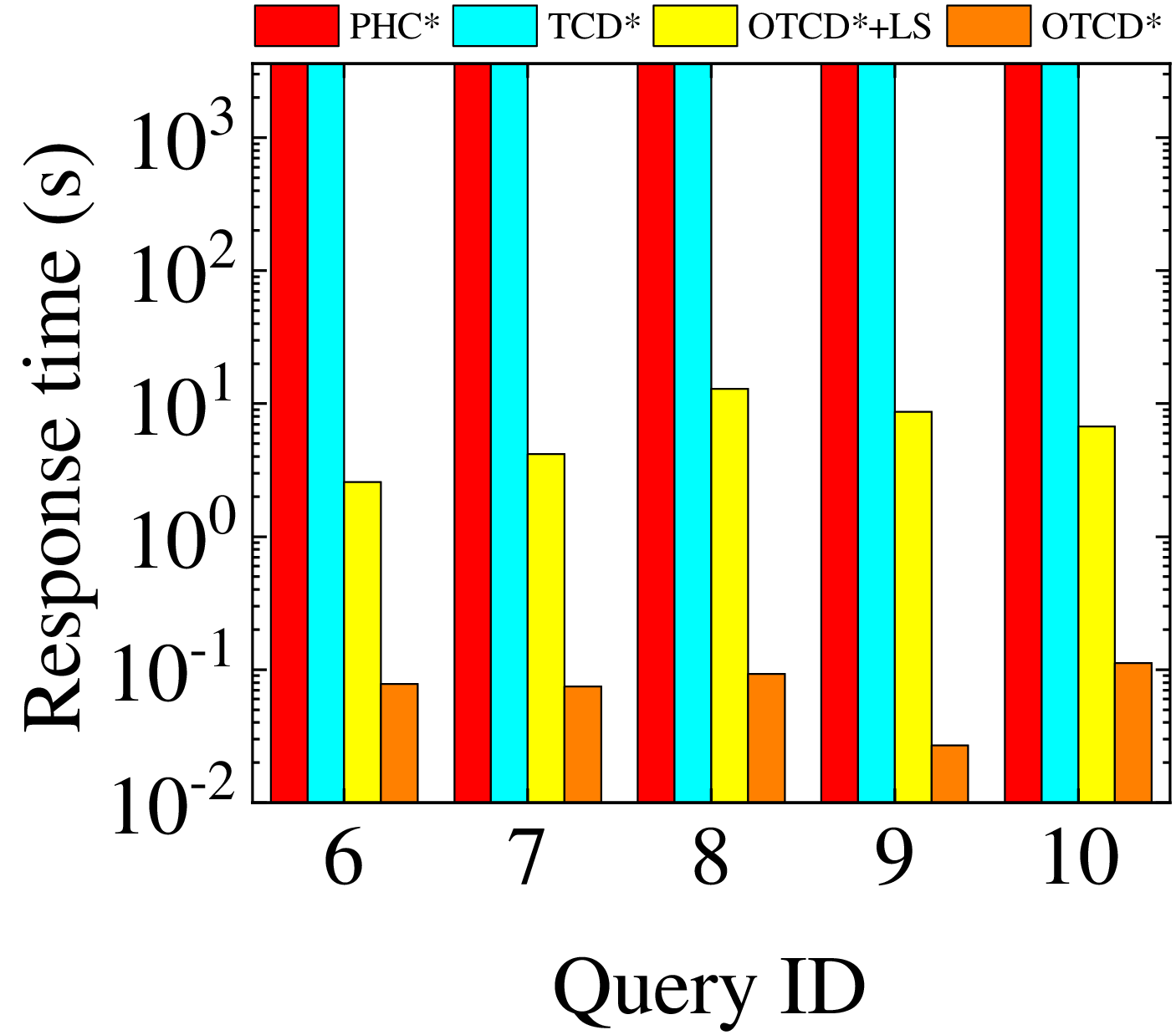}}
    \subfloat[TMC, math]{\label{subfig:math-c}
        \includegraphics[width=0.16\textwidth]{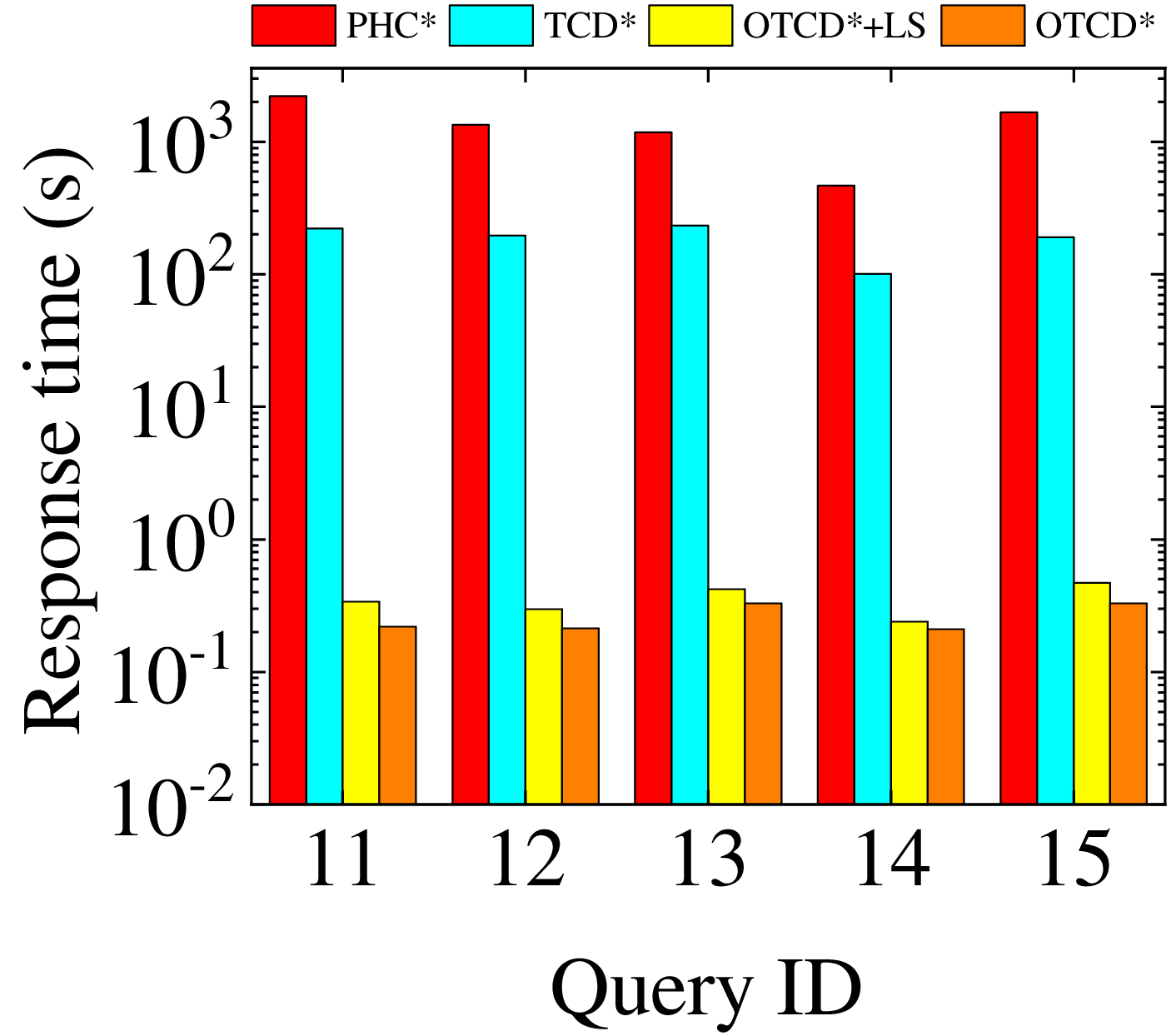}}
    \subfloat[TMC, stack]{\label{subfig:stack-c}
        \includegraphics[width=0.16\textwidth]{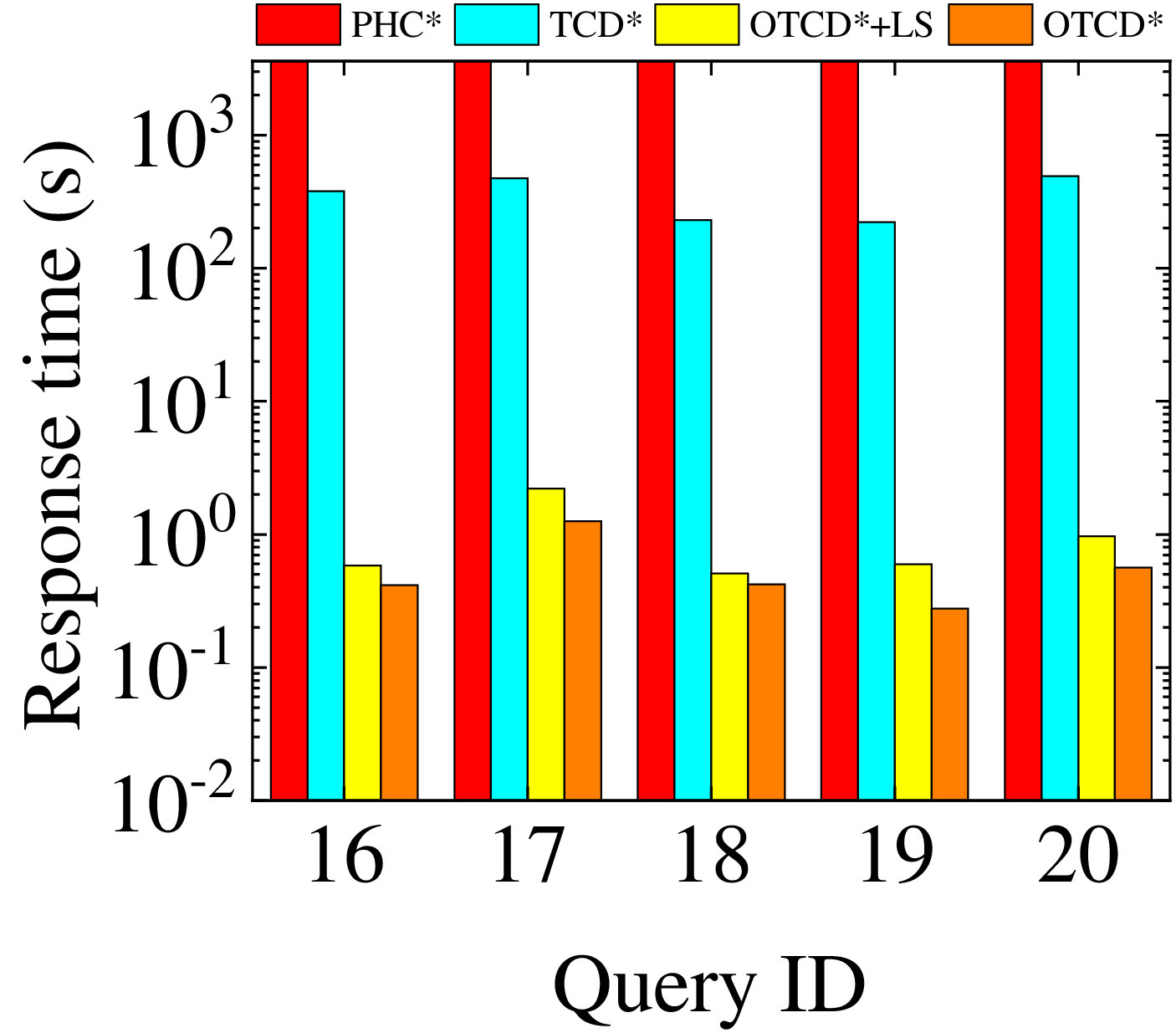}}
    \caption{Efficiency comparison for sixty TXCQ instances with specific $\mathcal{G}$, $k$, $[Ts, Te]$ and $\mathcal{X}(\cdot)$.}\label{fig:col}
\end{figure*}

To verify the efficiency of TXCQ processing, we test PHC*, TCD*, OTCD* and OTCD*+LS for above TXCQ instances on SNAP datasets respectively. OTCD*+LS is the proposed two-phase framework with local search strategies corresponding to different kinds of TXCQ instances. OTCD* is just the Phase 1 algorithm and used to compare the time costs of two phases. Baselines TCD* and PHC* use TCD algorithm and PHC-Index to obtain temporal $k$-cores respectively and both evaluate $\mathcal{X}$ for each subinterval. The results are shown in Fig~\ref{fig:col}.

For time-insensitive and time-monotonic optimizing queries, OTCD*+LS usually finishes in less than one second, as efficient as OTCD*. Because TI-LS or TMO-LS only have one or few times of $\mathcal{X}$ evaluation for each time zone. Compared with OTCD*+LS, TCD* is at least two orders of magnitude slower, due to expensive $\mathcal{X}$ evaluation for all subintervals. 

For time-monotonic constraining queries, OTCD*+LS spends at most 137x of time to process them than time-monotonic optimizing queries because TMC-LS revisits much more subintervals than TMO-LS. The time cost of OTCD* is certain for the same $\mathcal{G}$, $k$ and $[Ts,Te]$ (each column). Thus, if a query will mostly return many small time zones, TMC-LS only takes limited advantage from butterfly-pruning and thus becomes time consuming, like Q1$\sim$Q10 shown in Figs~\ref{subfig:college-c} and~\ref{subfig:email-c}. Otherwise, TMC-LS is very fast by skipping most subintervals, like Q11$\sim$Q20 shown in Figs~\ref{subfig:math-c} and~\ref{subfig:stack-c}. Even though, OTCD*+LS still beats TCD* by a large margin.

As expected, PHC* is even slower than TCD*. Because PHC* also evaluates $\mathcal{X}$ for all subintervals, and is more costly than TCD algorithm on inducing temporal $k$-cores.

Moreover, we have an ``abnormal'' observation that reveals the impact of measurement $\mathcal{X}(\cdot)$. The difference between the time cost of TCD* to address queries like Q6$\sim$Q10 is huge for the two metrics size and engagement, by comparing Fig~\ref{subfig:email-i} and~\ref{subfig:email-o}. However, such a huge difference is not found on all other queries. The reason is that, the computational complexity of engagement is $O(|\mathcal{E}_{[ts,te]}|)$ for a subinterval $[ts,te]$, and the computational complexity of size is always $O(1)$, with respect to our implementation of $\mathcal{X}(\cdot)$. Thus, the time cost of TCD* increases dramatically for the queries that will return $k$-cores with massive temporal edges. While, email-Eu-core-tempora is such a temporal graph in which lots of edges have a same timestamp. Considering the above results, we remark that even though OTCD*+LS is not so sensitive to the implementation of $\mathcal{X}(\cdot)$ since it avoids all unnecessary evaluations, a better implementation for $\mathcal{X}(\cdot)$ computation still benefit the process.

  	\begin{figure}[t!]
		\centering
            \captionsetup[subfloat]{labelfont=scriptsize,textfont=scriptsize}
		\subfloat[TI]{\label{subfig:insensitive-k}
			\includegraphics[width=0.33\linewidth]{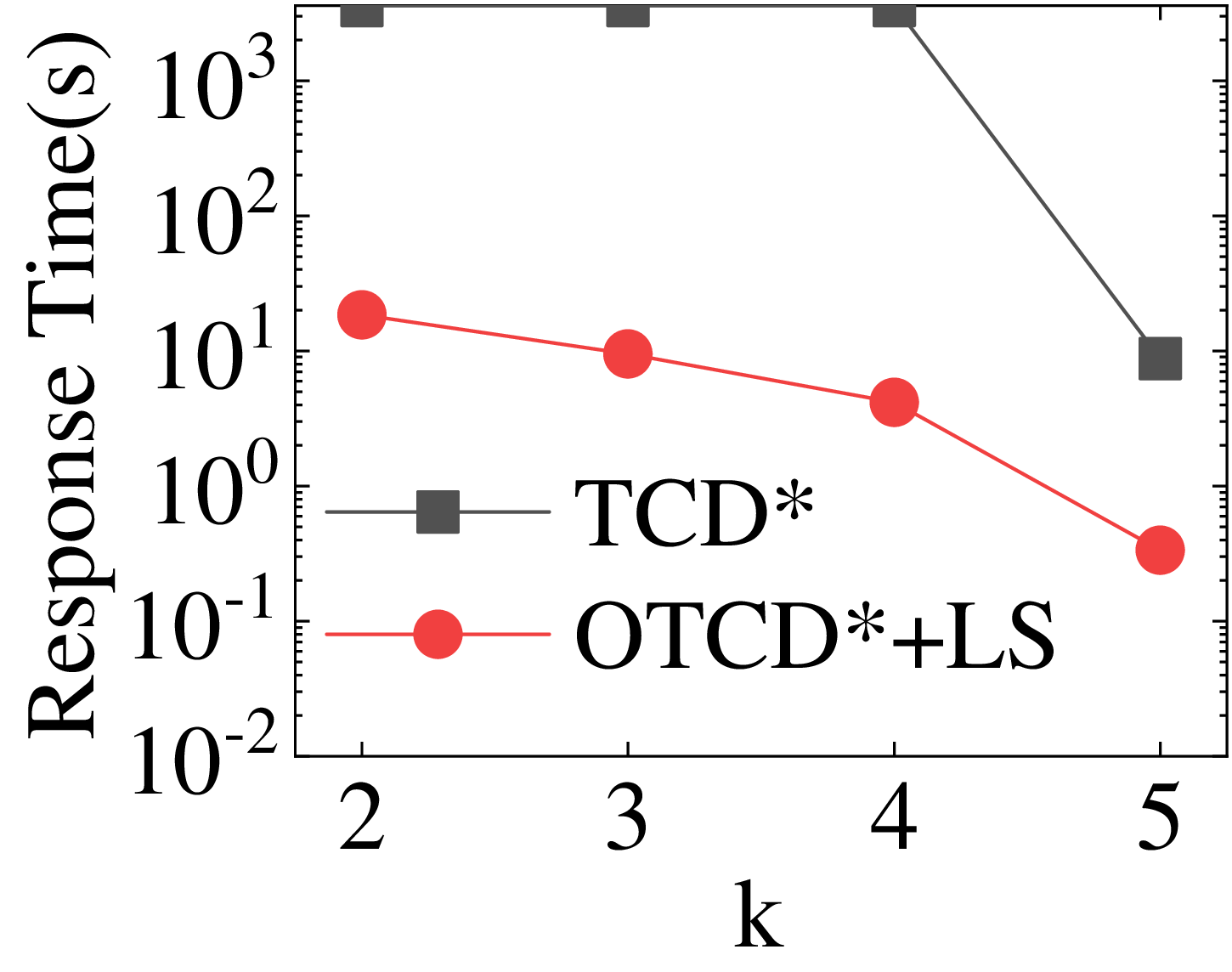}}
		\subfloat[TMO]{\label{subfig:optimizing-k}
			\includegraphics[width=0.33\linewidth]{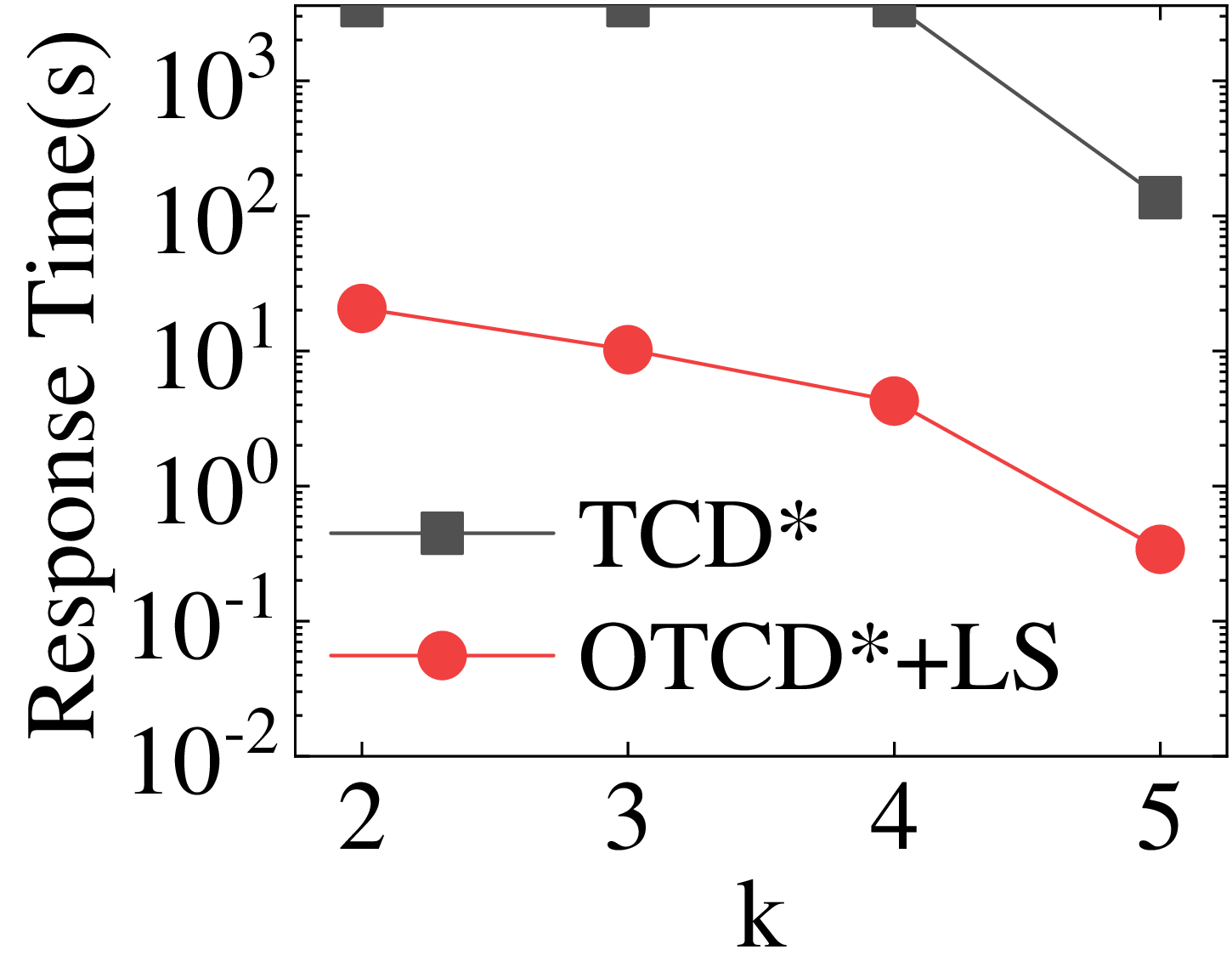}}
   		\subfloat[TMC]{\label{subfig:constraining-k}
			\includegraphics[width=0.33\linewidth]{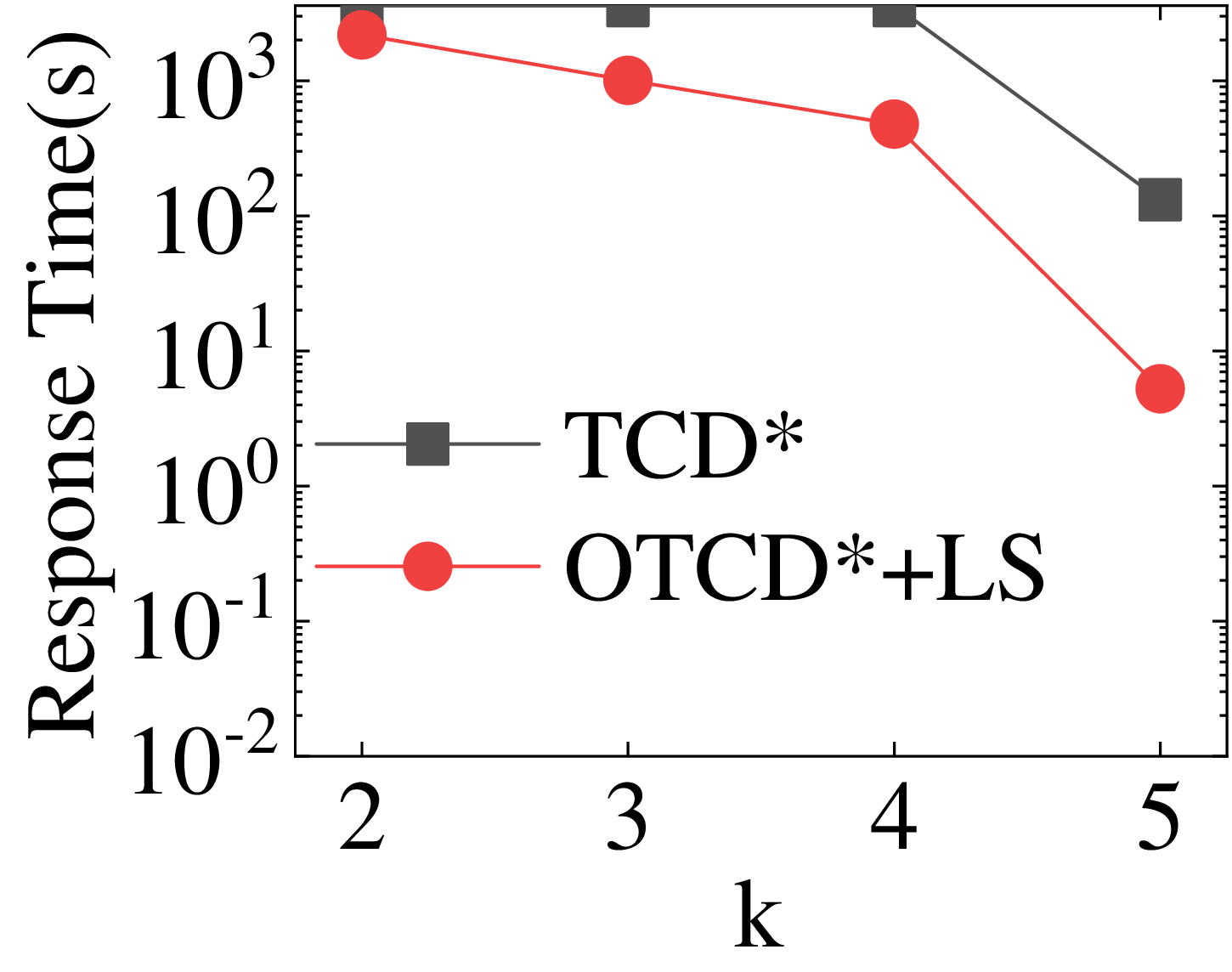}}
		\caption{The TXCQ response time with respect to varying $k$.}\label{fig:k-chart}
	\end{figure}

   	\begin{figure}[t!]
		\centering
            \captionsetup[subfloat]{labelfont=scriptsize,textfont=scriptsize}
		\subfloat[TI]{\label{subfig:insensitive-span}
			\includegraphics[width=0.33\linewidth]{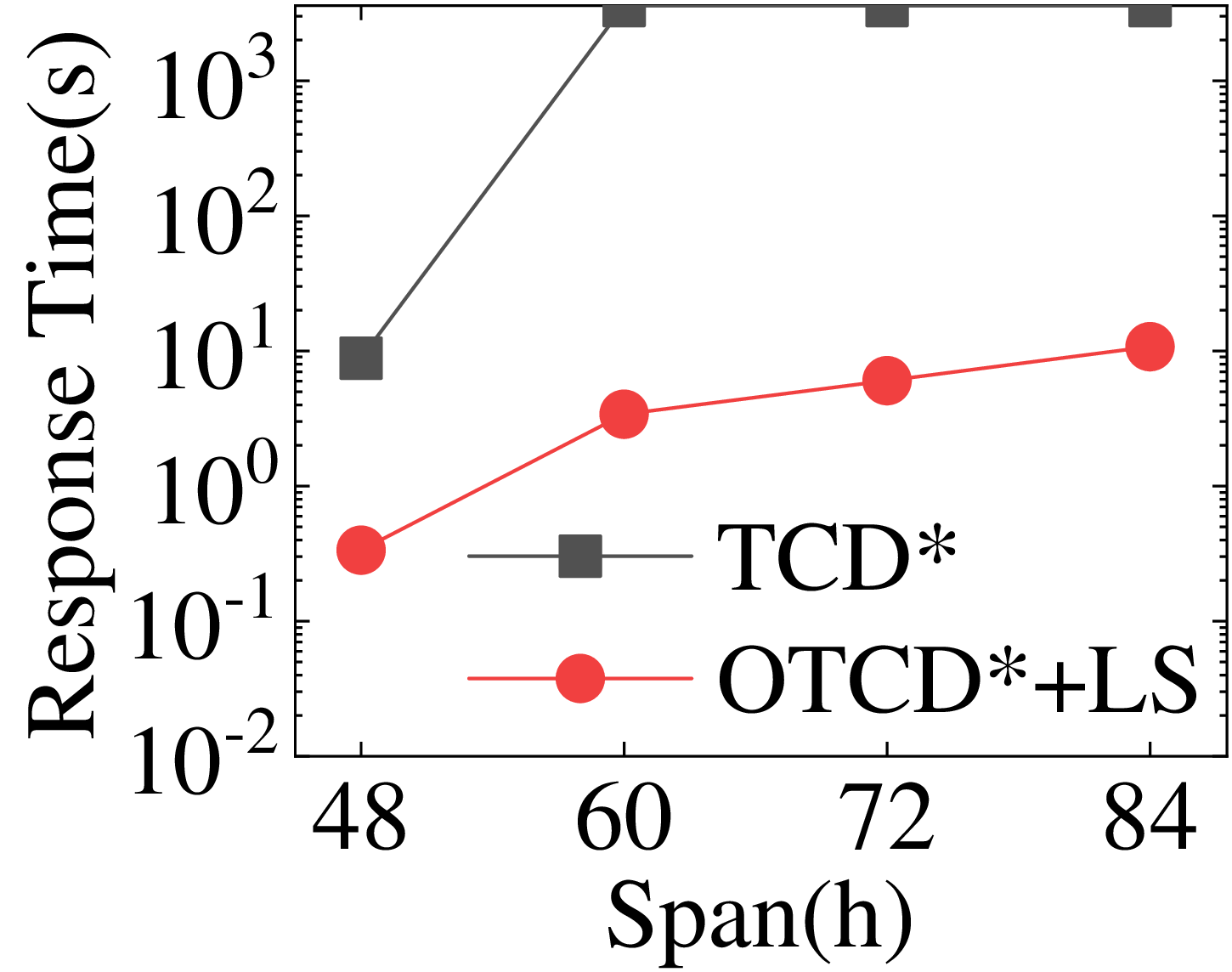}}
		\subfloat[TMO]{\label{subfig:optimizing-span}
			\includegraphics[width=0.33\linewidth]{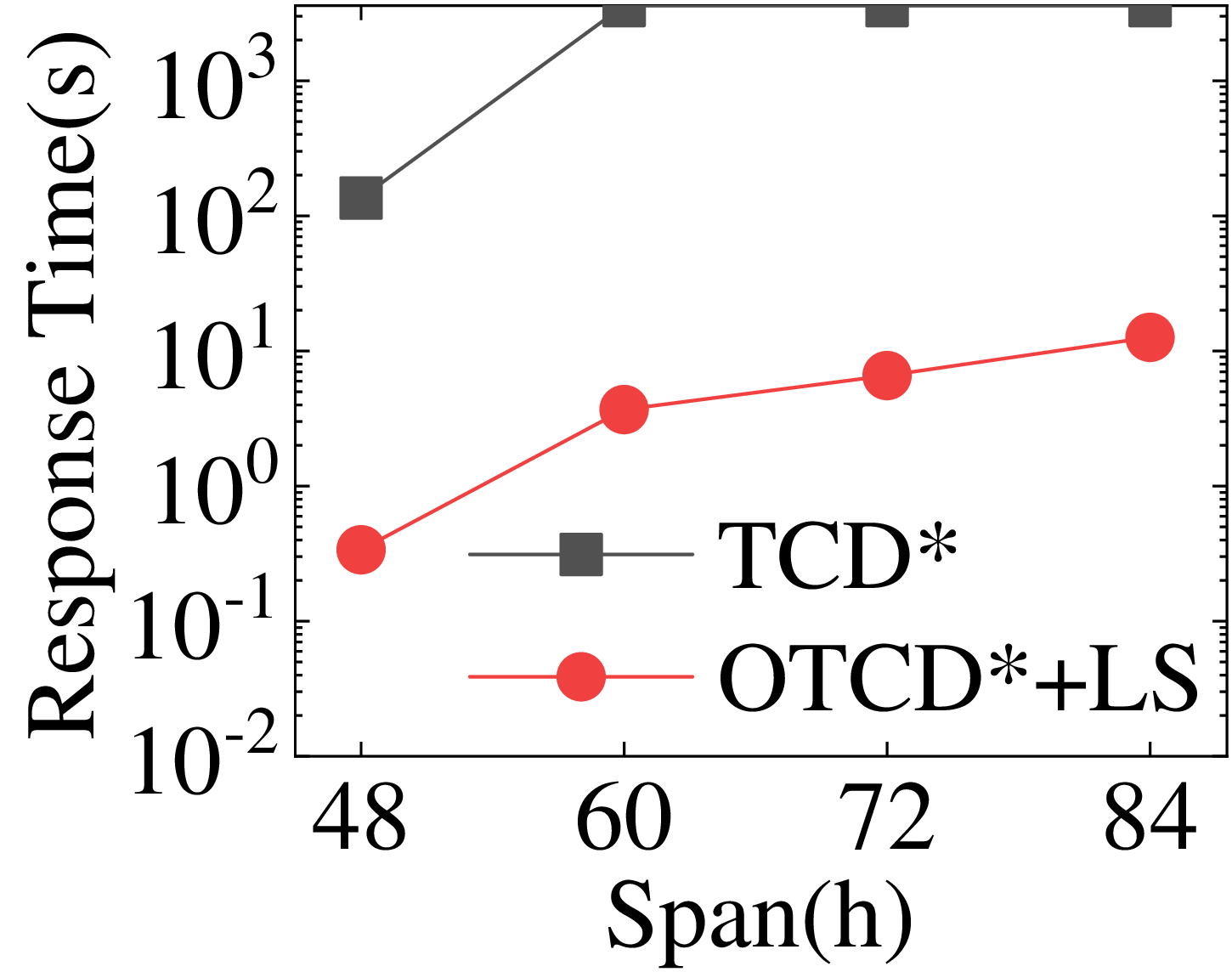}}
   		\subfloat[TMC]{\label{subfig:constraining-span}
			\includegraphics[width=0.33\linewidth]{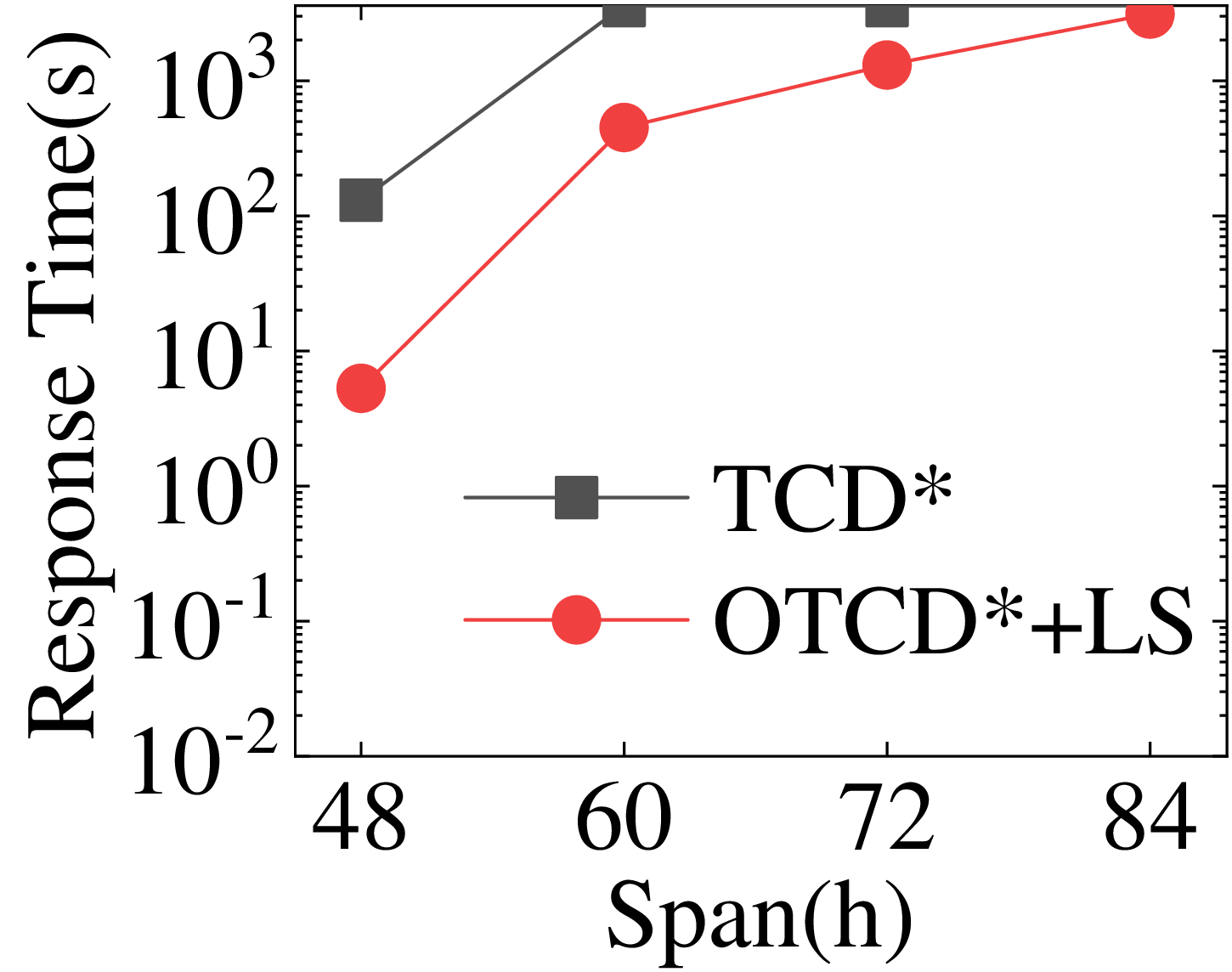}}
		\caption{The TXCQ response time with respect to varying time span.}\label{fig:span-chart}
	\end{figure}

To investigate the impact of $k$ and the span of $[Ts,Te]$ to efficiency, we test twenty-four other queries on mathoverflow with varying values of $k$ or span. The results are presented in Fig~\ref{fig:k-chart} and~\ref{fig:span-chart} respectively. Firstly, the time cost of OTCD*+LS decreases gradually with the increase of $k$. Because the number of distinct cores and zones is less for a greater $k$, and thereby both OTCD* and local search become faster. Secondly, the time cost of OTCD*+LS increases gradually with the increase of span, since the number of subintervals is quadratic to the span. Even for the time consuming TMC query, OTCD*+LS still scales well. 

To investigate the number of LTIs in each time zone, we also conduct an empirical study in mathoverflow for five different values of $k$. The results are shown in Table~\ref{tab:lti-num}. We can observe that, the total number of LTIs is a bit more than that of time zones. It means most time zones have only one rectangle and thereby one LTI in them. Such an observation ensures that the OTCD* has almost none redundant computation in practice.

\begin{table}
    \centering
    \caption{An empirical study for the number of LTIs on mathoverflow. The query time interval is fixed as $[1,5000]$.}

    \begin{tabular}{ccc}
    \hline
     $k$ & time zone \# & LTI \#\\
    \hline
     15 & 57314 & 57351\\
     14 & 111165 & 111197\\
     12 & 333832 & 334033\\
     10 & 684674 & 685139\\
     8 & 1331560 & 1332384\\
    \hline
    \end{tabular}
	\label{tab:lti-num}
\end{table}

\subsection{Case Study}

\subsubsection{TCQ Case Study}

\begin{figure}[t]
		\centering
		\includegraphics[width=\linewidth]{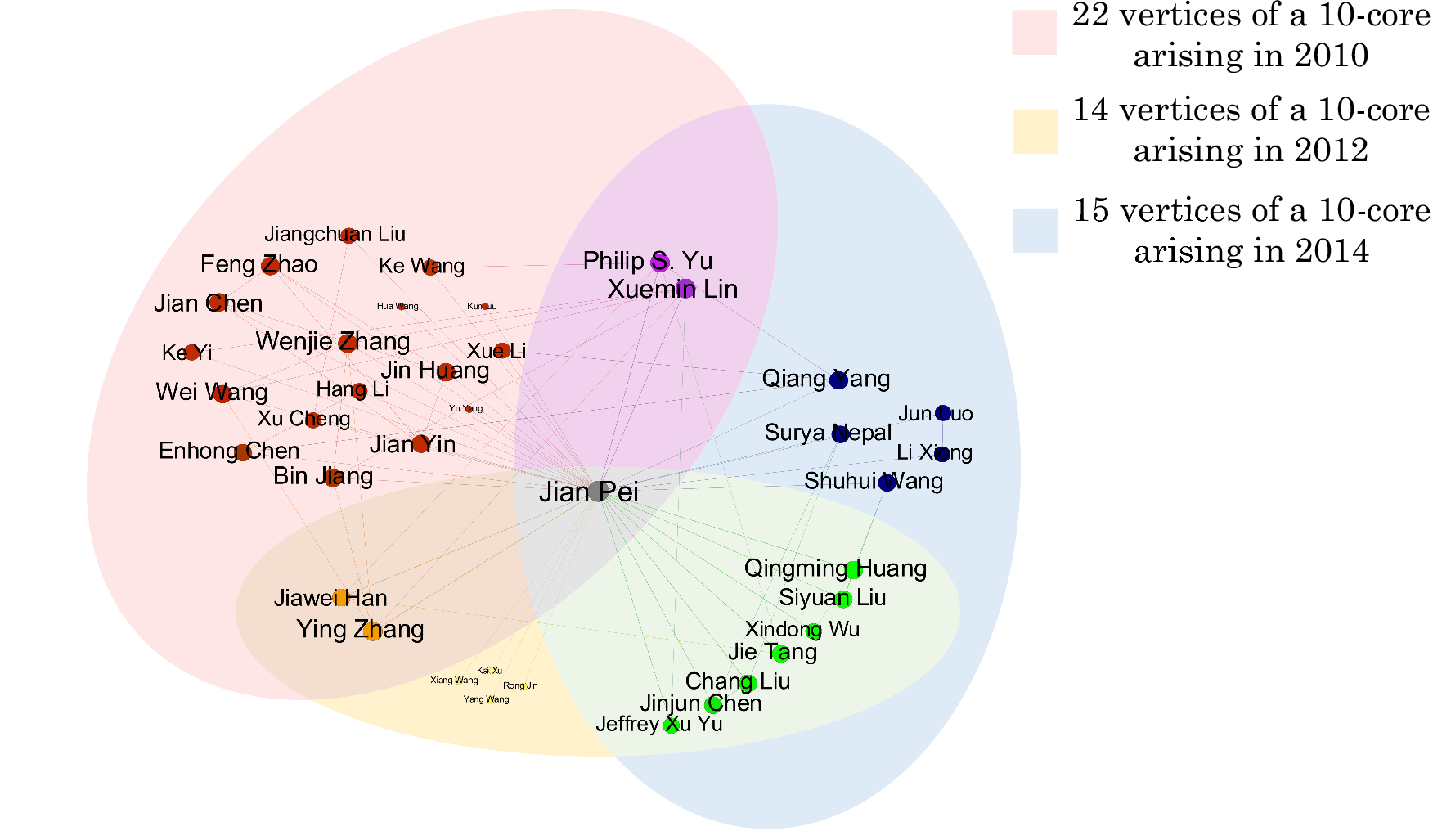}
		\caption{A case study of temporal $k$-core in DBLP, where the size of vertices and labels are automatically generated based on degree.}\label{fig:cstudydblp}
	\end{figure}

For case study, we employ OTCD algorithm to query temporal 10-cores on DBLP coauthorship graph. The query range is set as from 2010 to 2018, which spans over 8 years. By statistics, there exist 43 temporal 10-cores during that period, and 39 of them contain the author Jian Pei, for whom we further build an ego network from three selected cores in different years. Figure~\ref{fig:cstudydblp} shows the ego network. The authors in the three cores emerged in 2010, 2012 and 2014 are shaded by red, yellow and blue respectively. By observing the evolution of ego network over years, we can infer the change of author's research interests or affiliations.

\begin{figure}[t]
		\centering
		\includegraphics[width=0.8\linewidth]{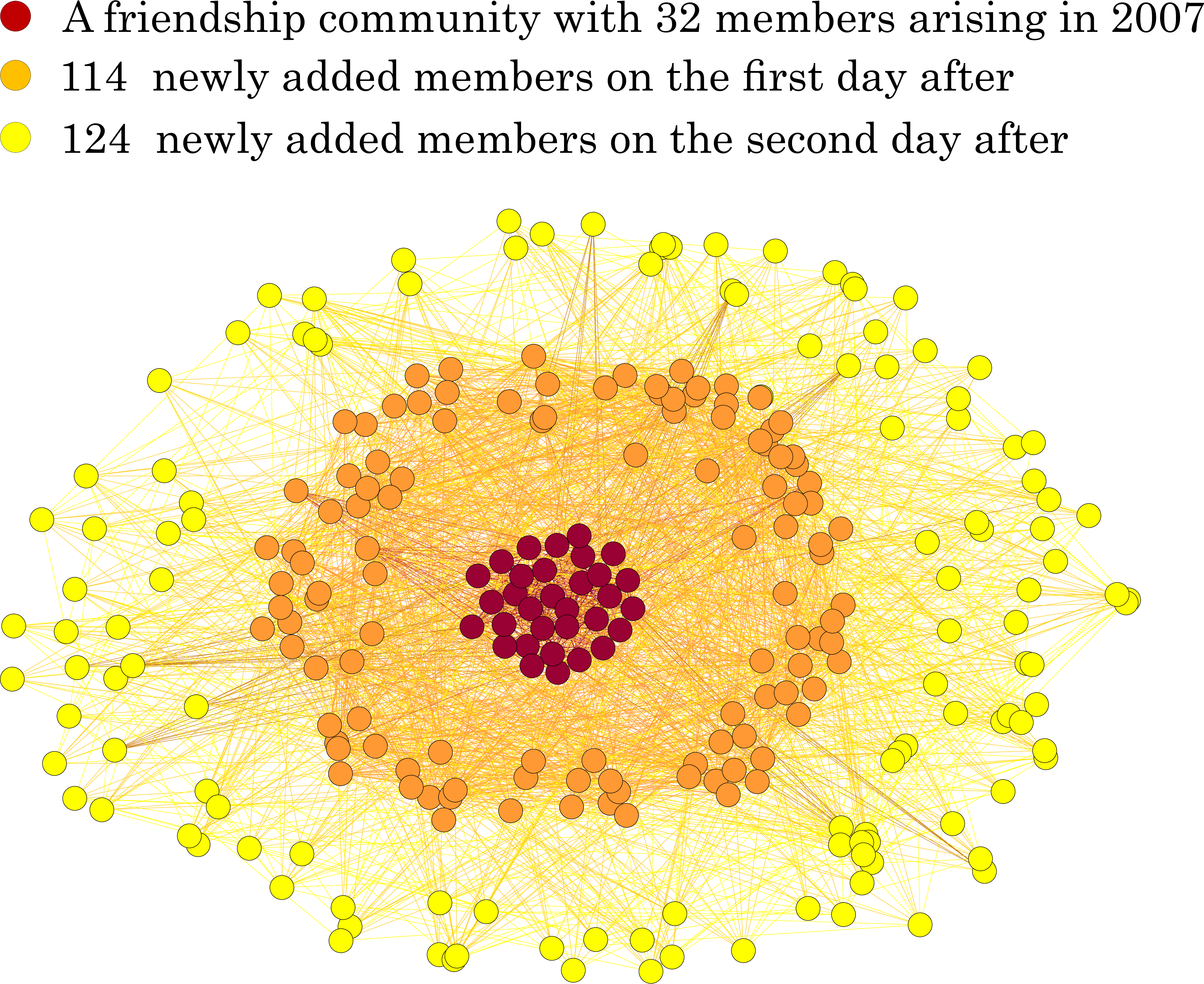}
		\caption{A case study of mining bursting $k$-core in Youtube, which is implemented by TCQ.}\label{fig:cstudyytb}
	\end{figure}

Besides, we also adopt TCQ to search for bursting communities on Youtube friendship network. Figure~\ref{fig:cstudyytb} shows such a bursting community. The 32 central vertices colored in red comprise an initial temporal 10-core within two days. This core is contained by another core about four times larger, while the TTI of the larger core only expands by one day. The new vertices in the larger core are colored in orange. Then, the new vertices colored in yellow join them to comprise a twice larger new core in the next day. However, we have to post-process the results of TCQ to get such a more specific community.

\subsubsection{TXCQ Case Study}

\begin{figure}[t!]
		\centering
            \captionsetup[subfloat]{labelfont=scriptsize,textfont=scriptsize}
		\subfloat[2010]{\label{subfig:hiro-10}
			\includegraphics[width=0.32\linewidth]{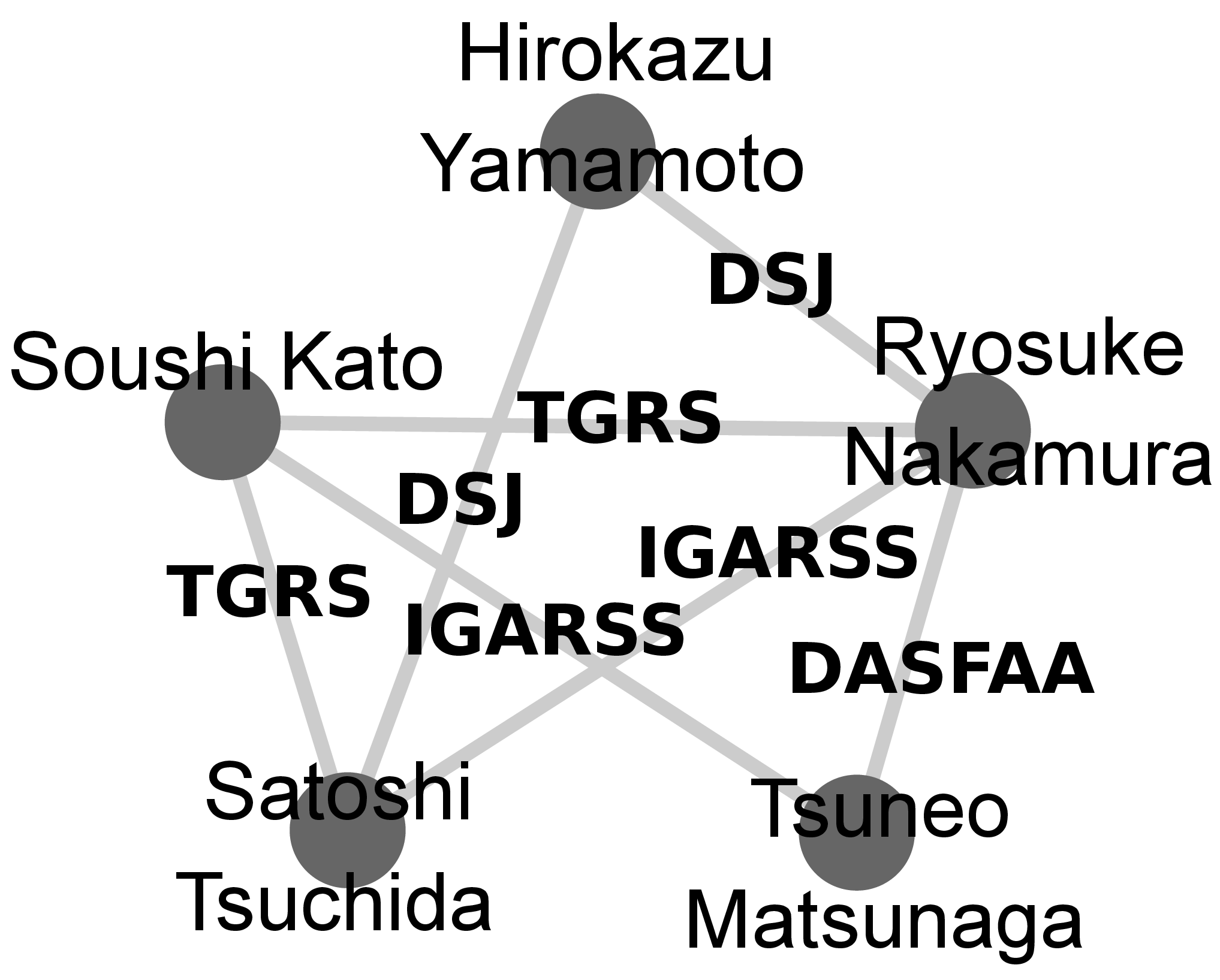}}
		\subfloat[2013]{\label{subfig:hiro-13}
			\includegraphics[width=0.32\linewidth]{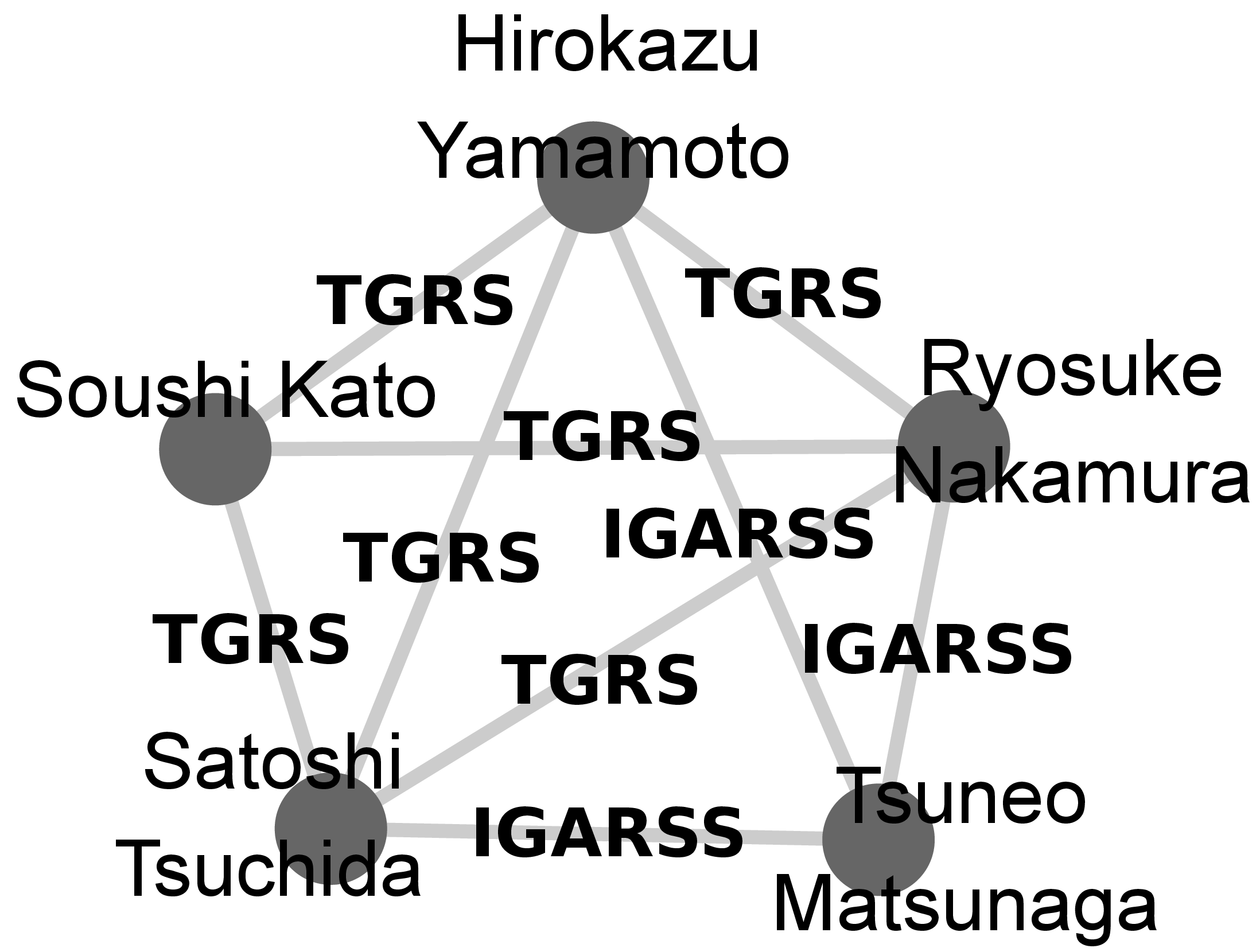}}
   	    \subfloat[2017]{\label{subfig:hiro-17}
			\includegraphics[width=0.32\linewidth]{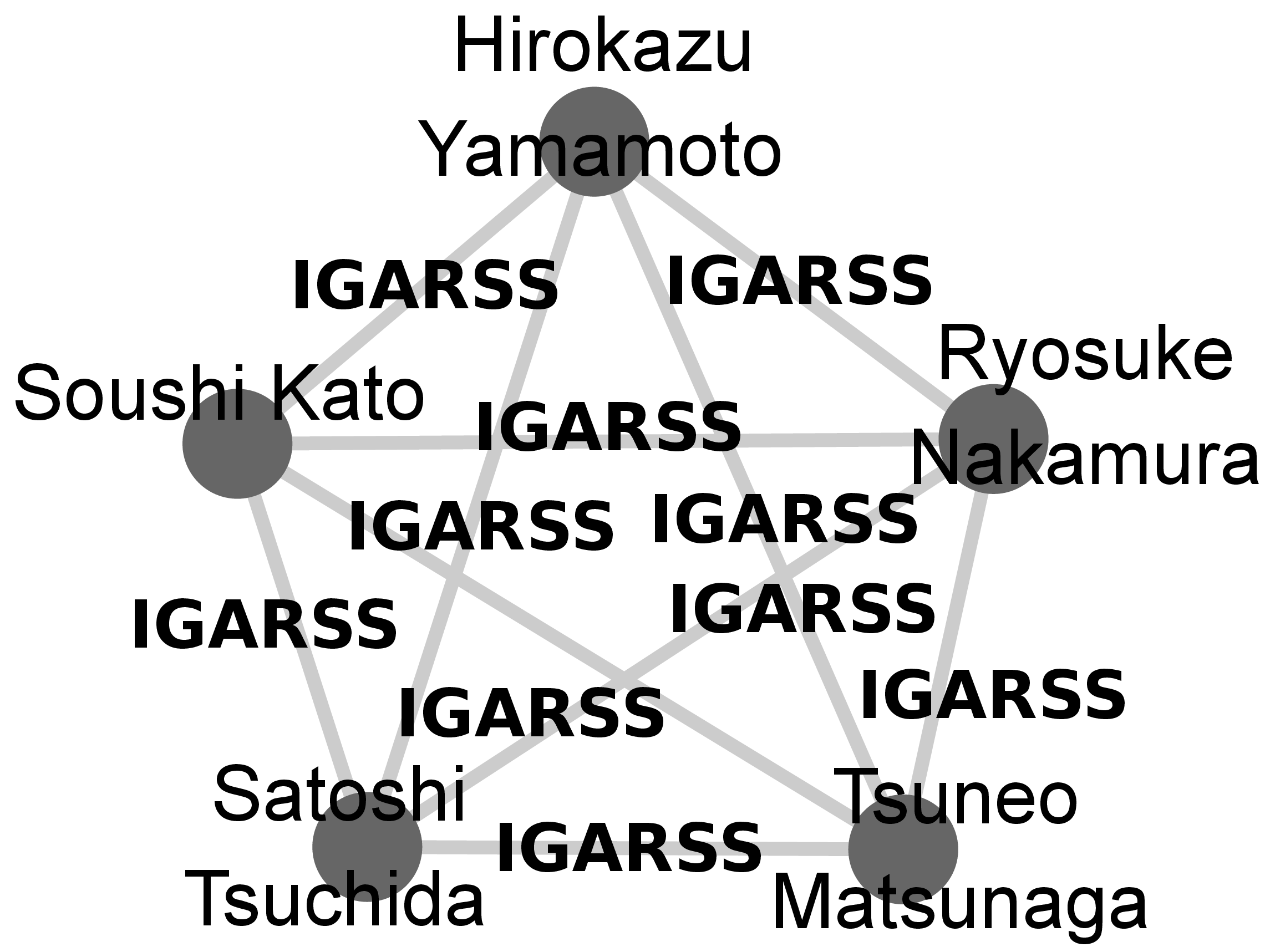}}
        \caption{A case study of periodic $k$-core in DBLP.}\label{fig:cs}
	\end{figure}

We firstly use TXCQ to explore the periodic $k$-cores on the DBLP coauthorship graph. For a temporal $k$-core, we use $\mathcal{X}(\cdot)$ to evaluate the maximum number of non-overlapped TTIs during which the same set of vertices can comprise a $k$-core, as shown in Table~\ref{tab:attr}. One of the results is illustrated in Fig~\ref{fig:cs}. The five authors comprise a $k$-core during three different periods respectively. Such a periodic community represents strong bond between members in the dimension of time and is valuable in many application scenarios. 

\begin{figure*}[t!]
	\centering
        \captionsetup[subfloat]{labelfont=scriptsize,textfont=scriptsize}
	\subfloat[10 days, 1220 interactions.]{\label{subfig:tcc0}
			\includegraphics[width=0.23\linewidth]{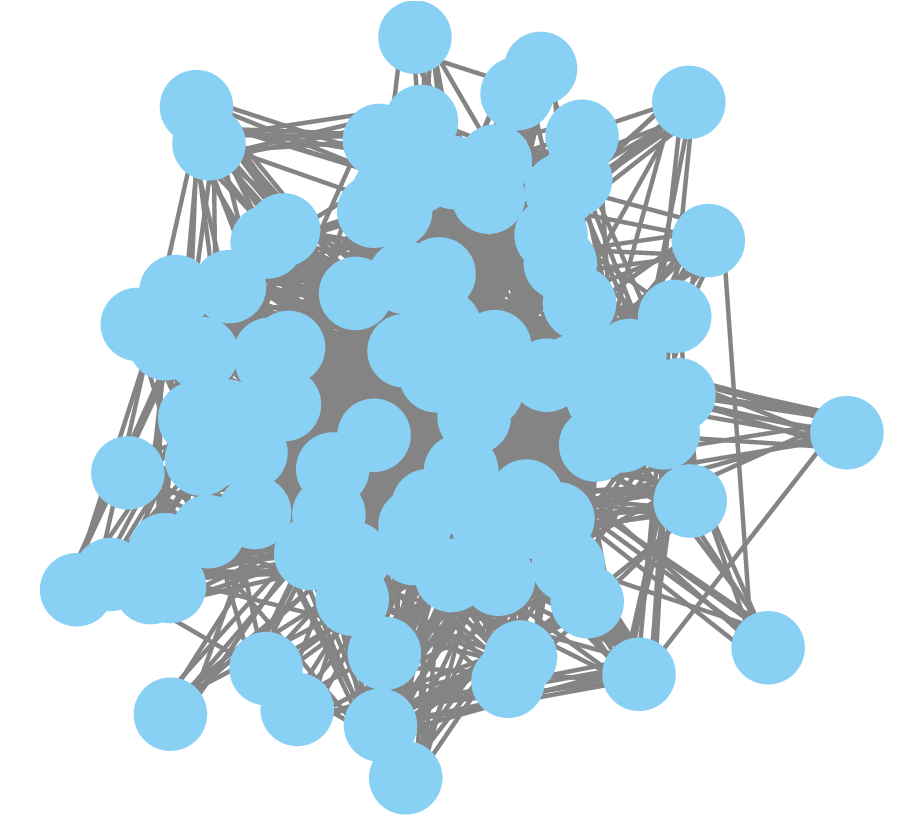}}
	\subfloat[16 days, 2391 interactions.]{\label{subfig:tcc1}
			\includegraphics[width=0.23\linewidth]{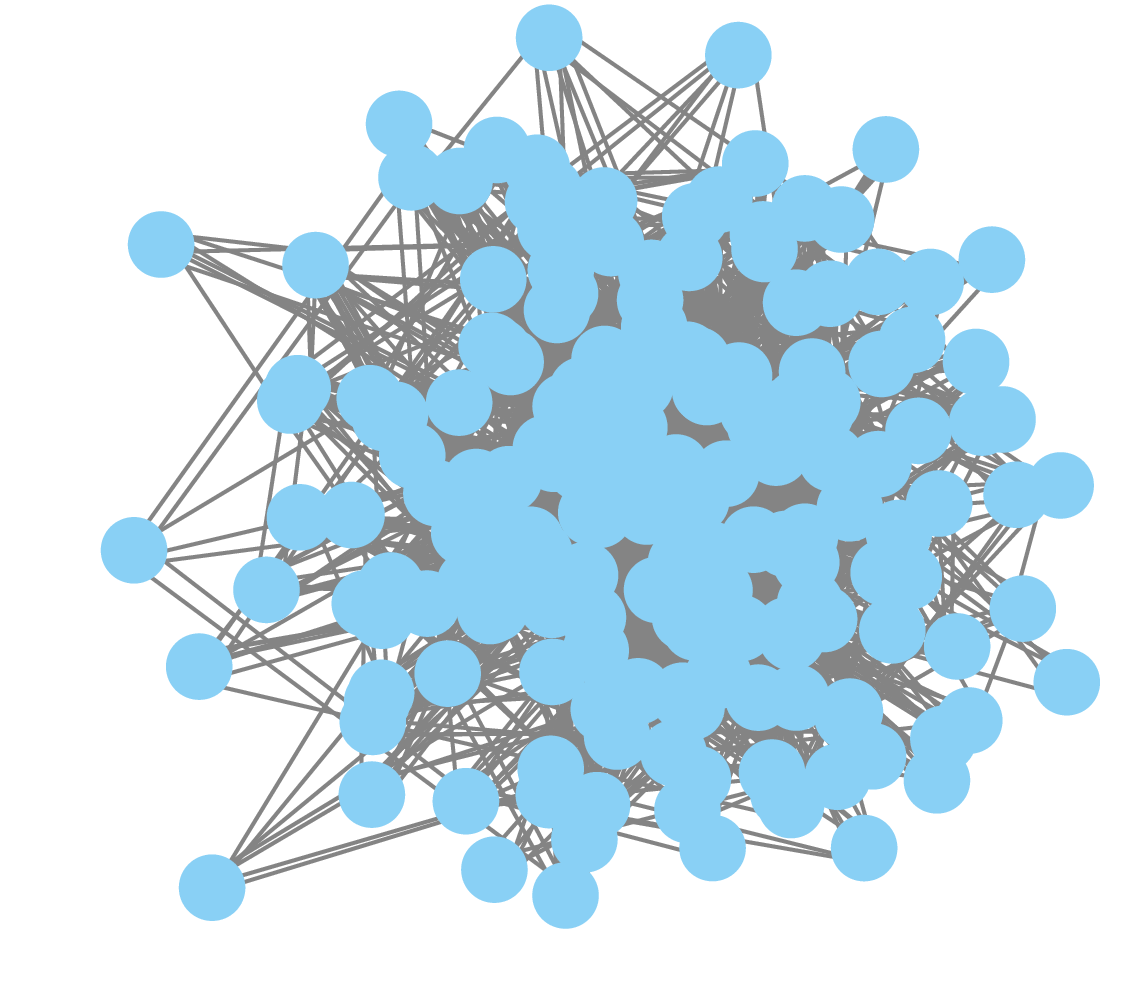}}
   	\subfloat[17 days, 2105 interactions.]{\label{subfig:tcc2}
			\includegraphics[width=0.23\linewidth]{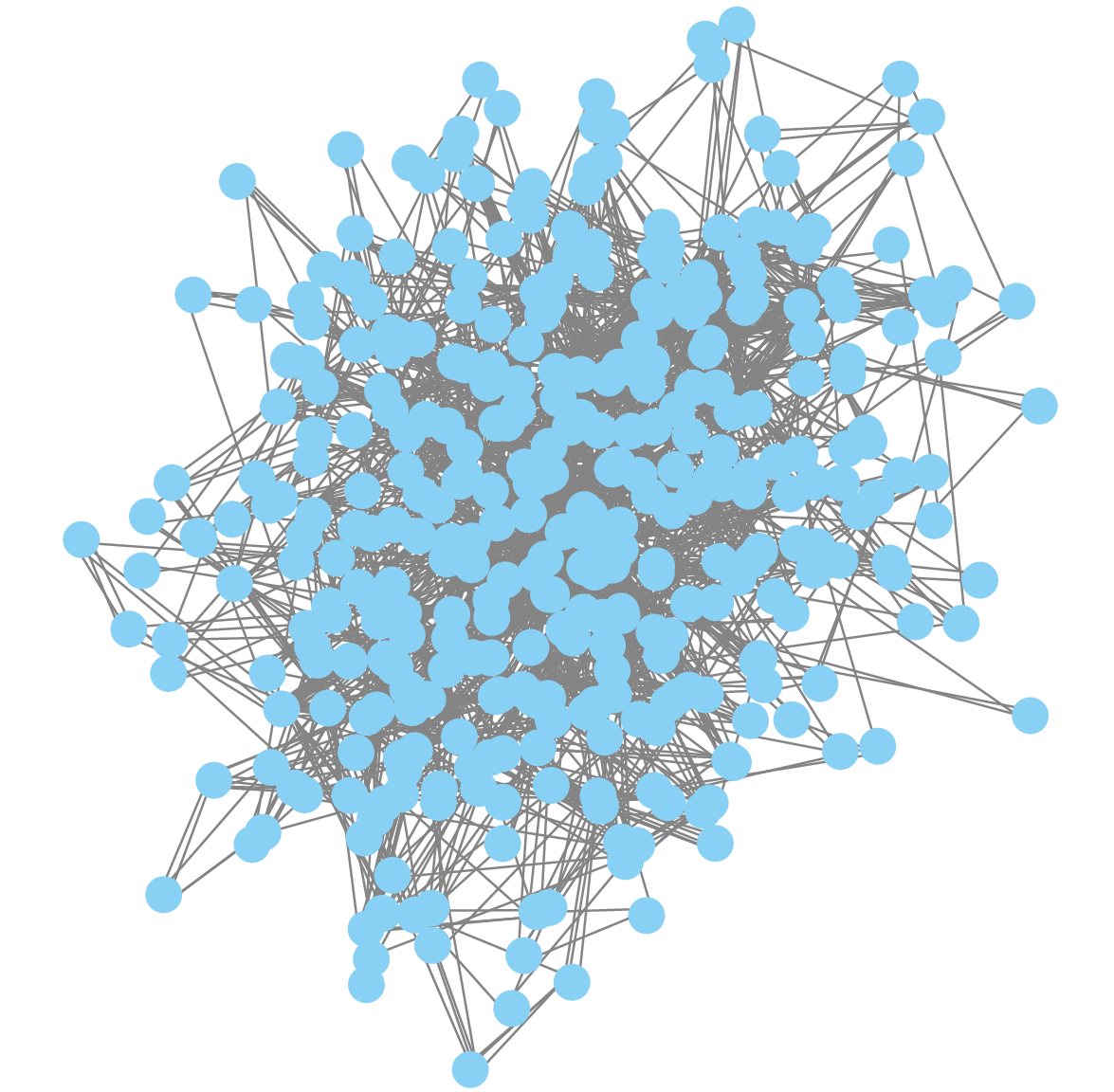}}
        \subfloat[14 days, 1253 interactions.]{\label{subfig:tcc3}
			\includegraphics[width=0.23\linewidth]{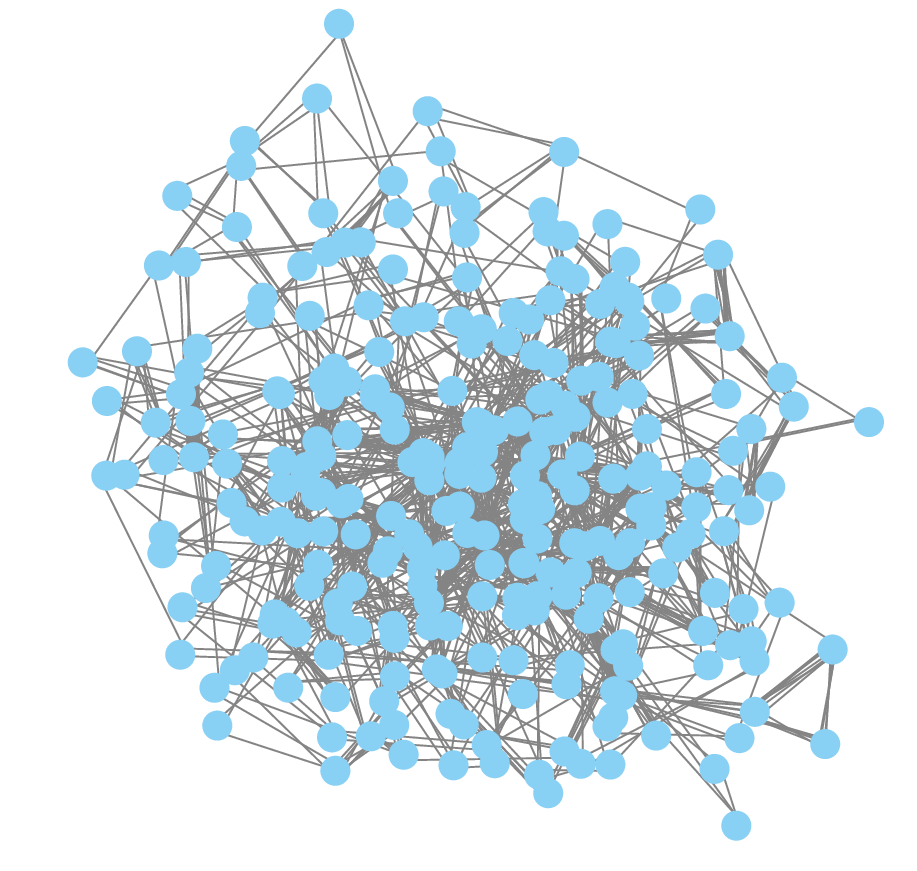}}
        \caption{A case study of mining bursting $k$-core in MathOverflow, which is implemented by TXCQ.}\label{fig:tcc}
\end{figure*}

In addition, we apply TXCQ on a social network with $\mathcal{X}(\cdot)$ measuring the burstiness. Specifically, we compose sixty TXCQ instances to optimize burstiness on MathOverflow, each of which has a random non-overlapping time interval with a span of one month. For each query, the $k$ value is set as half of the largest valid $k$ for temporal $k$-cores over the month. Then, four typical results are visualized in Figure~\ref{fig:tcc}. We can see that all results have a short time span but a large number of interactions, which indicates that they emerge quickly and tend to keep growing. Such a TXCQ instance can facilitate applications like network monitoring, recommendation, disease control, etc. More importantly, compared to TCQ, it does not need to post-process a possibly massive set of intermediate results.

\section{Related Work}\label{sec:relwork}

Querying cohesive subgraph has a wide spectrum of applications in various domains, such as economy~\cite{gould1989structures, kondor2014rich, le2010application}, biology~\cite{ideker2008protein, tucker2001towards, typas2015bacterial}, and chemistry~\cite{nitschke2009molecular, han2008understanding, bray2003molecular, schadt2009molecular}. Take economy as an example. In a transaction network where each vertex represents an account and each edge represents a transaction between two accounts, finding a group of densely connected vertices can help to detect some criminal activities like money laundrying. Similarly, in a protein-to-protein network, a cohesive subgraph may reveal a group of proteins that participate in a common physical process.

To better capture the properties of graph, various cohesive subgraph models have been proposed. In most cases, a model is devised on top of a fundamental cohesive subgraph pattern, such as $k$-core~\cite{khaouid2015k, kong2019k, montresor2011distributed, sariyuce2013streaming, wang2022generalized}, $k$-truss~\cite{huang2014querying, zheng2017finding, chen2014distributed, sun2023efficient}, and $k$-clique~\cite{tsourakakis2015k, li2020ordering, gregori2012parallel}. Among them, $k$-core is widely adopted since it has properties like uniqueness and containment, and can be computed in linear time. The efficient retrieval of $k$-core based model has been widely studied, where most existing works focus on non-temporal graphs such as undirected graph~\cite{huang2014querying, bonchi2019distance, liu2021local, zhang2020exploring, fang2020survey, yao2021efficient}, directed graph~\cite{sozio2010community, ma2020efficient, chen2021efficient}, labeled graph~\cite{sun2020stable, li2015influential, dong2021butterfly}, attributed graph~\cite{islam2022keyword, huang2017attribute, matsugu2019flexible, fang2017effectiveattr}, spatial graph~\cite{fang2017effectivespat, fang2018spatial, zhu2017geo}, heterogeneous graph~\cite{jiang2022effective, zhou2023influential}, bipartite graph~\cite{wang2021efficient, zhang2021pareto, wang2021discovering, liu2019efficient}, etc. However, graphs are naturally equipped with time features, since each edge has a timestamp to indicate when it occurs. To fill the gap, a variety of $k$-core query problems~\cite{wu2015core, li2018persistent, galimberti2018mining, chu2019online, ma2019efficient, qin2020periodic, bai2020efficient, li2021efficient, yu2021querying, tang2022reliable} have also been studied on temporal graphs recently (see Section~\ref{sec:introduction}). 

Specifically, Wu et al~\cite{wu2015core} proposed $(k,h)$-core and studied its decomposition algorithm, which gives an additional constraint on the number of parallel edges between each pair of linked vertices in the $k$-core, namely, they should have at least $h$ parallel edges. Li et al~\cite{li2018persistent} proposed the persistent community search problem and gives a complicated instance called $(\theta,\tau)$-persistent $k$-core, which is a $k$-core persists over a time interval whose span is decided by the parameters. Similarly, Li et al~\cite{li2021efficient} proposed the continual cohesive subgraph search problem. Chu et al~\cite{chu2019online} studied the problem of finding the subgraphs whose density accumulates at the fastest speed, namely, the subgraphs with bursting density. Qin et al~\cite{qin2020periodic, qin2019mining} proposed the periodic community problem to reveal frequently happening patterns of social interactions, such as periodic $k$-core. Wen et al~\cite{bai2020efficient} relaxed the constraints of $(k,h)$-core and proposed quasi-$(k,h)$-core for better support of maintenance. Lastly, Ma et al~\cite{ma2019efficient} studied the problem of finding dense subgraph on weighted temporal graph. These works all focus on some specific patterns of cohesive substructure on temporal graphs, and propose sophisticated models and methods.

Compared with them, our work addresses a fundamental querying problem TCQ, which finds the most general $k$-cores on temporal graphs with respect to two basic conditions, namely, $k$ and time interval. Moreover, we extend TCQ to TXCQ such that queries under various user-defined conditions can be resolved in a unified framework, so that different query models can be potentially fulfilled by TXCQ.

\section{Conclusion}\label{sec:conc}

We propose TCQ and TXCQ as two general $k$-core query models on temporal graphs. Given a time range, TCQ returns all distinct $k$-cores that emerge during the period. Given another optimizing or constraining condition on a user-defined metric of $k$-core, TXCQ further returns the $k$-cores satisfying the condition. By introducing the crucial concepts such as TCD, TTI and LTI that reveal the laws of $k$-core evolution, we establish a unified algorithm framework with guaranteed scalability, which is a ``master key'' that opens the doors to various temporal $k$-core analysis tasks.

In the future, we will try to further relax the query models of TCQ and TXCQ, in order to improve the ease-of-use. For example, we would like to allow users to give such a query time range $[ts,ts']\sim[te,te']$, where both start time and end time fall in time ranges $[ts, ts’]$ and $[te, te’]$ respectively.



%
\bibliographystyle{IEEEtran}
\bibliography{tkde-ext.bib}





\end{document}